\documentclass[aps,prd,reprint,superscriptaddress,twocolumn,nofootinbib]{revtex4-2}

\usepackage{amsmath,amssymb,amsthm,amstext}
\usepackage{graphicx}
\usepackage[dvipsnames]{xcolor}
\usepackage{array, enumerate}
\usepackage{bm}
\usepackage{multirow}
\usepackage[breaklinks,colorlinks,citecolor=blue]{hyperref}
\usepackage{braket}
\usepackage{txfonts}
\usepackage{textcomp,gensymb}
\usepackage{enumitem}
\usepackage{orcidlink}
\usepackage[normalem]{ulem}

\graphicspath{{figure/}}

\def\be{\begin{equation}}
\def\ee{\end{equation}}

\usepackage{tikz}
\usetikzlibrary{shadows,arrows,positioning}

\pgfdeclarelayer{background}
\pgfdeclarelayer{foreground}
\pgfsetlayers{background,main,foreground}

\tikzset{
    etape/.style={
        rectangle, draw, thin, fill=white, text centered,
        rounded corners, drop shadow,
        inner sep=14pt,
        minimum width=3.2cm,
        minimum height=1.5cm,
        font=\small
    },
    line/.style={draw, thick, color=black!50, -latex'}
}

\newcommand{\etape}[2]{node (p#1) [etape] {#2}}
\newcommand{\UIUC}{Illinois  Center  for  Advanced  Studies  of  the  Universe \& Department of Physics, 
\\University of Illinois Urbana-Champaign, Urbana, Illinois 61801, USA}

\allowdisplaybreaks
\begin{document}
\title{Modified Teukolsky Formalism for Extreme Mass-Ratio Inspirals in Higher-Derivative Gravity}
\author{Chaoyi Yang }
\affiliation{Department of Astronomy, Tsinghua University, Beijing 100084, China}

\author{Neev Khera \orcidlink{0000-0003-3515-2859}}
\affiliation{Department of Astronomy, Tsinghua University, Beijing 100084, China}

\author{Dongjun Li \orcidlink{0000-0002-1962-680X}}
\thanks{Contact author: \href{dongjun@illinois.edu}{dongjun@illinois.edu}}
\affiliation{\UIUC}

\author{Huan Yang \orcidlink{0000-0002-9965-3030}}
\thanks{Contact author: \href{hyangdoa@tsinghua.edu.cn}{hyangdoa@tsinghua.edu.cn}} 
\affiliation{Department of Astronomy, Tsinghua University, Beijing 100084, China}

\begin{abstract}
In this work, we study a model problem involving a point particle spiraling into a non-rotating black hole in higher-derivative theories of gravity. In such theories, both the background spacetime and the generation and propagation of gravitational waves differ from those in General Relativity. We develop a modified Teukolsky formalism to describe gravitational waves sourced by the point particle and, as an illustrative example, compute the resulting fluxes to the black hole horizon and null infinity for a cubic gravity theory. The formalism is constructed in a way that can be naturally extended to rotating black holes. These results represent essential steps to build extreme mass-ratio-inspiral waveforms in modified gravity theories, which may also be rescaled to approximate waveforms from comparable-mass binary black hole systems, analogous to existing approaches in General Relativity.

\end{abstract}

\maketitle

\section{Introduction}
There is a long-standing challenge in gravitational-wave astronomy that the evolution equations of many modified theories of gravity fail to admit a well-posed initial value formulation in the strong-field, dynamical regime relevant for binary black hole mergers \cite{Papallo:2017qvl,Ripley:2019hxt,Papallo:2017ddx,Cayuso:2017iqc}, with only a few notable exceptions, including scalar-tensor theories and selected subclasses of Horndeski gravity \cite{Kovacs:2020pns,Papallo:2017qvl,East:2020hgw}. As a result, obtaining complete inspiral-merger-ringdown waveforms through direct numerical evolution of the nonlinear field equations remains challenging in these theories. Tests of gravity therefore often rely on post-Newtonian inspiral waveforms, ringdown waveforms with deformed quasinormal modes, or approximate inspiral-merger-ringdown models~\cite{Maenaut:2024oci,Liu:2024atc, Perkins:2021mhb}.

Various approaches have been proposed to ``cure'' the problem of ill-posedness, including the introduction of viscosity terms \cite{LattesLions1969}, damping unstable modes through Israel-Stewart-type approaches \cite{Israel1979}, imposing analytic restrictions \cite{Treves1967}, and applying spectral filters \cite{GottliebOrszag1977,Boyd2001}. These methods aim to stabilize the evolution by modifying the original equations of motion. However, validating such approaches still relies crucially on comparison with solutions of the underlying original theories.

In recent years, the surprising observation that several gauge-invariant quantities computed within gravitational self-force theory remain accurate even in the comparable-mass regime has motivated extensive efforts to use extreme mass-ratio inspiral (EMRI) and black-hole-perturbation-theory calculations to inform comparable-mass binary dynamics and waveforms \cite{Damour:2009sm, LeTiec:2011bk, Akcay:2012ea, Rifat:2019ltp}. The basic idea is to first study an EMRI system and expand the waveform in the symmetric mass ratio. Waveforms at different orders in the mass-ratio expansion can, in principle, be obtained by combining gravitational self-force calculations with the long-term evolution of the EMRI system, for example through a two-timescale formalism. Then, the small mass ratio appearing in the EMRI expansion can be substituted with the symmetric mass ratio of a comparable-mass binary to approximate the corresponding waveform. Practical implementations have shown that even the leading-order EMRI waveform can already reproduce comparable-mass binary waveforms with phase errors of only $\mathcal{O}(1)$ \cite{vandeMeent:2020xgc}. The precise reason why this mass-ratio expansion exhibits such remarkable convergence remains unclear, but it is recognized that the symmetric mass ratio seems to be a better re-summation parameter than the normal mass ratio \cite{vandeMeent:2020xgc}. Nevertheless, the success of this strategy has already motivated several EMRI- and black-hole-perturbation-theory-based approaches for generating waveforms of comparable-mass binary black holes \cite{Islam:2022laz, Wardell:2021fyy, Rink:2024swg}.

Because EMRI calculations only require perturbation theory around a single black hole background, rather than solving the full nonlinear two-body problem, this framework is particularly attractive for generating binary black hole waveforms in modified theories of gravity. Schematically, we may write the frequency-domain waveform as
\begin{align}\label{eq:phase}
h_{\rm bGR}(\omega)
=\sum_{i=1}^{\infty}\eta^{i} h^{(0,i)}_{\rm GR}(\omega)+\zeta\sum_{i=1}^{\infty}\eta^{i}h^{(1,i)}(\omega)
+\mathcal{O}(\zeta^2)\,,
\end{align}
where $\eta$ is the symmetric mass ratio, and $\zeta$ is some dimensionless coupling constant characterizing deviations from General Relativity (GR) at the level of the action. In this work, we focus on computing the leading-order beyond-GR correction $h^{(1,1)}(\omega)$ to the waveform in the EMRI limit by assuming $\eta=\epsilon\equiv m_p/M\ll 1$, where $M$ and $m_p$ are the masses of the central supermassive black hole and its companion. Considering that the current observations constrain $\zeta\ll 1$ \cite{LIGOScientific:2026qni}, the corrections at $\mathcal{O}(\zeta^n)$, $n\geq2$, potentially get highly suppressed and require substantially more complicated calculations involving gravitational self-force effects in modified gravity theories, so they are beyond the scope of the present work. Nevertheless, if the connection between the extreme and comparable mass-ratio regimes in GR still extends to beyond-GR theories, the leading-order term $h^{(1,1)}(\omega)$ in the above expansion may already capture the dominant part of the beyond-GR corrections to the waveform across a broad range of mass ratios, which we will examine in our future work.

One standard way to obtain the leading piece $h_{\rm GR}(\omega)$ for an EMRI system is to expand the Einstein equations in the mass ratio $\epsilon$ and adiabatically evolve the point particle, in which the full inspiral is decomposed into a sequence of geodesics \cite{Hughes:1999bq, Hinderer:2008dm}. Due to gravitational radiation, the point particle loses energy and angular momentum and spirals in, the former of which could be computed from the perturbations of two Weyl scalars $\Psi_{0}$ and $\Psi_{4}$, which are governed by Teukolsky equations for Kerr black holes in GR \cite{Teukolsky:1973ha, Press:1973zz, Teukolsky:1974yv} under the Newman-Penrose (NP) formalism \cite{Newman:1961qr}. Such a procedure can be iterated to second and higher orders in the mass ratio \cite{Campanelli:1998jv, Barack:2005nr, Barack:2007tm, Wardell:2021fyy, Bourg:2024vre, PanossoMacedo:2024pox}, and it was shown that the next-to-leading order correction in the mass ratio could be important for the waveform accuracy required by future space-based detectors, such as LISA, and modeling the comparable-mass mergers by exploiting the above connection of waveforms across mass ratios \cite{Wardell:2021fyy}. 

Despite all the progress being made in GR, such a procedure has had a hard time being extended to EMRI systems in beyond-GR theories or matter environments, mainly due to a lack of Teukolsky-like formalism for describing radiation reaction in these more complicated scenarios. For example, black holes in many beyond-GR theories might no longer be Petrov type D under Petrov classification \cite{Petrov:2000bs} and non-Ricci-flat \cite{Owen:2021eez}, which have been necessary conditions for deriving the Teukolsky equation in GR \cite{Teukolsky:1973ha}. This main obstacle has been overcome by the recent development of the \textit{modified Teukolsky formalism} (MTF) \cite{Li:2022pcy, Hussain:2022ins}, where a set of decoupled equations of $\Psi_{0}$ and $\Psi_{4}$ are found for gravitational perturbations of algebraically generically black hole spacetimes that perturbatively deviate from Petrov type D black holes. This extension is particularly useful for effective-field-theory (EFT) extensions of GR, such as higher-derivative gravity \cite{Cardoso:2018ptl, deRham:2020ejn, Cano:2019ore, Cano:2020cao, Cano:2021myl, Cano:2023tmv, Cano:2023jbk}, scalar/Einstein-dilaton Gauss-Bonnet gravity \cite{Alty:1994xj, Pani:2009wy, Blazquez-Salcedo:2016enn, Blazquez-Salcedo:2017txk, Chung:2024vaf}, and dynamical Chern-Simons gravity \cite{Jackiw:2003pm, Yunes:2007ss, Alexander:2009tp, Yunes:2009hc, Cardoso:2009pk, Molina:2010fb, Pani:2011xj, Wagle:2021tam, Srivastava:2021imr, Wagle:2023fwl, Li:2025fci}, and has been substantially applied to model black hole ringdown in some of these theories \cite{Cano:2023tmv, Cano:2023jbk, Li:2023ulk, Wagle:2023fwl, Li:2025fci, Aly:2026otj}. Despite its success in beyond-GR ringdown, the application of the MTF to EMRIs is preliminary, with recent examples only for EMRIs embedded in ultralight scalar clouds \cite{Li:2025ffh,Keijzer:2026vul} and ring-like structures \cite{Polcar:2025yto}. Here, we present the first example of extending the MTF to EMRIs in beyond-GR theories.

\begin{figure}[tb]
    \centering
    \includegraphics[width=\columnwidth]{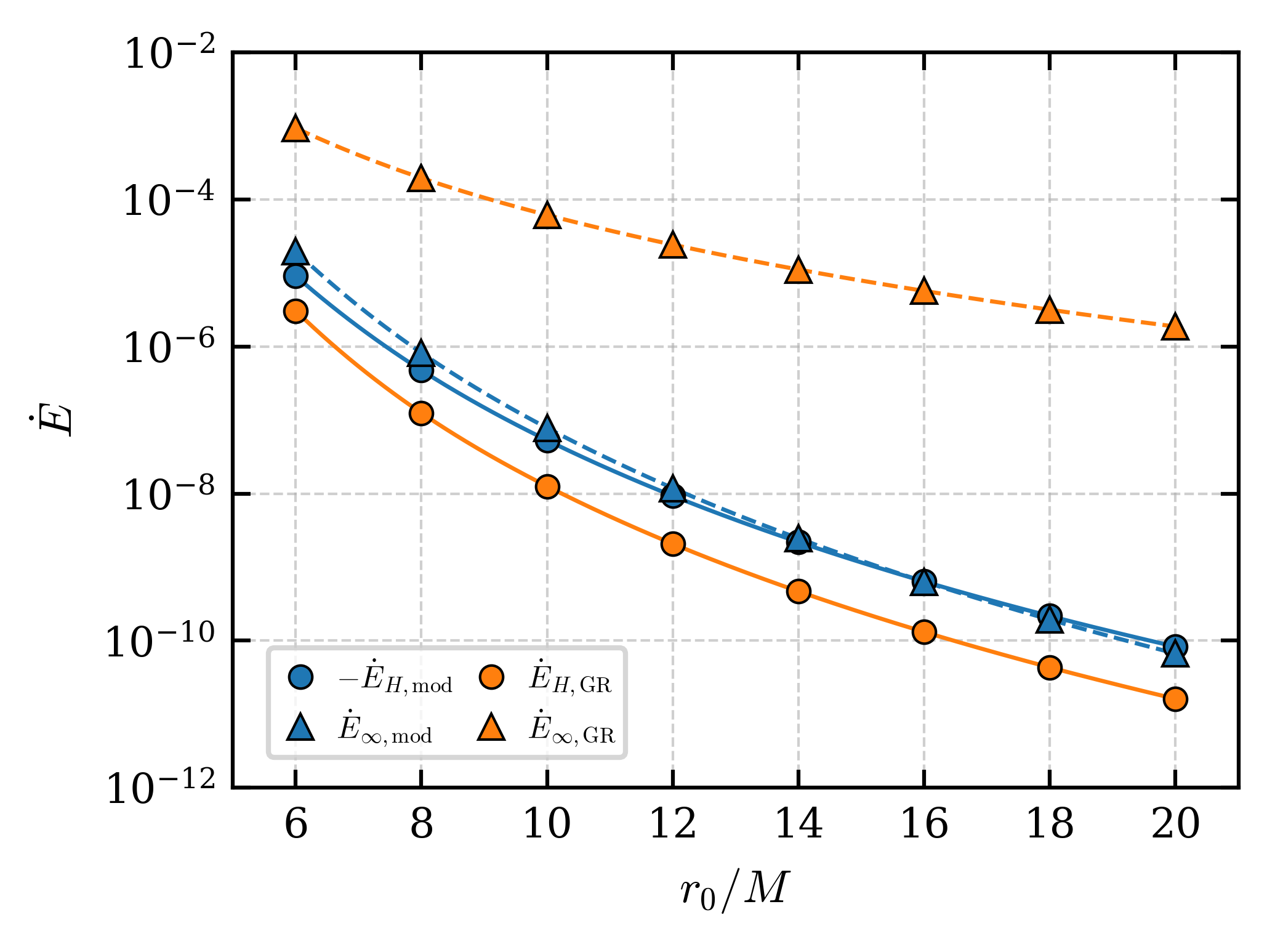}
    \caption{Comparison of the GR energy fluxes and the leading cubic-gravity corrections at the horizon ($H$) and at null infinity ($\infty$) for circular equatorial EMRIs around a non-rotating black hole, shown as functions of the orbital radius $r_0$. The overall factors $\eta^2$ and $\zeta\eta^2$ are removed from the GR and modified contributions, respectively. Circles connected by solid lines denote the horizon contributions, while triangles connected by dashed lines denote those at null infinity. The cubic-gravity correction to the horizon flux is negative throughout the displayed range, so $-\dot E_{H,\rm mod}$ is shown on the logarithmic scale.}
\label{fig:energy_flux}
\end{figure}

Specifically, we apply the MTF to an EMRI system consisting of a small compact object on a circular equatorial orbit around a non-rotating central black hole in parity-preserving cubic gravity \cite{Cano:2019ore}, following procedures similar to those developed in \cite{Li:2022pcy, Li:2025ffh, LaHaye:2025ley}. As described in detail below, the modified Teukolsky equations take the form of inhomogeneous Teukolsky equations sourced by the coupling between the beyond-GR correction to the background geometry and the GR gravitational perturbation generated by the secondary. For the former, we use the known background solution in \cite{Cano:2019ore}, supplemented by coordinate transformations that remove coordinate singularities from the source terms at the horizon and at infinity. To maintain this regularity, we work in the ingoing Eddington-Finkelstein coordinates and with the Hawking-Hartle tetrad \cite{HawkingHartle1972} throughout this work. For the latter, we use metric data generated with the codes developed in \cite{Barack:2005nr, Barack:2007tm, BHPToolkit}, which directly solve the Einstein equations for perturbations of a Schwarzschild black hole by a point particle. In extensions of this procedure to rotating black holes, one could, in principle, first solve the GR Teukolsky equations and then obtain the corresponding metric perturbation using metric-reconstruction techniques, such as those in \cite{Aksteiner:2016pjt, Wardell:2024yoi, Green:2019nam, Toomani:2021jlo, Bourg:2024vre, Hollands:2024iqp, Dolan:2021ijg, Dolan:2023enf, Nasipak:2025tby, Li:2026rkf}. Combining these two ingredients and performing the angular projection, we obtain two inhomogeneous second-order ordinary differential equations for $\Psi_0$ and $\Psi_4$ in the radial direction. We solve these equations using the Green's function techniques and regularize the relevant integrals with lower incomplete Gamma functions.

After obtaining the solutions of $\Psi_0$ and $\Psi_4$, we extend the method in \cite{HawkingHartle1972, Chandrasekhar_1983} for computing the horizon flux in GR to the parity-preserving cubic gravity. The method in \cite{HawkingHartle1972, Chandrasekhar_1983} essentially relates the energy flux across the horizon to the expansion of horizon area due to absorption of gravitational waves, the latter of which can be reconstructed from $\Psi_0$. As parity-preserving cubic gravity and many beyond-GR theories have nonzero effective stress-energy tensors and modify black hole geometry, the relations among the energy flux, horizon area change, and $\Psi_0$ get modified. For the energy flux at null infinity, similar arguments following \cite{Stein:2010pn} show that the Isaacson stress-energy tensor of gravitational waves \cite{Isaacson:1968zza} reduces to the GR one at null infinity, so the modified infinity flux is related to $\Psi_{4}$ in the standard way.

We find that, for positive $\zeta$, the leading cubic-gravity correction decreases the horizon flux and increases the flux at null infinity, as shown in Fig.~\ref{fig:energy_flux}. Relative to the corresponding GR flux, the correction is much larger at the horizon than at null infinity, although the two modified contributions have comparable absolute magnitudes and opposite signs. More broadly, our framework provides a systematic method for computing both horizon and infinity fluxes in a broad class of modified-gravity theories, paving the way for future EMRI waveform modeling beyond GR and for extensions of these waveforms toward comparable mass ratios.

This paper is organized as follows. In Sec.~\ref{sec:EMRI dynamics}, we briefly review the adiabatic evolution of an EMRI system. In Sec.~\ref{sec:cubic gravity}, we briefly review the EFT extensions of GR, including the parity-preserving cubic gravity considered in this work. In Sec.~\ref{sec:MTF}, we then show how to calculate the modified Teukolsky equations in this particular theory for a non-rotating central black hole with a companion in the circular equatorial orbits, the procedure of which could be directly extended to rotating black holes with more generic orbits. Once the equations are obtained, we show in Sec.~\ref{sec:solve equation} how to handle the high-order derivative terms in the source and regularize the integrals when solving the equations with Green's function techniques. With the solution at hand, we calculate the energy fluxes to the horizon and null infinity in Sec.~\ref{sec:gravitational wave flux}. Finally, Sec.~\ref{sec:conclusion} summarizes our work and discusses future avenues.  

Throughout this paper, we work in four spacetime dimensions and use
units with $G=c=1$, with metric signature $(-,+,+,+)$. Apart from the signature, all NP quantities follow the notation and conventions of Chandrasekhar~\cite{Chandrasekhar_1983}.

\section{EMRI dynamics}\label{sec:EMRI dynamics}

EMRIs exhibit a strong separation between the orbital timescale and the radiation-reaction timescale. This hierarchy underlies the adiabatic approximation and, more generally, the two-timescale description of EMRI evolution \cite{Poisson:1993vp, Cutler:1993vq, Ryan:1995wh, Tanaka:1996lfd, Hinderer:2008dm, Hughes:2016xwf, Hughes:2021exa}. In a generic Kerr spacetime, the motion of the secondary is characterized by three constants of motion: orbital energy $E_p$, azimuthal angular momentum $L_z$, and Carter constant $Q$ \cite{Hughes:1999bq, Mino:1996nk}. Radiation reaction drives the slow evolution of these quantities and hence determines the inspiral trajectory and the accumulated gravitational-wave phase.

In this work, we focus on circular equatorial orbits around a non-rotating black hole. In this case, the Carter constant is trivial, and the energy evolution determines the angular momentum evolution via
\begin{align}
    \dot{E}_p=\omega_z\dot{L}_z\,,
\end{align}
where $\dot{E}_p\equiv dE_p/dt$, and $\omega_z$ is the orbital frequency of the particle, so the inspiral can be completely determined from the energy evolution alone. Furthermore, energy conservation relates the change in the secondary's orbital energy to the energy carried away by the gravitational radiation it sources through the balance law
\begin{align}
    \dot{E}_p=-(\dot{E}_{\infty}+\dot{E}_{H})\,,
\end{align}
where $\dot{E}_{H}$ and $\dot{E}_{\infty}$ denote the gravitational-wave energy fluxes into the horizon and out to null infinity, respectively. For a circular orbit, the gravitational-wave frequency $\omega_{\ell m}$ of each $(\ell,m)$ mode is completely determined by the secondary's orbital frequency $\omega_z$, i.e., $\omega_{\ell m}=m\omega_{z}$, while $\omega_z$ is determined by the secondary's orbital radius $r_0$ via $\omega_z=\sqrt{M/r_0^3}$. Thus, to the leading order in the adiabatic evolution, the evolution of the gravitational-wave phase satisfies
\begin{align} 
    \phi_{\ell m}
    =-m\int\omega_z\frac{dE_p/dr}{\dot{E}_{\infty}+\dot{E}_{H}}dr\,, \label{eq:phase_evolution}
\end{align}

In GR, the quantity $dE_p/dr$ can be obtained from the geodesic equation, while the energy fluxes are determined by the Weyl scalar perturbations $\Psi_{0,4}^{(0,1)}$
\cite{Chandrasekhar_1983, Teukolsky:1974yv, HawkingHartle1972}, i.e.,
\begin{subequations} \label{eq:GR_fluxes}
\begin{align} 
    \frac{d^2 E_H}{d\Omega dt} 
    &=\frac{M}{8\pi\varepsilon_H} 
    \left|\frac{1}{i\omega_{\ell m}+2\varepsilon_H}\Psi_0^{(0,1)}\right|^2\,, \\
    \frac{d^2 E_\infty}{d\Omega dt} 
    &=\frac{r^2}{4\pi\omega_{\ell m}^2}\left|\Psi_4^{(0,1)}\right|^2\,,
\end{align}    
\end{subequations}
where $\varepsilon_H$ is the value of the spin coefficient $\varepsilon$ on the horizon of a Kerr black hole.

Our goal in this work is to modify the procedure above for evolving an EMRI system in some generic beyond-GR theories, and we take the parity-preserving cubic gravity as an example. As the first step, we will calculate the modifications to the secondary's orbital energy and the gravitational-wave energy fluxes in this particular theory, which can be schematically expanded as
\begin{align}
    E_p 
    &= \eta E_{p,\rm GR} +\eta \zeta E_{p,\rm mod}+\mathcal{O}(\eta^2 \zeta)\,, \nonumber \\
    \dot{E}_{\infty} 
    &=\eta^2\dot{E}_{\infty,\rm GR}+\eta^2\zeta\dot{E}_{\infty,\rm mod}
    +\mathcal{O}(\eta^3 \zeta)\,, \nonumber \\
    \dot{E}_{H} 
    &=\eta^2\dot{E}_{H,\rm GR}+\eta^2\zeta\dot{E}_{H,\rm mod}
    +\mathcal{O}(\eta^3\zeta)\,,
\end{align}
where $E_{p,\rm mod}$, $\dot{E}_{H,\rm mod}$, and $\dot{E}_{\infty,\rm mod}$ all contribute to the leading beyond-GR waveform correction $h^{(1,1)}$ in Eq.~\eqref{eq:phase}. To get $\dot{E}_{H,\rm mod}$ and $\dot{E}_{\infty,\rm mod}$, one can observe from Eq.~\eqref{eq:GR_fluxes} that we likely need to compute the perturbations to the Weyl scalars $\Psi_{0,4}$ in the chosen beyond-GR theory. For this reason, we will apply the MTF developed in \cite{Li:2022pcy, Li:2025fci, LaHaye:2025ley} to derive a set of modified Teukolsky equations for $\Psi_{0,4}$ in Sec.~\ref{sec:MTF} and solve them using Green's function techniques in Sec.~\ref{sec:solve equation}. Furthermore, the precise relation between the energy fluxes and $\Psi_{0,4}$ in Eq.~\eqref{eq:GR_fluxes} will also get modified, as we will show in detail in Sec.~\ref{sec:gravitational wave flux}. In the next section, let us first introduce the beyond-GR theory we focus on in this work.

\section{Effective field theory extensions of GR and cubic gravity} \label{sec:cubic gravity}

In this section, we introduce the EFT extensions of GR and a specific case of this class: the parity-preserving cubic gravity. Within the EFT framework, deviations from GR are described by higher-curvature operators compatible with diffeomorphism invariance. Following Ref.~\cite{Cano:2019ore}, we consider the EFT action
\begin{widetext}
\begin{align} \label{eq:EFTaction}
\begin{split}
    S=\frac{1}{16\pi }\int d^4x\sqrt{|g|}
    \Bigg\{&R +\alpha_{1}\phi_{1}l_c^{2}\mathcal{X}_{4}+\alpha_{2}
    \left(\phi_{2}\cos\theta_{m}+\phi_{1}\sin\theta_{m}\right)
    l_c^{2}R_{\mu\nu\rho\sigma}\tilde{R}^{\mu\nu\rho\sigma}+\lambda_{\rm ev}l_c^{4}R_{\mu\nu}{}^{\rho\sigma} R_{\rho\sigma}{}^{\delta\gamma}R_{\delta\gamma}{}^{\mu\nu}\\&+\lambda_{\rm odd}l_c^{4}R_{\mu\nu}{}^{\rho\sigma}R_{\rho\sigma}{}^{\delta\gamma}\tilde{R}_{\delta\gamma}{}^{\mu\nu}-\frac{1}{2}(\partial\phi_{1})^{2}-\frac{1}{2}(\partial\phi_{2})^{2}\Bigg\}\,,
\end{split}
\end{align}
\end{widetext}
where $R_{\mu\nu\alpha\beta}$ is the Riemann tensor, $\tilde{R}_{\mu\nu\alpha\beta}\equiv\epsilon_{\mu\nu\rho\sigma}
R^{\rho\sigma}{}_{\alpha\beta}/2$ is its dual, and 
\begin{align} \label{eq:GBdensity}
    \mathcal{X}_{4}
    =R_{\mu\nu\rho\sigma}R^{\mu\nu\rho\sigma}
    -4R_{\mu\nu}R^{\mu\nu}+R^{2}\,,
\end{align}
is the Gauss-Bonnet invariant, while $\phi_{1}$ and $\phi_{2}$ are dynamical scalar fields. The constants $\alpha_{1,2}$, $\theta_m$, $\lambda_{\rm ev}$, and $\lambda_{\rm odd}$ parameterize different sectors of the EFT corrections, and $l_c$ characterizes the length scale associated with new physics. Here, we also ignore any self-interactions of the scalar fields.

The EFT framework provides a systematic way to parametrize deviations from GR in the strong-field regime, making EMRIs an ideal laboratory for testing such corrections through their long-lived gravitational-wave signals. Along this line, previous work studied gravitational radiation from EMRIs with a slowly rotating central black hole in dynamical Chern-Simons gravity [i.e., retaining only $\alpha_2$ and setting $\theta_m=0$ in Eq.~\eqref{eq:EFTaction}] \cite{Sopuerta:2009iy}, using a semi-relativistic geodesic-limit treatment that did not include radiation reaction. For non-rotating black holes in dynamical Chern-Simons gravity, Ref.~\cite{Pani:2011xj} developed a fully relativistic treatment by deriving the modified Regge-Wheeler, Zerilli, and scalar perturbation equations; the resulting horizon and infinity fluxes were then used to estimate the accumulated dephasing. Later, Refs.~\cite{Maselli:2020zgv, Barsanti:2022ana, Spiers:2023cva} showed that in EFT extensions of GR with a non-minimally coupled scalar field, the dimensionless coupling has an intrinsic mass-ratio dependence for EMRIs. As a result, at leading adiabatic order, most of the observable effects are captured by an additional scalar-radiation channel sourced by the secondary, whose strength is controlled by the scalar charge dressing the secondary. More recently, Ref.~\cite{Roy:2025kra} extended the treatment in \cite{Maselli:2020zgv, Barsanti:2022ana, Spiers:2023cva} to a complete self-force framework, incorporating for the first time the effects on the plunge and transition to plunge.

Despite all the progress made, the approaches used by most of the previous studies only apply in the semi-relativistic limit or for non-rotating (and possibly slowly rotating) black holes. The self-force approach developed in \cite{Maselli:2020zgv, Barsanti:2022ana, Spiers:2023cva, Roy:2025kra} should, in principle, be extendable to rotating black holes and generic orbits. However, when using such an approach to explore possible connections between extreme and comparable mass-ratio waveforms, some care may be needed: the EFT power counting adopted in these works is naturally tailored to the small mass-ratio regime, where higher-curvature corrections can be organized as subleading in the mass ratio, while the EFT coupling need not enter with the same mass-ratio suppression in the comparable mass-ratio regime. Since our ultimate goal is to extend the EMRI-limit description developed in this work to comparable-mass-ratio mergers in EFT extensions of GR, we will not assume any degeneracy between the beyond-GR coupling $\zeta$ and the symmetric mass ratio $\eta$. It is therefore natural to perform a two-parameter expansion of the field equations, whose EMRI limit is well captured by the MTF \cite{Li:2022pcy, Li:2025ffh, LaHaye:2025ley}. Furthermore, our MTF-based framework can be naturally extended to rotating black holes and generic orbits in a broad class of EFT extensions of GR. 

As a concrete example, we consider the parity-preserving cubic-curvature sector of the EFT extension of GR in Eq.~\eqref{eq:EFTaction} by setting $\alpha_1=\alpha_2=\lambda_{\rm odd}=0$ and ignoring all the scalar fields, such that the action becomes
\begin{equation} \label{eq:action}
    S=\frac{1}{16\pi}\int d^4x\,\sqrt{|g|}\left(R+\lambda_{\rm ev}l_c^4 
    R_{\mu\nu}{}^{\rho\sigma}R_{\rho\sigma}{}^{\alpha\beta}R_{\alpha\beta}{}^{\mu\nu}\right)-m_p\int d\tau\,,
\end{equation}
which constitutes one of the simplest EFT extensions of GR involving cubic curvature corrections. We have also included the point-particle action in the last term of Eq.~\eqref{eq:action}. In the following, we refer to this theory simply as \emph{cubic gravity} or parity-preserving cubic gravity. Varying the action in Eq.~\eqref{eq:action} yields the modified Einstein equation:
\begin{equation} \label{eq:mod_einstein}
    G_{\mu\nu}=8\pi T^{p}_{\mu\nu}+T^{\rm cubic}_{\mu\nu}\,,
\end{equation}
 where $G_{\mu\nu}$ is the Einstein tensor, $T^{p}_{\mu\nu}$ is the stress-energy tensor of a point particle, and the effective stress-energy tensor $T^{\rm cubic}_{\mu\nu}$ associated with the cubic-gravity correction is
\begin{equation} \label{eq:T_cubic}
\begin{aligned}
    T^{\rm cubic}_{\mu\nu}
    =& \;\lambda_{\rm ev}l_c^4
    \Bigl[3 R_{\mu}{}^{\sigma\alpha\beta} R_{\alpha\beta}{}^{\rho\lambda} 
    R_{\rho\lambda\sigma\nu}
    + \frac{1}{2} g_{\mu\nu} R_{\alpha\beta}{}^{\rho\sigma} 
    R_{\rho\sigma}{}^{\delta\gamma} R_{\delta\gamma}{}^{\alpha\beta} \\
    & \;-6 \nabla^{\alpha}\nabla^{\beta} 
    \bigl(R_{\mu\alpha}{}^{\rho\lambda} R_{\nu\beta\rho\lambda}\bigr)
    \Bigr]\,.
\end{aligned}
\end{equation}

For the non-rotating black hole considered in this work, the corresponding spacetime metric takes the form \cite{Cano:2019ore}:
\begin{align} \label{eq:bg_metric_original}
\begin{split}
    ds^2=& \;-f(r)(1-\zeta H_1(r))dt^2+f^{-1}(r)(1+\zeta H_3(r))dr^2 \\
    & \;+r^2(1+\zeta H_3(r))d\Omega^2\,,
\end{split}
\end{align}
where we have defined
\begin{equation}
    \zeta\equiv\lambda_{\mathrm{ev}}l_c^4/M^4\,,
\end{equation}
for this particular theory, $d\Omega^2$ is the solid-angle element, and $H_{1,3}(r)$ encode the leading-order deviations from the Schwarzschild geometry induced by the cubic-curvature interaction. The explicit forms of $H_{1,3}(r)$ are given by
\begin{widetext}
\begin{align}
    & H_1(r)=\frac{8 M\left(63 M^5+35 M^4 r+20 M^3 r^2+12 M^2 r^3+8 M r^4+8 r^5\right)}{231 r^6}\,, \\
    & H_3(r)=-\frac{8 \left(1029 M^6-7 M^5 r-5 M^4 r^2-4 M^3 r^3-4 M^2 r^4-8 M r^5+8 r^6\right)}{231 r^6}\,.
\end{align}
\end{widetext}
Although we only focus on non-rotating black holes in the parity-preserving cubic gravity in this work as a simple demonstration of our framework, the geometry of rotating black holes in this theory has been studied in \cite{Cano:2019ore, Cano:2023tmv, Cano:2023jbk} with a high-order slow-rotation expansion. For black holes with general spin, one can also employ these spectral or pseudospectral approaches in \cite{Lam:2025fzi, Lam:2025elw, Fernandes:2025vxg} for solving modified black hole geometries.

With the theory prescribed and the expansion parameters defined, we then introduce a two-parameter expansion of the NP quantities in $\zeta$ and $\eta$:
\begin{align} \label{eq:psi_expansion}
    \Psi=\Psi^{(0,0)}+\zeta\Psi^{(1,0)}+\eta\Psi^{(0,1)} 
    +\zeta\eta\Psi^{(1,1)}+\cdots\,,
\end{align}
where the symmetric mass ratio $\eta$ becomes the standard mass ratio $m_p/M$ in the case of EMRIs. The field $\Psi$ stands schematically for any of the NP quantities. The first and second superscript indices refer to the order in the beyond-GR correction and the mass ratio, respectively. For example, $\Psi^{(0,0)}$ represents NP quantities evaluated on a Schwarzschild or Kerr background in GR, while $\Psi^{(1,0)}$ are background corrections in cubic gravity. $\Psi^{(0,1)}$ is associated with the gravitational radiation driven by the secondary in GR. For simplicity, we will omit the superscript $(0,0)$ of all the quantities at $\mathcal{O}(\zeta^0,\eta^0)$ in expressions with perturbative expansion in the remainder of this work. We aim to calculate the corrections to gravitational radiation represented by $\Psi_0^{(1,1)}$ and $\Psi_4^{(1,1)}$ in this work, from which we can extract the corrections to the energy fluxes in the parity-preserving cubic gravity. In the next section, we will then derive the modified Teukolsky equations governing $\Psi_0^{(1,1)}$ and $\Psi_4^{(1,1)}$ following the prescription in \cite{Li:2022pcy, Li:2025ffh, LaHaye:2025ley}.

\section{Modified Teukolsky formalism} \label{sec:MTF}

In this section, we first review the EMRI-adapted MTF developed in Refs.~\cite{Li:2022pcy, Li:2025ffh, LaHaye:2025ley}. We then derive the modified Teukolsky equations of the Weyl scalar perturbations $\Psi_{0,4}^{(1,1)}$ for EMRIs in parity-preserving cubic gravity. Under the MTF, $\Psi_{0,4}^{(1,1)}$ are driven by the deformation of the central black hole's geometry $h_{\mu\nu}^{(1,0)}$ and the gravitational radiation in GR  $h_{\mu\nu}^{(0,1)}$, so we show in detail how to calculate these two types of terms. Finally, we perform an angular projection of the resulting equations and present the inhomogeneous second-order ordinary differential equations governing $\Psi_{0,4}^{(1,1)}$ along the radial direction. Since the equations of $\Psi_{0}^{(1,1)}$ and $\Psi_{4}^{(1,1)}$ are largely symmetric, we will focus on $\Psi_{0}^{(1,1)}$ and present the result of $\Psi_{4}^{(1,1)}$ at the end of this section.

\subsection{The modified Teukolsky equations}
\label{sec:Teukolsky equation}

\begin{figure*}[htbp]
    \centering
    \begin{tikzpicture}[scale=1.0, transform shape]

  \path (-3.5,0) \etape{1}{$ H_0\Psi_0^{(1,1)}=\mathcal{S}_{geo}^{(1,1)}+\mathcal{S}^{(1,1)}$};

  \path (p1.south)+(-6.0,-1.8) \etape{2}{$ H_0\Psi_0^{(1,1)}$};
  \path (p1.south)+(-1.5,-1.8) \etape{3}{$\mathcal{S}_{geo}^{(1,1)}=-H_0^{(1,0)}\Psi_0^{(0,1)}$};
  \path (p1.south)+(3.5,-1.8)  \etape{4}{$\mathcal{S}^{(1,1)}$};

  \path (p4.south)+(-5.5,-1.5) \etape{5}{$\mathcal{E}_2S_2^{(1,1)}-\mathcal{E}_1S_1^{(1,1)}$};
  \path (p4.south)+(-0.5,-1.5) \etape{6}{$\mathcal{E}_2^{(1,0)}S_2^{(0,1)}-\mathcal{E}_1^{(1,0)}S_1^{(0,1)}$};
  \path (p4.south)+(4.5,-1.5)  \etape{7}{$\mathcal{E}_2^{(0,1)}S_2^{(1,0)}-\mathcal{E}_1^{(0,1)}S_1^{(1,0)}$};

  \path (p5.south)+(-4.5,-1.3) \etape{8}{$\mathcal{E}_2S_{2A}^{(1,1)}-\mathcal{E}_1S_{1A}^{(1,1)}$};
  \path (p5.south)+(0.0,-1.3)  \etape{9}{$ \mathcal{E}_2S_{2B}^{(1,1)}-\mathcal{E}_1S_{1B}^{(1,1)}$};
  \path (p5.south)+(4.5,-1.3)  \etape{10}{$ \mathcal{E}_2S_{2C}^{(1,1)}-\mathcal{E}_1S_{1C}^{(1,1)}$};

  \path (p7.south)+(0.0,-1.5)  \etape{11}{$\mathcal{E}_2^{(0,1)}S_2^{(1,0)}-\mathcal{E}_1^{(0,1)}S_1^{(1,0)}=0$};

  \path [line] (p1.south) -- (p2);
  \path [line] (p1.south) -- (p3);
  \path [line] (p1.south) -- (p4);

  \path [line] (p4.south) -- (p5);
  \path [line] (p4.south) -- (p6);
  \path [line] (p4.south) -- (p7);

  \path [line] (p5.south) -- (p8);
  \path [line] (p5.south) -- (p9);
  \path [line] (p5.south) -- (p10);

  \draw[->] (p7.south) to node[right] {$S_2^{(1,0)}=S_1^{(1,0)}=0$} (p11.north);

\end{tikzpicture}
    \caption{Hierarchical structure of the source terms for the modified Teukolsky equation of $\Psi_0^{(1,1)}$.}
    \label{fig:flowchart of psi0}
\end{figure*}
It was first found in \cite{Li:2022pcy, Hussain:2022ins} that for beyond-GR theories that are EFT extensions of GR, the Weyl scalar perturbations $\Psi_{0,4}^{(1,1)}$ in the expansion of Eq.~\eqref{eq:psi_expansion} satisfy the following modified Teukolsky equations
\begin{equation} \label{eq:modified_teuk_eq}
    H_0 \Psi_0^{(1,1)}
    =\mathcal{S}_{\rm geo}^{(1,1)}+\mathcal{S}^{(1,1)}\,,
\end{equation}
where $\Psi_{i}$, $i\in\{0,\cdots,4\}$, are Weyl scalars, and
\begin{align}
    & \mathcal{S}_{\rm geo}^{(1,1)}
    =-H_0^{(1,0)}\Psi_0^{(0,1)}
    -H_0^{(0,1)}\Psi_0^{(1,0)}
    +H_1^{(0,1)}\Psi_1^{(1,0)}\,, \label{eq:S_geo}\\
   & \mathcal{S}^{(1,1)}
    =\left[\mathcal{E}_2S_{2}-\mathcal{E}_1S_{1}\right]^{(1,1)}\,, \label{eq:S}
\end{align}
with the operators $H_{0}$, $H_1$, $\mathcal{E}_1$, and $\mathcal{E}_2$ defined as
\begin{equation} \label{eq:def_operators}
    \begin{aligned}
        & H_0=\mathcal{E}_2F_2-\mathcal{E}_1F_1-3\Psi_2\,, \\
        & H_1=\mathcal{E}_2J_2-\mathcal{E}_1J_1\,, \\
        & \mathcal{E}_1=E_1-\Psi_2^{-1}\delta\Psi_2\,,\quad
        \mathcal{E}_2=E_2-\Psi_2^{-1}D\Psi_2\,,
    \end{aligned}
\end{equation}
and
\begin{align} \label{eq:auxiliary_operators}
    & F_1\equiv\bar{\delta}_{[-4,0,1,0]}\,,
    && F_2\equiv\Delta_{[1,0,-4,0]}\,, \nonumber\\
    & J_1\equiv D_{[-2,0,-4,0]}\,,
    && J_2\equiv\delta_{[0,-2,0,-4]}\,, \nonumber\\
    & E_1\equiv\delta_{[-1,-3,1,-1]}\,,
    && E_2\equiv D_{[-3,1,-1,-1]}\,.
\end{align}
Here, we have adopted the convenient notations in \cite{Wagle:2023fwl}:
\begin{subequations}
\begin{align}
    & D_{[a,b,c,d]}
    \equiv D+a\varepsilon+b\bar{\varepsilon}
    +c\rho+d\bar{\rho}\,, \\
    & \Delta_{[a,b,c,d]}
    \equiv \Delta+a\mu+b\bar{\mu}+c\gamma+d\bar{\gamma}\,, \\
    & \delta_{[a,b,c,d]}
    \equiv \delta+a\bar{\alpha}+b\beta
    +c\bar{\pi}+d\tau\,, \\
    & \bar{\delta}_{[a,b,c,d]}
    \equiv \bar{\delta}+a\alpha+b\bar{\beta}
    +c\pi+d\bar{\tau}\,,
\end{align}    
\end{subequations}
where $\{a,b,c,d\}$ are integers, $\{D,\Delta,\delta,\bar{\delta}\}$ are directional derivatives along the null NP tetrad $e^{\mu}_{a}=\{l^{\mu},n^{\mu},m^{\mu},\bar{m}^{\mu}\}$, and the other quantities, such as $\{\varepsilon,\rho,\cdots\}$, are NP spin coefficients. The left-hand side of Eq.~\eqref{eq:modified_teuk_eq} is the homogeneous Teukolsky equation in GR \cite{Teukolsky:1973ha}, so Eq.~\eqref{eq:modified_teuk_eq} is essentially an inhomogeneous Teukolsky equation with a complicated source driven by background geometry corrections $h_{\mu\nu}^{(1,0)}$ and gravitational radiation $h_{\mu\nu}^{(0,1)}$ in GR. Furthermore, the source  terms $S_{1,2}$ are defined as
\begin{subequations} \label{eq:source_bianchi}
\begin{align}
\label{eq:source_bianchi_1}
\begin{split} 
    S_1\equiv& \;\delta_{[-2,-2,1,0]}\Phi_{00}
    -D_{[-2,0,0,-2]}\Phi_{01} \\
    & \;+2\sigma\Phi_{10}-2\kappa\Phi_{11}-\bar{\kappa}\Phi_{02}\,,
\end{split} \\
\label{eq:source_bianchi_2}
\begin{split} 
    S_2\equiv& \;\delta_{[0,-2,2,0]}\Phi_{01}
    -D_{[-2,2,0,-1]}\Phi_{02} \\
    & \;-\bar{\lambda}\Phi_{00}
    +2\sigma\Phi_{11}-2\kappa\Phi_{12}\,,
\end{split}
\end{align}
\end{subequations}
where $\Phi_{ij}$, $i,j\in\{0,1,2\}$ are NP Ricci scalars. An analogous equation for $\Psi_4^{(1,1)}$ can be found in Appendix~\ref{app:Sourceterms}. For a review of the NP formalism and the definition of all the NP quantities, we refer the reader to \cite{Newman:1961qr, Chandrasekhar_1983, Pound:2019lzj, Loutrel:2020wbw, Li:2022pcy}. 

As we will show in the Appendix~\ref{app:Sourceterms} , non-rotating black holes in parity-preserving cubic gravity are Petrov type D, i.e., $\Psi_{0,1,3,4}^{(1,0)}=0$. Thus, the last two terms in Eq.~\eqref{eq:S_geo} vanish, so $\mathcal{S}_{\rm geo}^{(1,1)}=-H_0^{(1,0)}\Psi_0^{(0,1)}$. Furthermore, from the stress-energy tensor $T^{\rm cubic}_{\mu\nu}$ in Eq.~\eqref{eq:T_cubic}, one can directly find that the only nonzero NP Ricci scalars are $\Phi_{00}^{(1,0)}$, $\Phi_{11}^{(1,0)}$, $\Phi_{22}^{(1,0)}$, and $\Lambda^{(1,0)}$, and they are purely radial. This further simplifies the source term $\mathcal{S}^{(1,1)}$ in Eq.~\eqref{eq:S} driven by the stress energy tensor, i.e., $S_{1,2}^{(1,0)}=0$. In total, for non-rotating BHs in parity-preserving cubic gravity, Eqs.~\eqref{eq:S_geo} and \eqref{eq:S} reduce to
\begin{align}
    & \mathcal{S}_{\rm geo}^{(1,1)}
    =-H_0^{(1,0)}\Psi_0^{(0,1)}\,, \label{eq:S_geo_nonrot}\\
     & \mathcal{S}^{(1,1)}
    =\mathcal{E}_2S_{2}^{(1,1)}
    -\mathcal{E}_1S_{1}^{(1,1)}
    +\mathcal{E}_2^{(1,0)}S_{2}^{(0,1)}
    -\mathcal{E}_1^{(1,0)}S_{1}^{(0,1)}\,, \label{eq:S_nonrot}
\end{align}
as also summarized in Fig.~\ref{fig:flowchart of psi0}. For convenience, we further decompose $S_{1,2}^{(1,1)}$ into three pieces based on their dependence on $\Phi_{ij}$: $\{S_{1A}^{(1,1)},S_{2A}^{(1,1)}\}$, $\{S_{1B}^{(1,1)},S_{2B}^{(1,1)}\}$, and $\{S_{1C}^{(1,1)},S_{2C}^{(1,1)}\}$ are driven by $\Phi_{ij}^{(1,0)}$, $\Phi_{ij}^{(0,1)}$, and $\Phi_{ij}^{(1,1)}$, respectively. We can analyze the equation for $\Psi_4$ in an analogous manner, as illustrated in Fig.~\ref{fig:flowchart of psi4}. The complete expressions for all source terms appearing in Eq.~\eqref{eq:S_nonrot} and \eqref{eq:S_geo_nonrot} in terms of $\Phi_{ij}$ are provided in Appendix~\ref{app:Sourceterms}. For completeness, we also present the corresponding source terms for $\Psi_4^{(1,1)}$. As one can directly observe from the results in Appendix~\ref{app:Sourceterms}, all the source terms here only involve geometrical quantities (i.e., tetrad, spin coefficients, and Weyl scalars) at $\mathcal{O}(\zeta^1,\eta^0)$ and $\mathcal{O}(\zeta^0,\eta^1)$, which we will compute in the next subsection.

\begin{figure*}[ht]
    \centering
    \begin{tikzpicture}[scale=1.0, transform shape]

  \path (-3.5,0) \etape{1}{$ H_4\Psi_4^{(1,1)}=\mathcal{T}_{geo}^{(1,1)}+\mathcal{T}^{(1,1)}$};

  \path (p1.south)+(-6.0,-1.8) \etape{2}{$ H_4\Psi_4^{(1,1)}$};
  \path (p1.south)+(-1.5,-1.8) \etape{3}{$\mathcal{T}_{geo}^{(1,1)}=-H_4^{(1,0)}\Psi_4^{(0,1)}$};
  \path (p1.south)+(3.5,-1.8)  \etape{4}{$\mathcal{T}^{(1,1)}$};

  \path (p4.south)+(-5.5,-1.5) \etape{5}{$\mathcal{E}_4S_4^{(1,1)}-\mathcal{E}_3S_3^{(1,1)}$};
  \path (p4.south)+(-0.5,-1.5) \etape{6}{$\mathcal{E}_4^{(1,0)}S_4^{(0,1)}-\mathcal{E}_3^{(1,0)}S_4^{(0,1)}$};
  \path (p4.south)+(4.5,-1.5)  \etape{7}{$\mathcal{E}_4^{(0,1)}S_4^{(1,0)}-\mathcal{E}_3^{(0,1)}S_3^{(1,0)}$};

  \path (p5.south)+(-4.5,-1.3) \etape{8}{$\mathcal{E}_4S_{4A}^{(1,1)}-\mathcal{E}_3S_{3A}^{(1,1)}$};
  \path (p5.south)+(0.0,-1.3)  \etape{9}{$ \mathcal{E}_4S_{4B}^{(1,1)}-\mathcal{E}_3S_{3B}^{(1,1)}$};
  \path (p5.south)+(4.5,-1.3)  \etape{10}{$ \mathcal{E}_4S_{4C}^{(1,1)}-\mathcal{E}_3S_{3C}^{(1,1)}$};

  \path (p7.south)+(0.0,-1.5)  \etape{11}{$\mathcal{E}_4^{(0,1)}S_4^{(1,0)}-\mathcal{E}_3^{(0,1)}S_3^{(1,0)}=0$};

  \path [line] (p1.south) -- (p2);
  \path [line] (p1.south) -- (p3);
  \path [line] (p1.south) -- (p4);

  \path [line] (p4.south) -- (p5);
  \path [line] (p4.south) -- (p6);
  \path [line] (p4.south) -- (p7);

  \path [line] (p5.south) -- (p8);
  \path [line] (p5.south) -- (p9);
  \path [line] (p5.south) -- (p10);

  \draw[->] (p7.south) to node[right] {$S_4^{(1,0)}=S_3^{(1,0)}=0$} (p11.north);

\end{tikzpicture}
    \caption{Hierarchical structure of the source terms for the modified Teukolsky equation of $\Psi_4^{(1,1)}$.}
    \label{fig:flowchart of psi4}
\end{figure*}

\subsection{NP quantities at $\mathcal{O}(\zeta^1,\eta^0)$}

To compute the NP quantities at $\mathcal{O}(\zeta^1,\eta^0)$ and $\mathcal{O}(\zeta^0,\eta^1)$, let us first specify our tetrad and coordinate choices. At $\mathcal{O}(\zeta^0,\eta^0)$, we adopt the Hawking-Hartle tetrad in the ingoing Eddington-Finkelstein coordinates $(v,r,\theta,\phi)$, where $v$ is related to $t$ in Schwarzschild coordinates by $dv=dt+(1-2M/r)^{-1}dr$. The Hawking-Hartle tetrad is given by \cite{Chandrasekhar_1983}:
\begin{subequations}\label{eq:HHtetrad}
\begin{align}
    l_{\rm HH}^{\mu} &= \left(1,\;\frac{1}{2}f(r),\;0,\;0\right)\,, \\
    n_{\rm HH}^{\mu} &= \left(0,\;-1,\;0,\;0\right)\,, \label{eq:n00}\\
    m_{\rm HH}^{\mu} &= \frac{1}{\sqrt{2}\,r}\left(0,\;0,\;1,\;i\csc\theta\right)\,,
\end{align}   
\end{subequations}
where we define $f(r)=1-2M/r$. In Eddington-Finkelstein coordinates, the Hawking-Hartle tetrad is related to the usual Kinnersley tetrad $\left(l_{\rm K}^{\mu},n_{\rm K}^{\mu}, m_{\rm K}^{\mu},\bar{m}_{\rm K}^{\mu}\right)$ by an overall boost:  $l_{\rm HH}^{\mu}=A^{-1}l_{\rm K}^{\mu}$, $n_{\rm HH}^{\mu}=A n_{\rm K}^{\mu}$ with $A=2f^{-1}(r)$. One can see that the Hawking-Hartle tetrad is regular throughout the entire spacetime, and so are the NP quantities computed from it. This feature will greatly simplify the regularization of divergences near the horizon when we solve the modified Teukolsky equations of $\Psi_0^{(1,1)}$ using Green's function in Sec.~\ref{sec:solve equation} and the computation of horizon flux in Sec.~\ref{sec:gravitational wave flux}.

At $\mathcal{O}(\zeta^1,\eta^0)$, we work with the background metric correction provided in \cite{Cano:2019ore, Cano:2020cao} and Eq.~\eqref{eq:bg_metric_original}. However, this metric is not asymptotically flat at infinity, so we make an additional coordinate transformation at $\mathcal{O}(\zeta^1,\eta^0)$ as below:
\begin{widetext}
\begin{align}
    r\rightarrow r
    -\zeta\frac{4(r-2M)\bigl(63M^5+35M^4r+20M^3r^{2}
     + 12 M^2r^{3} + 8 Mr^{4}+8r^{5}\bigr)}{231r^{5}}
\end{align}
\end{widetext}
such that Eq.~\eqref{eq:bg_metric_original} becomes
\begin{align} \label{eq:bg_metric_regular}
    ds^2=-f(r)dv^2+2\bigl(1+\zeta\mathcal{H}_1(r)\bigr)drdv 
    +r^{2}\bigl(1+\zeta\mathcal{H}_3(r)\bigr)d\Omega^2,
\end{align}
where
\begin{align}
    \mathcal{H}_1(r)
    =-\frac{8 M^6}{r^{6}}\,, \quad
    \mathcal{H}_3(r)
    =-\frac{40 M^6}{r^{6}}\,.
\end{align}
The resulting metric in Eq.~\eqref{eq:bg_metric_regular} is regular at the horizon and asymptotically flat at infinity, which removes the
coordinate singularities of the source terms at both boundaries, as we will show in the next subsection. Given the background metric correction in Eq.~\eqref{eq:bg_metric_regular}, one can then expand the orthogonality conditions defining the NP tetrad \cite{Newman:1961qr, Chandrasekhar_1983} to $\mathcal{O}(\zeta^1,\eta^0)$ and find the corrected tetrad following \cite{Li:2022pcy}. One valid choice is
\begin{subequations} \label{eq:tetrad_10}
\begin{align}
   & l^{\mu(1,0)}
   =\left(0,-\frac{1}{2}f(r)\mathcal{H}_1(r),0,0\right)\,, \label{eq:10l} \\
   & n^{\mu(1,0)}
   =\left(0,\mathcal{H}_1(r),0,0\right)\,, \\
   & m^{\mu(1,0)}
   =\frac{1}{2\sqrt{2}r}\mathcal{H}_3(r)
   \left(0,0,1,i \csc\theta\right)\,.
\end{align}  
\end{subequations}
The spin coefficients and Weyl scalars at $\mathcal{O}(\zeta^1,\eta^0)$ can then be directly calculated from Eqs.~\eqref{eq:bg_metric_regular} and \eqref{eq:tetrad_10}, which are provided in Appendix~\ref{app:Sourceterms}. Alternatively, one can follow the procedure in \cite{Loutrel:2020wbw, Wagle:2023fwl, LaHaye:2025ley} to calculate the perturbed spin coefficients and Weyl scalars from the linearized commutation relations and Ricci identities, respectively, which is the approach we employ at $\mathcal{O}(\zeta^0,\eta^1)$. 

\subsection{NP quantities at $\mathcal{O}(\zeta^0,\eta^1)$}
\label{sec:np_01}

To calculate the metric perturbation $h_{\mu\nu}^{(0,1)}$ driven by the point particle in GR, one usually has to perform metric reconstruction, as solving the Teukolsky equations only provides us the Weyl scalars $\Psi_{0,4}^{(0,1)}$. In vacuum GR, the most widely used metric reconstruction approach is the one developed by Chrzanowski, Cohen, and Kegeles (CCK) \cite{Chrzanowski:1975wv, Cohen_Kegeles_1975}. This approach expresses $h_{\mu\nu}^{(0,1)}$ in terms of a single scalar potential (the Hertz potential) and relies on the radiation gauges, which set $h_{\mu\nu}^{(0,1)}e_{a}^{\mu}=0$ and $h^{(0,1)}\equiv h_{\mu\nu}^{(0,1)}g^{\mu\nu}=0$, with $e_{a}^{\mu}$ being the ingoing (outgoing) principal null direction $n^{\mu}$ ($l^{\mu}$) for outgoing (ingoing) radiation gauges. The Hertz potential can then be calculated from either $\Psi_{0}^{(0,1)}$ or $\Psi_{4}^{(0,1)}$ via a fourth-order differential equation, which can be easily inverted with the Teukolsky-Starobinsky identities in vacuum GR \cite{Teukolsky:1974yv, Starobinsky:1973aij, Ori:2002uv}. 

Despite its wide usage in studying beyond-GR ringdown via the MTF \cite{Cano:2023tmv, Cano:2023jbk, Li:2023ulk, Wagle:2023fwl, Li:2025fci} and nonlinearities of GR ringdown \cite{Ma:2024qcv, Khera:2024bjs}, the standard CCK approach cannot be directly applied in the existence of a source (i.e., the point-particle source of an EMRI system). One main reason is that the radiation gauges used by the CCK approach are only valid for vacuum perturbations or restricted sources. For this reason, several approaches have been developed over the past few years for non-vacuum metric reconstruction, including deriving an operator identity expressing $h_{\mu\nu}^{(0,1)}$ in terms of $\Psi_{0,4}^{(0,1)}$ and the stress-energy tensor \cite{Aksteiner:2016pjt, Wardell:2024yoi}, adding a corrector tensor to restore the radiation gauge conditions \cite{Green:2019nam, Toomani:2021jlo, Bourg:2024vre, Hollands:2024iqp}, or systematically solving a subset of NP equations after relaxing the tracefree condition $h^{(0,1)}=0$ in the radiation gauges \cite{Li:2026rkf}, the last of which avoids the use of a Hertz potential by extending the direct reconstruction approach in \cite{Chandrasekhar_1983, Loutrel:2020wbw, Ripley:2020xby}. Besides these approaches aiming for general sources, one can also glue vacuum CCK solutions along the particle's worldline in the case of an EMRI system \cite{Dolan:2021ijg, Dolan:2023enf, Nasipak:2025tby}. 
 
Although in this work, we do not use the metric reconstruction approaches above due to the simplicity provided by a non-rotating central black hole, these approaches will be extremely useful when extending our method to rotating primaries and generic orbits, as demonstrated in the study of EMRIs embedded in ultralight scalar clouds \cite{Dyson:2025dlj, Li:2025ffh}. Instead, we use the metric perturbation data provided by \cite{Barack:2005nr, Barack:2007tm, BHPToolkit}, which directly solve the Einstein equations associated with a point-particle source in the Lorenz gauge. Reconstructing or solving $h_{\mu\nu}^{(0,1)}$ in the Lorenz gauge is usually helpful, as this gauge removes certain distributional singularities associated with the radiation gauges and some other gauges \cite{Barack:1999wf, Pound:2013faa, Dolan:2023enf}. We have also validated our $h_{\mu\nu}^{(0,1)}$ data against the ones from other approaches, such as the Lorenz-gauge extension of the CCK approach in \cite{Dolan:2021ijg, Dolan:2023enf}, and we get consistent results.

Given $h_{\mu\nu}^{(0,1)}$, we then follow \cite{Campanelli:1998jv, Loutrel:2020wbw, Li:2022pcy, Wagle:2023fwl, LaHaye:2025ley} to calculate the perturbed tetrad from the orthogonality conditions, and one widely used choice is
\begin{subequations} \label{eq:tetrad_01}
\begin{align}
    {l^{\mu}}^{(0,1)} 
    &= \frac{1}{2}h_{ll}^{(0,1)}n^{\mu}\,, \label{eq:01l} \\
	{n^{\mu}}^{(0,1)} 
    &= \frac{1}{2}h_{nn}^{(0,1)}l^{\mu}+h_{ln}^{(0,1)}n^{\mu}\,, \\
    {m^{\mu}}^{(0,1)} 
    &= h_{nm}^{(0,1)} l^{\mu} + h_{lm}^{(0,1)} n^{\mu} -\frac{1}{2} h_{m\bar{m}}^{(0,1)} m^{\mu} -\frac{1}{2} h_{mm}^{(0,1)} \bar{m}^{\mu}\,. 
\end{align}    
\end{subequations}
Following \cite{Campanelli:1998jv, Loutrel:2020wbw, Wagle:2023fwl, LaHaye:2025ley}, we further compute the spin coefficients and Weyl scalars at $\mathcal{O}(\zeta^0,\eta^1)$ from the linearized commutation relations [i.e., Eq.~(B4) in \cite{Wagle:2023fwl}] and linearized Ricci identities [i.e., Eq.~(B5) in \cite{Wagle:2023fwl}], respectively.

In deriving the modified Teukolsky equation of $\Psi_0^{(1,1)}$ in Eq.~\eqref{eq:modified_teuk_eq} and similarly of $\Psi_4^{(1,1)}$, we have made the gauge choices that
\begin{equation}
    \Psi_1^{(0,1)}=\Psi_3^{(0,1)}=0\,.
\end{equation}
As shown in \cite{Li:2022pcy, Wagle:2023fwl}, we have to make additional type I and type II tetrad rotations after computing the $\mathcal{O}(\zeta^0,\eta^1)$ NP quantities using the tetrad in Eq.~\eqref{eq:tetrad_01}, i.e.,
\begin{equation} \label{eq:tetrad_rotation}
\begin{aligned}
    l^{\mu(0,1)} 
    & \rightarrow l^{\mu(0,1)}+\bar{b}^{(0,1)}m^{\mu}
    +b^{(0,1)}\bar{m}^{\mu}\,, \\
    n^{\mu(0,1)} 
    & \rightarrow n^{\mu(0,1)}+\bar{a}^{(0,1)}m^{\mu}
    +a^{(0,1)}\bar{m}^{\mu}\,, \\
    m^{\mu(0,1)} 
    & \rightarrow m^{\mu(0,1)}+a^{(0,1)}l^{\mu}
    +b^{(0,1)}n^{\mu}\,,
\end{aligned} 
\end{equation}
where the rotation parameters are
\begin{align}
    a^{(0,1)}=-\frac{\bar{\Psi}_3^{(0,1)}}{3\Psi_2}\,,\qquad
    b^{(0,1)}=-\frac{\Psi_1^{(0,1)}}{3\Psi_2}\,, \label{eq:psi13rotation}
\end{align}
with $\Psi_{1,3}^{(0,1)}$ calculated in the tetrad of Eq.~\eqref{eq:tetrad_01}. The transformation rules of the spin coefficients and Weyl scalars at $\mathcal{O}(\zeta^0,\eta^1)$ given Eq.~\eqref{eq:tetrad_rotation} are provided in Eqs.~(B9) and (B10) of \cite{Wagle:2023fwl}. At this point, we have obtained all the necessary geometrical quantities at $\mathcal{O}(\zeta^1,\eta^0)$ and $\mathcal{O}(\zeta^0,\eta^1)$.

\subsection{$\Phi_{ij}^{(0,1)}$, $\Phi_{ij}^{(1,0)}$, and $\Phi_{ij}^{(1,1)}$}
\label{sec:11phiab}

Besides the geometrical quantities, we also need to compute the NP Ricci scalars $\Phi_{ij}$, $i,j\in\{0,1,2\}$, driven by the point-particle stress-energy tensor $T_{\mu\nu}^p$ and the effective stress-energy tensor $T_{\mu\nu}^{\rm cubic}$ of the parity-preserving cubic gravity in Eq.~\eqref{eq:T_cubic}, with $\Phi_{ij}$ defined in \cite{Chandrasekhar_1983, Wagle:2023fwl}. These two contributions are computed using different methods.

For the contribution from $T_{\mu\nu}^{\mathrm{cubic}}$, we first project $T_{\mu\nu}^{\mathrm{cubic}}$ in Eq.~\eqref{eq:T_cubic} onto the NP basis and express everything in terms of the NP quantities. We then expand the resulting expression to $\mathcal{O}(\zeta^1,\eta^1)$ such that the $\mathcal{O}(\zeta^1,\eta^0)$ contribution is completely determined by the NP quantities on the Schwarzschild background (i.e., quantities at $\mathcal{O}(\zeta^0,\eta^0)$), and we provide the complete expression of $\Phi_{ij}^{(1,0)}$ in Appendix~\ref{app:Sourceterms}. The $\mathcal{O}(\zeta^1,\eta^1)$ contribution also involves the gravitational radiation in GR (i.e., quantities at $\mathcal{O}(\zeta^0,\eta^1)$), which we have computed in Sec.~\ref{sec:np_01} following \cite{Barack:2005nr, Barack:2007tm, BHPToolkit}. Due to its complication, we provide the part of $\Phi_{ij}^{(1,1)}$ driven by $T_{\mu\nu}^{\mathrm{cubic}}$ in the supplementary Mathematica notebook \cite{Yang2026SourceTerm}. Since $T_{\mu\nu}^{\rm cubic}$ is proportional to $\zeta$, there are no contributions from it to $\Phi_{ij}$ at $\mathcal{O}(\zeta^0)$. 

For the contribution from $T_{\mu\nu}^{p}$, some care is required when expressing the same circular orbit in the ingoing Eddington--Finkelstein coordinates. We choose to keep the advanced-time coordinate introduced above and absorb the constant shift induced by the coordinate transformation into the orbital phase, $\phi_p(v)\equiv\omega_z\left[v-r_\star(r_0)\right]$, where $r_\star=r+2M\log(r/2M-1)$ is the Schwarzschild tortoise coordinate. After performing the proper-time integral, the point-particle stress-energy tensor on the equatorial plane can then be written compactly as
\begin{equation} \label{eq:T_p}
   T_{\mu\nu}^{p}= \frac{m_p u_{\mu}u_{\nu}}{u^t\sqrt{-g}}
   \,\delta(r-r_0)\, 
   \delta \left[\phi-\phi_p(v)\right]\,\delta(\theta-\pi/2)\,.
\end{equation}
Equivalently, since $v=t+r_\star(r)$ up to an additive constant, one may shift the origin of the advanced-time coordinate by $r_\star(r_0)$. With this convention, $\phi_p(v)=\omega_z v$. The two conventions differ only by a constant time shift and give identical energy fluxes when implemented consistently.

For a circular orbit, $u^v=u^t$, and the four-velocity therefore retains the form $u^\mu = u^t(1,0,0,\omega_z)$, where $u^t = 1/\sqrt{1-3M/r_0}$ and $\omega_z=\sqrt{M/r_0^3}$ for Schwarzschild black holes in GR. At $\mathcal{O}(\zeta^1,\eta^1)$, we need to include the background correction to this four-velocity expression. Since $\omega_z$ is observable whereas $r_0$ is not, we keep $\omega_z$ as the independent variable and instead express $r_0$ and $u^t$ in terms of $\omega_z$. We assume that the particle remains on a circular orbit at this order, neglecting stability; consequently, the form of the four-velocity does not need to be modified. To determine the corrections to $r_0$ and $u^t$, we use the normalization condition $u^\mu u_\mu=-1$ together with the geodesic equation on the modified background in Eq.~\eqref{eq:bg_metric_regular}. In the end, we find the corrected orbital radius $\tilde{r}_0$ to be
\begin{equation} \label{eq:r0_corrected}
    \tilde{r}_0=r_0
    -\zeta\frac{1}{6}r_0\left(2\mathcal{H}_3(r_0)
    +r_0\partial_r\mathcal{H}_3(r_0)\right)\,,
\end{equation}
where $r_0$ denotes the original orbital radius in GR. Since we choose to keep the orbital angular frequency unchanged, it is convenient to express the radius as $r_0=(M/\omega_z^2)^{1/3}$. The time component $u^t$ of the four-velocity is correspondingly modified to 
\begin{equation} \label{eq:ut_corrected}
    \tilde{u}^t=u^t+\zeta \frac{M}{2r_0}\left(u^t\right)^3\mathcal{H}_3(r_0)\,,
\end{equation}
where $u^t$ is the GR value, and $\tilde{u}^t$ denotes the one in the parity-preserving cubic gravity. Since $T^{p}_{\mu\nu}$ is proportional to $\eta$, it only contributes to $\Phi_{ij}$ starting from $\mathcal{O}(\eta^1)$. Thus, contracting $T_{\mu\nu}^{p}$ in Eq.~\eqref{eq:T_p} with the $\mathcal{O}(\zeta^0,\eta^0)$ tetrad in Eq.~\eqref{eq:HHtetrad} and the $\mathcal{O}(\zeta^1,\eta^0)$ tetrad in Eq.~\eqref{eq:tetrad_10} and using the corrected $r_0$ and $u^t$ in Eqs.~\eqref{eq:r0_corrected} and \eqref{eq:ut_corrected}, we obtain $\Phi_{ij}^{(0,1)}$ and the part of $\Phi_{ij}^{(1,1)}$ driven by $T_{\mu\nu}^{p}$. The complete expression of $\Phi_{ij}^{(0,1)}$ is provided in Appendix~\ref{app:Sourceterms}, while the result of $\Phi_{ij}^{(1,1)}$ is provided in the supplementary Mathematica notebook \cite{Yang2026SourceTerm}.

\subsection{Extraction of the radial part}\label{sec:radial part}

After obtaining all the NP quantities necessary for computing the modified Teukolsky equations, we now assemble all the terms and extract the radial part of the equations in this section. Since all the $\mathcal{O}(\zeta^1,\eta^0) $ quantities are purely radial, and $\Phi_{ij}^{(1,1)}$ can be separated into two parts based on their dependence on $T_{\mu\nu}^p$ and $T_{\mu\nu}^{\rm cubic}$, as discussed in Sec.~\ref{sec:11phiab}, there is no coupling between the spin-weighted spherical harmonics and the Dirac delta function. This feature makes extracting the radial part of the modified Teukolsky equations straightforward.

For convenience, we decompose the source $\mathcal{S}^{(1,1)}$ driven by the stress-energy tensor in Eq.~\eqref{eq:S_nonrot} into two parts: the regular piece $\mathcal{S}_A^{(1,1)}$, with its angular part expanded in spin-weighted spherical harmonics, and the singular piece $\mathcal{S}_B^{(1,1)}$, which is proportional to the Dirac delta function $\delta^{(3)}(x-z(\tau))$ at the particle's position $z(\tau)$. Together with the geometrical source $\mathcal{S}_{\rm geo}^{(1,1)}$, we refer to $\mathcal{S}_{\rm geo}^{(1,1)}+\mathcal{S}_A^{(1,1)}$ and $\mathcal{S}_B^{(1,1)}$ as the \textit{regular} and \textit{singular} pieces of the final source, respectively. This decomposition is formal: although $\mathcal{S}_{\rm geo}^{(1,1)}+\mathcal{S}_{A}^{(1,1)}$ contains no explicit Dirac delta functions in its present form, the Ricci identities and Bianchi identities can be used to expose terms proportional to Dirac delta functions and their derivatives at $r=r_0$. These terms vanish away from the particle and therefore do not modify the regular part of the source there, but their contributions must be included together with $\mathcal{S}_{B}^{(1,1)}$ when evaluating the Green's-function solution.

As discussed in Sec.~\ref{sec:solve equation}, the regular part away from the particle is treated numerically, while the Dirac-delta terms and their derivatives are evaluated analytically using the properties of the Dirac delta function. Before performing the harmonic decomposition and separating the radial and angular dependence, it is convenient to simplify the angular structure of each piece with several useful identities.

For the formally regular piece, the angular derivatives $\delta$ and $\bar{\delta}$ acting on the spin-weighted spherical harmonics behave as raising and lowering operators of the spin weight, respectively, i.e.,
\begin{widetext}
\begin{subequations}
\begin{align}
    \delta\bigl[{}_s Y_{\ell m}(\theta,\phi)\bigr] 
    &= -\frac{1}{\sqrt{2}\,r}\left(\sqrt{(\ell-s)(\ell+s+1)}\;{}_{s+1}Y_{\ell m}(\theta,\phi)- s\cot\theta\,{}_s Y_{\ell m}(\theta,\phi)\right)\,, \\
    \bar{\delta}\bigl[{}_s Y_{\ell m}(\theta,\phi)\bigr] 
    &= -\frac{1}{\sqrt{2}\,r}\left(-\sqrt{(\ell+s)(\ell-s+1)}\;{}_{s-1}Y_{\ell m}(\theta,\phi)+ s\cot\theta\,{}_s Y_{\ell m}(\theta,\phi)\right)\,.
\end{align}    
\end{subequations}
\end{widetext}

For the singular piece, the angular dependence can be simplified using the identities below, which follow directly from the properties of the Dirac delta function:
\begin{align}
    f(x)\,\delta^{(n)}(x-x_0)=\sum_{k=0}^{n}(-1)^k \binom{n}{k}\delta^{(n-k)}(x-x_0)\,\left.\frac{d^k f}{dx^k}\right|_{x=x_0}\label{eq:deltaidentity}\,,
\end{align}
The identity above can be applied independently to the variables $r$, $\theta$, and $\phi$ by replacing $x$ and $x_0$ with the corresponding coordinate and its source location.

We can then express the source terms of the modified Teukolsky equations purely in terms of spin-weight-$2$ (or $-2$) spherical harmonics for $\Psi_0^{(1,1)}$ (or $\Psi_4^{(1,1)}$), i.e.,
\begin{subequations} \label{eq:mtf_source_decompose}
\begin{align}
  & H_0\Psi_0^{(1,1)}
  =\mathcal{S}_{\rm geo}^{(1,1)}+\mathcal{S}_A^{(1,1)}
  +\mathcal{S}_B^{(1,1)}\,, \\
  & \mathcal{S}_{\rm geo}^{(1,1)}+\mathcal{S}_A^{(1,1)}
  \sim e^{-i\omega v}q(r){}_2Y_{\ell m}(\theta,\phi)\,, \\
  & \mathcal{S}_B^{(1,1)}\sim p_0(r)\delta\big(x-z(\tau)\big)
  +p_1(r)\delta^\prime\big(x-z(\tau)\big)+\cdots\,.
\end{align}    
\end{subequations}
Here and throughout this work, a prime denotes differentiation with respect to the argument of the corresponding function. For example,
$\delta'(x-z(\tau)) \equiv \partial_x \delta(x-z(\tau))$
and $p'(r) \equiv \partial_r p(r)$.

The equation of $\Psi_4^{(1,1)}$ has a similar structure, with ${}_2Y_{\ell m}(\theta,\phi)$ replaced by ${}_{-2}Y_{\ell m}(\theta,\phi)$. The radial function $q(r)$ contains the radial part of the metric $h_{\mu\nu}^{(0,1)}$ associated with the gravitational radiation in GR and its derivatives, while $p_i(r)$ ($i=0,1,2,3,\ldots$) contains contributions from the deformed background metric functions (i.e., $\mathcal{H}_{1,3}(r)$) and their derivatives. One may further apply the identity in Eq.~\eqref{eq:deltaidentity} along the radial direction, such that the functions $p_i(r)$ can be evaluated at the particle's orbital radius $r_0$. Due to their length, the explicit expressions for $q(r)$ and $p_i(r)$ are provided in the supplementary Mathematica notebook~\cite{Yang2026SourceTerm}.

Now, we can extract the radial part of the modified Teukolsky equations. Specifically, we multiply both sides of the equations by ${}_{2}\bar{Y}_{\ell m}(\theta,\phi)$ for $\Psi_0^{(1,1)}$ (or by ${}_{-2}\bar{Y}_{\ell m}(\theta,\phi)$ for $\Psi_4^{(1,1)}$) and integrate them over $\theta$ and $\phi$, using the following orthogonality condition of the spin-weighted spherical harmonics and the property of the Dirac delta function:
\begin{align}
    &\int_{0}^{2\pi} d\phi\int_{-1}^{1} d\cos\theta\, 
    {}_s\bar{Y}_{\ell m}(\theta,\phi){}_sY_{\ell'm'}(\theta,\phi)
    = \delta_{\ell \ell'} \delta_{mm'}\,, \\
    &\int_{-\infty}^{+\infty}\delta^{(n)}(x-x_0)\,f(x)\,dx
    =(-1)^n f^{(n)}(x_0)\,.\label{eq:detla integral}
\end{align}
After obtaining the radial part of the equations, we notice that the highest derivative order of $h_{\mu\nu}^{(0,1)}$ in the source term is six. This is readily understood: the stress-energy tensor $T_{\mu\nu}^{\rm cubic}$ in Eq.~\eqref{eq:T_cubic} contains second derivatives of the Riemann tensor, and the source term involves two additional derivatives acting on it. In the absence of cancellations, the highest derivative order therefore reaches six. For $\mathcal{S}_{\rm geo}^{(1,1)}$, it is in the form of a second-order differential operator at $\mathcal{O}(\zeta^1,\eta^0)$ acting on $\Psi_0^{(0,1)}$, so it can contain up to four derivatives of $h_{\mu\nu}^{(0,1)}$. Since we choose to solve $\Psi_{0,4}^{(0,1)}$ directly from the Teukolsky equations via BHPToolkit \cite{BHPToolkit} instead of computing them from $h_{\mu\nu}^{(0,1)}$, this order reduces to two. We have verified that our $\Psi_{0,4}^{(0,1)}$ agree with the ones computed from the $h_{\mu\nu}^{(0,1)}$ data generated by the codes in \cite{Barack:2005nr, Barack:2007tm, BHPToolkit}. Given that we only know the numerical values of $h_{\mu\nu}^{(0,1)}$, a challenge arises: how can we accurately represent these high-order derivative terms in the source terms? We will discuss our strategy in Sec.~\ref{sec:solve equation}.

\section{Solving the modified Teukolsky equations}
\label{sec:solve equation}

In this section, we solve the radial modified Teukolsky equations obtained in Sec.~\ref{sec:MTF}. With our tetrad and coordinate choices, both sides of the equations are regular at the horizon and at infinity, so we can use Green's function to obtain the solution.

The first step is to obtain the homogeneous solutions. We choose to use the codes developed by BHPToolkit \cite{BHPToolkit} to obtain ${}_{\pm2}R^{\rm in}_{\ell m}(r)$ and ${}_{\pm2}R^{\rm up}_{\ell m}(r)$, which are homogeneous solutions in the Kinnersley tetrad, while we denote the solutions in the Hawking-Hartle tetrad as ${}_{\pm2}\mathcal{R}_{\ell m}(r)$. These two sets of solutions are related by
\begin{subequations}
\begin{align}
    & {}_{2}\mathcal{R}_{\ell m}(r)
    =\frac{1}{4}f^2(r){}_{2}R_{\ell m}(r)e^{i\omega r_{\star}}\,,\\
    & {}_{-2}\mathcal{R}_{\ell m}(r)
    = 4f^{-2}(r){}_{-2}R_{\ell m}(r)e^{i\omega r_{\star}}\,, \label{eq:Kin_to_HH_-2}
\end{align}
\end{subequations}
where $r_\star\equiv r+2M\log(r/2M-1)$ is the tortoise coordinate. The factor $e^{i\omega r_{\star}}$ comes from the transformation between Schwarzschild and ingoing Eddington-Finkelstein coordinates, and the factor $f^{\pm2}(r)$ comes from the tetrad rotations relating these two tetrads in the ingoing Eddington-Finkelstein coordinates.

For $\Psi_0^{(1,1)}$, the asymptotic forms of ${}_{2}\mathcal{R}^{\rm in}_{\ell m}(r)$ and ${}_{2}\mathcal{R}^{\rm up}_{\ell m}(r)$ are
\begin{subequations}
\begin{align}
  {}_{2}\mathcal{R}_{\ell m\omega}^{\mathrm{in}}(r) &\to
  \begin{cases}
    \mathcal{A}_{\ell m\omega}^{\mathrm{trans}} & \text{for } r\to 2M\\[4pt]
    r^{-5}\,\mathcal{A}_{\ell m\omega}^{\mathrm{ref}}\,e^{2i\omega r_\star}
    + r^{-1}\,\mathcal{A}_{\ell m\omega}^{\mathrm{inc}} & \text{for } r\to +\infty
  \end{cases}\,, \\[6pt]
  {}_{2}\mathcal{R}_{\ell m\omega}^{\mathrm{up}}(r) &\to
  \begin{cases}
    \mathcal{C}_{\ell m\omega}^{\mathrm{up}}\,(r-2M)^{2}\,e^{2i\omega r_\star}
    + \mathcal{C}_{\ell m\omega}^{\mathrm{ref}} & \text{for } r\to 2M\\[4pt]
    \mathcal{C}_{\ell m\omega}^{\mathrm{trans}}\, r^{-5}\,e^{2i\omega r_\star}
    & \text{for } r\to +\infty
  \end{cases}\,, \label{eq:R_up_2}
\end{align}    
\end{subequations}
where both solutions are regular at the horizon. For $\Psi_4^{(1,1)}$, the corresponding asymptotic forms are
\begin{subequations}
\begin{align}
  {}_{-2}\mathcal{R}_{\ell m\omega}^{\mathrm{in}}(r) &\to
  \begin{cases}
    \mathcal{B}_{\ell m\omega}^{\mathrm{trans}} & \text{for } r\to 2M\\[4pt]
    r^{3}\,\mathcal{B}_{\ell m\omega}^{\mathrm{ref}}\,e^{2i\omega r_\star}
    + r^{-1}\,\mathcal{B}_{\ell m\omega}^{\mathrm{inc}} & \text{for } r\to +\infty
  \end{cases}\,, \\[6pt]
  {}_{-2}\mathcal{R}_{\ell m\omega}^{\mathrm{up}}(r) &\to
  \begin{cases}
    \mathcal{D}_{\ell m\omega}^{\mathrm{up}}\,(r-2M)^{-2}\,e^{2i\omega r_\star}
    + \mathcal{D}_{\ell m\omega}^{\mathrm{ref}} & \text{for } r\to 2M\\[4pt]
    \mathcal{D}_{\ell m\omega}^{\mathrm{trans}}\, r^{3}\,e^{2i\omega r_\star}
    & \text{for } r\to +\infty
  \end{cases}\,. \label{eq:RupPsi4}
\end{align}    
\end{subequations}
Although the part of ${}_{-2}\mathcal{R}_{\ell m\omega}^{\mathrm{up}}$ proportional to $\mathcal{D}_{\ell m\omega}^{\mathrm{up}}$ diverges at the horizon, this is harmless because the Green's function for $\Psi_4^{(1,1)}$ renders the integrand regular there.

Using the Green's function, the radial part of $\Psi_0^{(1,1)}$ is given by
\begin{widetext}
\begin{align}\label{eq:solutionpsi0}
    & {}_2\mathcal{R}_{\ell m \omega}(r)
    =\frac{{}_2\mathcal{R}^{\rm in}_{\ell m \omega}(r)}{{}_{2}W}
    \int_{r}^{\infty} -\frac{32r'^8}{(r'-2M)^{2}}e^{-2i\omega r_\star} {}_2\mathcal{R}^{\rm up}_{\ell m \omega}(r') {}_2S_{\ell m}(r') \, dr' 
    +\frac{{}_2\mathcal{R}^{\rm up}_{\ell m \omega}(r)}{{}_{2}W} 
    \int_{2M}^{r}-\frac{32r'^8}{(r'-2M)^{2}}e^{-2i\omega r_\star} {}_2\mathcal{R}^{\rm in}_{\ell m \omega}(r') {}_2S_{\ell m}(r') \, dr'\,, \nonumber\\
    & {}_{2}W=\Delta^{3}\left[{}_2 R^{\rm in}_{\ell m \omega}(r)\partial_r\left({}_2R^{\rm up}_{\ell m\omega}(r)\right)-{}_2R^{\rm up}_{\ell m\omega}(r)\partial_r\left({}_2R^{\rm in}_{\ell m\omega}(r)\right)\right]\,.
\end{align}
\end{widetext}
Here, ${}_{2}W$ is the Wronskian determinant for ${}_{2}R^{\rm in/up}_{\ell m \omega}(r)$, which is a constant. The term ${}_2S_{\ell m\omega}(r)$ denotes the source term in the radial modified Teukolsky equation of $\Psi_0^{(1,1)}$ obtained in Sec.~\ref{sec:radial part}. 

As noted earlier, the only divergence in Eq.~\eqref{eq:solutionpsi0} originates from the Green's function itself and occurs near the horizon. The integrand that requires regularization takes the form:
\begin{equation} \label{eq:integral_need_reg}
    \int_{2M}^{r_1} (r-2M)^{-2} e^{-2i\omega r_\star}g(r)\,dr\,,
\end{equation}
where $g(r)$ denotes a regular function, and $r_1$ is an arbitrary finite cutoff introduced to limit the regularization procedure to be near the horizon. The contribution to the integral from $r_1$ to $r$ can be evaluated using standard methods, so we focus on Eq.~\eqref{eq:integral_need_reg} for now.

For this integrand, one can treat the factor $(r-2M)^{-2}$ as arising 
from the derivative of $e^{-2i\omega r_\star}$ via
\begin{equation}
    \frac{e^{-2i\omega r_\star}}{(r-2M)^2} = 
    \frac{1}{4i\omega(M+ir^2\omega)}
    \frac{d^2}{dr^2}\bigl(e^{-2i\omega r_\star}\bigr)\,,
\end{equation}
and then integrate it by parts. However, implementing this procedure requires taking numerical derivatives of the source term, which already was computed by taking many numerical derivatives of the metric. This would significantly amplify the numerical noise in the source term, making the approach not practical. Thus, we use the lower incomplete Gamma functions to perform the integral:
\begin{align} \label{eq:incomplete gamma function}
    I &= \int_{2M}^{r_1} (r-2M)^{-2+n} e^{-2i\omega r_\star} dr \notag\\
    &= \Bigl(\frac{2M}{e}\Bigr)^{4i\omega M} 
    (2i\omega)^{1-n+4i\omega M} \,
    \gamma\bigl(n-1-4i\omega M,\, 2i\omega(r_1-2M)\bigr)\,.
\end{align}
To apply this method, we then need to expand the source term as a Taylor series in $(r-2M)$ accurately. To achieve this, we first take a short detour to explain how high-order derivative terms are represented in the source term.

Given that the highest-order derivative in the source term is six, it is challenging to numerically take these many derivatives without uncontrolled numerical noise. Continuously improving the precision of the metric data $h_{\mu\nu}^{(0,1)}$ is not an efficient approach. Instead, we choose to fit the numerical data $h_{\mu\nu}^{(0,1)}$ spectrally with Chebyshev polynomials $T_n(x)$ and represent the high-order derivatives on the source terms via the identities:
\begin{align}
    \frac{d}{dr}[T_n(x)] &= n\, G_{n-1}^{(1)}(x)\,\frac{dx}{dr}\,, \\
    \frac{d}{dr}[G_n^{(\lambda)}(x)] &= 2\lambda\, G_{n-1}^{(\lambda+1)}(x)\,\frac{dx}{dr}\,,
\end{align}
where $G_n^{(\lambda)}(x)$ are the ultraspherical polynomials. The functions $T_n(x)$ and $G_n^{(\lambda)}(x)$ satisfy the relation $T_n(x)=\left(G_n^{(1)}(x)-G_{n-2}^{(1)}(x)\right)/2$. Because the source term is discontinuous at the particle location, we split the spacetime into two domains, mapping the physical radial intervals $r\in[2M,r_0]$ and $r\in[r_0,r_{\max}]$ to the computational coordinate $x\in[-1,1]$, with $r_0$ being the radial position of the particle.

Having resolved the issue of representing high-order derivatives, we now return to the problem of solving the equation. In practice, only the part of the integral very near the horizon requires regularization. Moreover, the closer $r_1$ is to the horizon in Eq.~\eqref{eq:integral_need_reg}, the faster the lower incomplete Gamma function decays with increasing $n$. In other words, the sum over the lower incomplete Gamma functions converges more rapidly. According to the near-horizon asymptotic form of ${}_2\mathcal{R}^{\mathrm{up}}_{\ell m}(r)$ given in Eq.~\eqref{eq:R_up_2}, only the second term, which is proportional to ${}_2\mathcal{R}^{\mathrm{in}}_{\ell m}(r)$, requires regularization. Near the horizon, one can subtract the part proportional to ${}_2\mathcal{R}^{\mathrm{in}}_{\ell m}(r)$ using the identities:
\begin{align}
     & {}_sR^{\rm up}_{\ell m\omega}
     =A_1 {}_sR^{\rm in}_{\ell m\omega}
     +B_1 {}_sR^{\rm out}_{\ell m\omega}\,, \\
    & {}_sR^{\rm out}_{\ell m\omega}=(r^2-2Mr)^{-s}
    {}_{-s}R^{\rm in}_{\ell-m-\omega}\,,
\end{align}
which are in the Kinnersley tetrad. To apply these identities in the Hawking-Hartle tetrad we are using, we make additional transformations of them. Alternatively, since only the near-horizon region is needed, one may obtain the asymptotic form of ${}_2\mathcal{R}^{\mathrm{up}}_{\ell m\omega}(r)$ from the homogeneous Teukolsky equation and manually extract the contribution proportional to ${}_2\mathcal{R}^{\mathrm{in}}_{\ell m\omega}(r)$.

Since only the near-horizon region needs regularization, two methods can be employed to Taylor expand the source in powers of $(r-2M)$ such that we can use the lower incomplete Gamma functions in Eq.~\eqref{eq:incomplete gamma function} for regularization. The first method exploits the relation between the Chebyshev expansion and the Taylor expansion:
\begin{align}
    -32 r^8 {}_{2}\mathcal{R}^{in}_{\ell m\omega}(r) S_{\ell m}(r)
    = \sum_n c_n T_n(x) = \sum_n c_n \sum_k b_{nk} (r-2M)^k\,.
\end{align}
However, we find that this transformation is ill-conditioned and amplifies the errors of the Chebyshev expansion of the source term.

The second approach directly obtains the Taylor expansion from the asymptotic form of the Einstein field equations. Refs.~\cite{Barack:2005nr,Barack:2007tm} provide ten field equations together with four Lorenz-gauge conditions. From these fourteen equations, one can derive recurrence relations for the Taylor-expansion coefficients of the Barack-Lousto variables $h^{(i)}(r)$ ($i=1,\ldots,10$; denoted by $\bar h^{(i)}$ in Refs.~\cite{Barack:2005nr,Barack:2007tm,Dolan:2023enf}), which are particular linear combinations of the tensor-harmonic components of $h_{\mu\nu}^{(0,1)}$ and encode the 10 components of the metric perturbation. Here the superscript $(i)$ labels the tensor-harmonic amplitude and is unrelated to the perturbative-order notation used elsewhere in this work. The same Lorenz-gauge Einstein equations are also useful for treating the Dirac-delta contributions discussed in Sec.~\ref{sec:radial part}. Although these contributions do not affect the source away from the particle, they must be included when evaluating the Green's-function integrals. In practice, we use the Einstein equations of Refs.~\cite{Barack:2005nr,Barack:2007tm} to isolate the Dirac-delta contributions generated by the radial derivatives of the amplitudes $h^{(i)}$, and include them together with the explicitly singular source. The leading Taylor coefficient of each $h^{(i)}(r)$ is obtained numerically, while higher-order coefficients are generated recursively from the field equations and Lorenz-gauge conditions. A similar method can be applied to ${}_{2}\mathcal{R}_{\ell m}(r)$. In this way, we obtain a Taylor expansion of the source term, and consequently of the integrand. The details are presented in Appendix~\ref{ap:series_expansion}. The regularization method introduced here can also be extended to the case of a spinning black hole and to situations where the integrand of $\Psi_4^{(1,1)}$ diverges at infinity. All Dirac-delta contributions, including those isolated from the radial derivatives of $h^{(i)}$ and those appearing explicitly in the source, are then evaluated analytically in the Green's-function integrals using Eq.~\eqref{eq:detla integral}.

For the radial solution of $\Psi_4^{(1,1)}$, the relation in Eq.~\eqref{eq:Kin_to_HH_-2} causes part of ${}_{-2}\mathcal{R}^{\mathrm{up}}_{\ell m \omega}(r)$ to diverge at the horizon, as shown in Eq.~\eqref{eq:RupPsi4}. To make the divergence analysis more transparent, we write the integrand in terms of the radial functions ${}_{-2}R^{\rm in}_{\ell m\omega}(r)$ and ${}_{-2}R^{\rm up}_{\ell m\omega}(r)$, which are regular at the horizon, as defined in the Kinnersley tetrad. Thus, the solution of $\Psi_4^{(1,1)}$ is given by
\begin{widetext}
\begin{align}\label{eq:solutionpsi4}
    & {}_{-2}\mathcal{R}_{\ell m\omega}(r) 
    = \frac{ {}_{-2}\mathcal{R}^{\rm in}_{\ell m \omega}(r)}{{}_{-2}W} \int_{r}^{\infty} -\frac{1}2\,  r'^{2} e^{- i \omega r_\star}  {}_{-2}R^{\rm up}_{\ell m \omega}(r') {}_{-2}S_{\ell m\omega }(r') \, dr' 
    + \frac{{}_{-2}\mathcal{R}^{\rm up}_{\ell m \omega}(r)}{{}_{-2}W} \int_{2M}^{r} -\frac{1}2\,  r'^{2} e^{- i \omega r_\star}  {}_{-2}R^{in}_{\ell m \omega}(r') {}_{-2}S_{\ell m \omega}(r') \, dr'\,, \nonumber\\
    & {}_{-2}W=\Delta^{-1}\left[{}_{-2} R^{\rm in}_{\ell m \omega}(r)\partial_r\left({}_{-2}R^{\rm up}_{\ell m\omega}(r)\right)-{}_{-2}R^{\rm up}_{\ell m\omega}(r)\partial_r\left({}_{-2}R^{\rm in}_{\ell m\omega}(r)\right)\right]\,,
\end{align}
\end{widetext}
where ${}_{-2}W$ is the Wronskian determinant for ${}_{-2}R^{\rm in/up}_{\ell m \omega}(r)$, which is a constant. Here, ${}_{-2}S_{\ell m\omega}(r)$ denotes the source term in the radial modified Teukolsky equation of $\Psi_4^{(1,1)}$ obtained in Sec.~\ref{sec:radial part}.

This expression shows that the integrand is convergent at the horizon and that, after a $\mathcal{O}(\zeta^1,\eta^0)$ coordinate transformation, the source term has the required asymptotic behavior, decaying faster than $\mathcal{O}(1/r^5)$ at infinity. Thus, for $\Psi_4^{(1,1)}$, the calculation becomes simpler: it suffices to evaluate the integral in Eq.~\eqref{eq:solutionpsi4}. With the method described above, we have obtained the values of $\Psi_0^{(1,1)}$ at the horizon and $\Psi_4^{(1,1)}$ at null infinity.

As an independent consistency check of both the source construction and the solution procedure, we consider in Appendix~\ref{ap:Check_calculation} a simple higher-derivative benchmark theory, which we refer to as the $RK$ theory~\cite{Bernard:2025dyh,Wang:2026qst}. The theory is obtained by adding the interaction $\lambda_{\rm ev}l_c^4 R K$ to the Einstein-Hilbert action, where $K=R_{\mu\nu\rho\sigma}R^{\mu\nu\rho\sigma}$ is the Kretschmann scalar. At linear order in the coupling, this higher-curvature interaction can be removed from the bulk by a field redefinition, at the price of generating a curvature-dependent finite-size interaction in the point-particle action. The same physical correction is therefore represented in two equivalent descriptions: in the $RK$ description it appears through the deformed background and the higher-curvature effective source, whereas in the field-redefined ($FR$) description the bulk dynamics is that of GR and the correction is transferred to the worldline sector. We construct and solve the modified Teukolsky equations independently in the two descriptions and compare the resulting Weyl-scalar perturbations after accounting for the tetrad convention specified in Appendix~\ref{ap:Check_calculation}. The independently obtained solutions satisfy the relations implied by the field redefinition, providing a nontrivial check of the modified source construction and the solution procedure. With this consistency check in place, we now turn to the computation of the horizon and infinity fluxes.

\section{Gravitational Wave flux}
\label{sec:gravitational wave flux}

In this section, we show how to compute the energy fluxes from the solutions of $\Psi_{0,4}^{(1,1)}$ found in Sec.~\ref{sec:solve equation}. As we will show in this section, the flux is quadratic in the Weyl scalars, so the leading GR contribution scales as $\mathcal{O}(\eta^2)$. The metric perturbation at $\mathcal{O}(\zeta^1,\eta^0)$ is stationary and therefore satisfies $ \partial_t h^{(1,0)}=0$. As a result, the would-be cross terms contributing at $\mathcal{O}(\zeta^1,\eta^1)$ vanish identically. Consequently, the leading-order cubic-gravity correction to the energy fluxes arises at $\mathcal{O}(\zeta^1,\eta^2)$.

\subsection{Horizon flux}\label{sec:horizon flux}

For the horizon flux, the computation is more complicated than at null infinity because we must take into account the deformations of the horizon geometry. We adapt the approach used in GR~\cite{HawkingHartle1972, Chandrasekhar_1983} to the modified theory. Because the energy flux flowing into the event horizon is related to the change in the horizon surface area, the overall strategy is to express this area change in terms of the NP quantities.

On a spatial cross-section of the event horizon, the induced line element takes the form 
\begin{align}
  d\widetilde{s}^{2}=r^2(1+\zeta\mathcal{H}_3)d\Omega^2,
\end{align}
from which the horizon area is
\begin{align}
\Sigma=\int \sqrt{h} d\theta d\phi
=16\pi M^2\left(1+\zeta\mathcal{H}_3|_{r=2M}\right).\label{eq:H3-M-relation}
\end{align}

We note that the $\mathcal{O}(\zeta^0,\eta^1) $ perturbations do not affect the surface area of the event horizon. This can be seen from the Hawking-Hartle flux formula~\cite{HawkingHartle1972}, where the GR change in area is \emph{quadratic} in the metric perturbation and thus can be neglected at $\mathcal{O}(\zeta^0,\eta^1) $ order. Consequently, the expression~\eqref{eq:H3-M-relation} is valid to both $\mathcal{O}(\zeta^0,\eta^1) $ and $\mathcal{O}(\zeta^1,\eta^0)$ order. To convert the horizon-area variation into an energy flux, we differentiate the mass-area relation in Eq.~\eqref{eq:H3-M-relation} with respect to the black-hole mass. This derivative is taken at fixed theory, so the dimensionful coupling $\lambda_{\rm ev}l_c^4$, rather than $\zeta=\lambda_{\rm ev}l_c^4/M^4$, is held fixed. Hence,
$d\zeta/dM=-4\zeta/M$. Using $dM=dE$,
where $M$ is the ADM mass, gives
\begin{align}
    \label{eq:dSigma-dE-relation}
    \frac{d^2 \Sigma}{dtd\Omega}
    =32\pi M(1-\zeta\mathcal{H}_3|_{r=2M})\frac{d^2 E_{H}}{dtd\Omega}\,.
\end{align}

As shown in Eq.~\eqref{eq:HHtetrad}, the Hawking-Hartle tetrad is regular throughout the spacetime, and its null vector $l^\mu_{\rm HH}$ coincides with the generator of the event horizon. For this reason, we adopt the Hawking-Hartle tetrad for the computation of gravitational-wave energy fluxes, both at the horizon and at infinity. Our calculation is performed on the exact event horizon, and we choose $l^\mu$  as the generator of the event horizon, so $\kappa=0$; this simplifies the calculation of the horizon flux.

To determine the horizon area change $d\Sigma$ in terms of $\Psi_0$, we use the following relationship on the horizon:
\begin{align}
    \frac{d}{dt}d\Sigma=-2\rho d\Sigma\,.
    \label{eq:optical scalar}
\end{align}
This expression can be obtained by taking the Lie derivative of the area form ${}^{2}\epsilon_{\mu\nu}=2i m_{[\mu}\bar{m}_{\nu]}$ of the 2-surfaces on the horizon with respect to $\ell^\mu$, and then using that $\rm{Im}(\rho)=0$ on the horizon, which follows from the property that $\ell^\mu$ is hypersurface orthogonal on the event horizon. Thus, the optical scalar $\rho$ is the expansion of the event horizon. We then calculate $\rho$ using the following Ricci identity:
\begin{equation}\label{eq:Ricci1}
    D\rho=\left(\rho^{2}+\sigma\bar{\sigma}\right)
    +(\varepsilon+\bar{\varepsilon})\rho+\Phi_{00}\,,
\end{equation}
where we have used that the spin coefficient $\kappa=0$ on the event horizon. We also use the following Ricci identity relating $\sigma$ to $\Psi_0$:
\begin{equation}\label{eq:Ricci2}
    D\sigma = (\rho + \bar{\rho}) \sigma + (3\varepsilon - \bar{\varepsilon}) \sigma  + \Psi_0\,,
\end{equation}
where we have used $\kappa=0$ again. 

In GR, one expands Eq.~\eqref{eq:Ricci1} to $\mathcal{O}(\zeta^0,\eta^1)$ and $\mathcal{O}(\zeta^0,\eta^2)$ :
\begin{align}
    D\rho^{(0,1)} &= 2\rho^{(0,1)}\varepsilon \label{eq:01rho}\,, \\
    D\rho^{(0,2)} &= |\sigma^{(0,1)}|^2 + 2\rho^{(0,2)}\varepsilon \label{eq:02rho}\,.
\end{align}
Equation~\eqref{eq:01rho} yields exponentially growing solutions, but the event horizon corresponds to the solution with $\rho$ asymptotic to 0, so we set $\rho^{(0,1)}=0$. The $\mathcal{O}(\zeta^0,\eta^2)$ equation is more involved, since it contains both oscillatory and non-oscillatory contributions. However, for an EMRI system, the horizon area evolves on a timescale much longer than the orbital timescale $M$, as implied by our perturbative expansion. Consequently, $D\rho^{(0,2)}\sim\rho^{(0,2)}/T\ll\varepsilon\rho^{(0,2)}$, where $T$ denotes the characteristic timescale associated with the evolution of the horizon area and $\varepsilon=1/(8M)$. To leading order, the term $D\rho^{(0,2)}$ may therefore be neglected, so Eq.~\eqref{eq:02rho} reduces to $\rho^{(0,2)}=-\left|\sigma^{(0,1)}\right|^2/(2\varepsilon)$.

Considering the corrections to the above relations from cubic gravity, we need to further expand Eq.~\eqref{eq:Ricci1} to $\mathcal{O}(\zeta^1,\eta^1)$ and $\mathcal{O}(\zeta^1,\eta^2)$:
\begin{widetext}
\begin{align}
   D\rho^{(1,1)}
   =& \;2\rho^{(1,1)}\varepsilon+\Phi_{00}^{(1,1)}\,, \label{eq:11rho}\\
   D\rho^{(1,2)}
   =& \;-D^{(0,1)}\rho^{(1,1)}+\sigma^{(0,1)}\bar{\sigma}^{(1,1)}
   +\bar{\sigma}^{(0,1)}\sigma^{(1,1)}
   +2\rho^{(0,2)}\varepsilon^{(1,0)}
   +\rho^{(1,1)}\left(\varepsilon^{(0,1)}
   +\bar{\varepsilon}^{(0,1)}\right)
   +\Phi_{00}^{(1,2)}+2\rho^{(1,2)}\varepsilon\,. \label{eq:12rho}
\end{align}
\end{widetext}
To obtain $\rho^{(1,1)}$ and $\rho^{(1,2)}$, we need to compute $\Phi_{00}^{(1,1)}$ and $\Phi_{00}^{(1,2)}$. As discussed in Sec.~\ref{sec:11phiab}, $\Phi_{00}$ naturally splits into two contributions: the regular piece driven by the cubic-gravity stress-energy tensor $T_{\mu\nu}^{\rm cubic}$ and the singular piece driven by the point-particle stress-energy tensor $T_{\mu\nu}^{p}$.  Since the particle's orbit is outside the black hole, the singular piece has no support near the horizon and, thus, does not directly contribute to the horizon flux. One can then prove that $\Phi_{00}^{(1,1)}=0$ on the horizon using the Ricci and Bianchi identities. This implies $\rho^{(1,1)}=0$ on the horizon for the same reason that $\rho^{(0,1)}$ vanishes on the horizon. If $\Phi_{00}^{(1,1)}$ does not vanish in certain cases, we may replace $D\rho^{(1,1)}$ by $-i\omega\rho^{(1,1)}$ to express $\rho^{(1,1)}$ in terms of $\Phi_{00}^{(1,1)}$. On the other hand, $\Phi_{00}^{(1,2)}$ remains nonzero and takes the form of a quadratic functional in $\mathcal{O}(\zeta^0,\eta^1)$ quantities, as shown in Appendix~\ref{app:phi00} in detail. Solving $\rho^{(1,2)}$ from Eq.~\eqref{eq:12rho} in the same manner as for Eq.~\eqref{eq:02rho}, and using the fact that, in cubic gravity, $D^{(1,0)}$ vanishes on the horizon, we can express the horizon-area variation in terms of perturbations of the relevant spin coefficients.

Furthermore, Eq.~\eqref{eq:Ricci2} provides a relation between $\sigma$ and $\Psi_0$, so we can express the spin coefficients relevant for the horizon area change in terms of $\Psi_0$:
\begin{subequations} \label{eq:spin_coeff_horizon}
\begin{align}
    & \rho^{(0,2)}=-\frac{1}{2\varepsilon}|\sigma^{(0,1)}|^2 \label{eq:prerho02}\,, \\
    & \sigma^{(0,1)}
    =-\frac{1}{i\omega+2\varepsilon}\Psi_0^{(0,1)}\,, \\
    & \sigma^{(1,1)}
    =-\frac{1}{i\omega+2\varepsilon}
    \left(\Psi_0^{(1,1)}+2\varepsilon^{(1,0)}\sigma^{(0,1)}\right)\,,
\end{align}   
\end{subequations}
where $\varepsilon^{(1,0)}$ is provided in Appendix~\ref{app:Sourceterms}. Together with Eqs.~\eqref{eq:dSigma-dE-relation} and~\eqref{eq:optical scalar}, we obtain the horizon flux at $\mathcal{O}(\zeta^1,\eta^2)$:
\begin{widetext}
  \begin{align}
     \left(\frac{d^2 E}{d\Omega dt}\right)^{(1,2)}
     =-\frac{M}{4\pi}\left[2\mathcal{H}_3(2M)\rho^{(0,2)}+\rho^{(1,2)}\right]=-\frac{5M}{32\pi\varepsilon}|\sigma^{(0,1)}|^2+\frac{M}{8\pi\varepsilon}\left(\sigma^{(0,1)}\bar{\sigma}^{(1,1)}
     +\bar{\sigma}^{(0,1)}\sigma^{(1,1)}
     +2\rho^{(0,2)}\varepsilon^{(1,0)}
     +\Phi_{00}^{(1,2)}\right)\,.\label{eq:horizonflux}
\end{align}  
\end{widetext}
The two terms in the intermediate expression have distinct origins. The term proportional to $\mathcal{H}_3(2M)\rho^{(0,2)}$ is the background-geometry correction: it results from converting the GR $\mathcal{O}(\eta^2)$ horizon-area growth into an energy flux on the $\mathcal{O}(\zeta)$-deformed black-hole background. By contrast, $\rho^{(1,2)}$ is the genuine $\mathcal{O}(\zeta\eta^2)$ correction to the horizon expansion itself, determined by Eq.~\eqref{eq:12rho}. Using Eq.~\eqref{eq:spin_coeff_horizon} and \eqref{eq:phi0012}, the above expression can be expressed in terms of the Weyl scalars $\Psi_0^{(0,1)}$ and $\Psi_0^{(1,1)}$.  The above procedure can, in principle, be extended to other beyond-GR theories. In our calculation, however, the tetrad used to solve the modified Teukolsky equations does not satisfy the conditions required for the direct application of Eq.~\eqref{eq:horizonflux}. We therefore perform an additional tetrad rotation to bring the tetrad into the required form. The details of this transformation are provided in Appendix~\ref{app:gauge transformation}.

\subsection{Infinity flux}

For the stationary and asymptotically flat black-hole solutions considered here, the gravitational-wave energy flux at infinity can be computed using the Isaacson effective stress-energy tensor. Employing the asymptotic relation between $\Psi_4$ and the gravitational-wave strain $h_+$ and $h_\times$, this flux may be written as
\begin{align}
    \left(\frac{dE}{d\Omega \, dt}\right) 
    =\frac{\omega^2 r^2}{16\pi}
    \left(|h_+|^2+|h_\times|^2\right)
    =\frac{r^2}{64\pi \omega^2}|\Psi_4|^2\,.
    \label{eq:infinityflux_full}
\end{align}
Since our calculation is carried out in the Hawking-Hartle tetrad, the relation between $\Psi_4$ and the gravitational-wave strains 
$h_+$, $h_\times$ is modified. As a result, our convention yields a coefficient that is smaller by a factor of $16$ than the corresponding coefficient in Ref.~\cite{Chandrasekhar_1983}. Expanding Eq.~\eqref{eq:infinityflux_full} in $\zeta$ and $\eta$, we obtain the leading cubic-gravity correction at $\mathcal{O}(\zeta^1,\eta^2)$:
\begin{align}
    \left(\frac{dE}{d\Omega\,dt}\right)^{(1,2)} 
    =\frac{r^2}{64\pi \omega^2} 
    \left(\Psi_4^{(0,1)}\bar{\Psi}_4^{(1,1)}
    +\bar{\Psi}_4^{(0,1)}\Psi_4^{(1,1)}\right)\,.
    \label{eq:infinityflux}
\end{align}
It remains to verify that Eq.~\eqref{eq:infinityflux_full} continues to hold in parity-preserving cubic gravity.

To establish this result, we follow the approach of Ref.~\cite{Stein:2010pn}, which derives the effective stress-energy tensor from the second variation of the action. Applying this method, we can write the second variation $\delta S^{\rm eff(2)}$ of a beyond-GR effective action about the background solution in the form:
\begin{align}
    \delta S^{\rm eff(2)} 
    &= \int d^4x \, \sqrt{-g} \, \delta g^{\mu\nu}t_{\mu\nu}\,. 
\end{align}
The quantity $t_{\mu\nu}$ is an intermediate tensor arising in this construction, whose explicit form depends on the particular modified gravity theory under consideration. One can then define an effective stress-energy tensor for gravitational waves,
\begin{align}
    T^{\rm eff}_{\mu\nu} &= -2 \langle t_{\mu\nu} \rangle,
\end{align}
where $\langle\cdots\rangle$ denotes an average over spacetime. Although the action considered in this work complicates the effective stress-energy tensor, most terms decay faster than $\mathcal{O}(r^{-2})$. Substituting the cubic-gravity action in Eq.~\eqref{eq:action} into the above formalism yields additional contributions to the effective stress-energy tensor. However, these terms decay faster than $\mathcal{O}(r^{-2})$ at infinity and therefore do not contribute to the energy flux. Consequently, the asymptotic effective stress-energy tensor coincides with its GR counterpart:
\begin{align}
    T^{\rm eff}_{\rm CG, \mu\nu} = T^{\rm eff}_{\rm GR, \mu\nu} + \mathcal{O}\left(r^{-3}\right)\,.
\end{align}
This confirms that the expression for the flux at null infinity in terms of $\Psi_4$ in GR [i.e., Eq.~\eqref{eq:infinityflux_full}] remains valid in the parity-preserving cubic gravity.

\subsection{Results}
\label{sec:result}

In this section, we provide the technical details of implementing the formalism developed above for cubic gravity. We then present the resulting energy fluxes at the horizon and at infinity, quoted before reinstating the overall expansion parameter $\zeta$ and $\eta$. 

Since our calculation is performed in the ingoing Eddington-Finkelstein coordinates, some care is required when applying the approach in Refs.~\cite{Barack:2005nr, Barack:2007tm} to solve for $h_{\mu\nu}^{(0,1)}$. In particular, after transforming from Schwarzschild coordinates to ingoing Eddington-Finkelstein coordinates, the Fourier basis changes from $e^{-i\omega t}$ to $e^{-i\omega v}$. Consequently, an additional factor $e^{-i\omega r_\star}$ must be extracted from the $h^{(i)}(r)$  before applying the reconstruction procedure. 

When computing the source term $\mathcal{S}_A^{(1,1)}$ in Eq.~\eqref{eq:mtf_source_decompose}, we find that several coefficients multiplying $h^{(1,2,4,5,8,9)}$, as well as their first derivatives, diverge at the horizon. These divergences can be removed by rewriting the metric amplitudes in combinations that remain regular at the horizon. For example,
\begin{widetext}
\begin{align}
    B_1(r)h^{(1)}(r)+B_2(r)h^{(2)}(r)
    =\frac{B_1(r)+B_2(r)}{2}
    \left[h^{(1)}(r)+h^{(2)}(r)\right]
    +\frac{(B_1(r)-B_2(r))(r-2M)^2}{2}
    \frac{h^{(1)}(r)-h^{(2)}(r)}
    {(r-2M)^2}\,.
\end{align}
\end{widetext}
Here, $B_1(r)$ and $B_2(r)$ denote the coefficient functions multiplying $h^{(1)}(r)$ and $h^{(2)}(r)$, respectively, and depend only on the radial coordinate $r$. Their explicit expressions are provided in the supplementary Mathematica notebook~\cite{Yang2026SourceTerm}. Although $B_1(r)$ and $B_2(r)$ diverge individually at the horizon, the combinations $B_1(r)+B_2(r)$ and $(B_1(r)-B_2(r))(r-2M)^2$ remain finite. Using a near-horizon series expansion together with the Lorenz-gauge conditions, as described in Appendix~\ref{ap:series_expansion}, one finds that $\left(h^{(1)}(r)-h^{(2)}(r)\right)/(r-2M)^2$
is also regular at the horizon. We therefore replace $h^{(1)}$ and $h^{(2)}$ by the regular combinations $h^{(1)}+h^{(2)}$ and $\left(h^{(1)}-h^{(2)}\right)/(r-2M)^2$, and similarly for their first derivatives. An analogous treatment applies to the pairs $(h^{(4)},h^{(5)})$ and $(h^{(8)},h^{(9)})$, although in these cases, the regular quantities are $\left(h^{(4)}-h^{(5)}\right)/(r-2M)$ and $\left(h^{(8)}-h^{(9)}\right)/(r-2M)$. These apparent divergences are artifacts of the strategy to solve the Einstein equations in Ref.~\cite{Barack:2005nr}, which was formulated in Schwarzschild coordinates. They disappear when the reconstruction is performed directly in the ingoing Eddington-Finkelstein coordinates. 

After convolving the source terms with the Green's function, one may evaluate the resulting integrals using the lower incomplete gamma function series by extracting the Taylor-series coefficients of the source term from the numerical solution. Throughout this work, we adopt $r_1=2.1M$ in Eq.~\eqref{eq:incomplete gamma function} and truncate the series after the first nine terms. By increasing
the truncation order $N$ and comparing the solution, we find that this choice yields an accuracy better
than $1\%$. The convergence further improves as $r_1$ is taken closer to the horizon, reducing the number of terms required for a given accuracy.

Having established the solution procedure and assessed the numerical accuracy of the series expansion, we now turn to the resulting fluxes. We first discuss the multipolar convergence and then present the mode-summed fluxes and their weak-field behavior.
\begin{figure}[tb]
  \centering
  \includegraphics[width=\columnwidth]{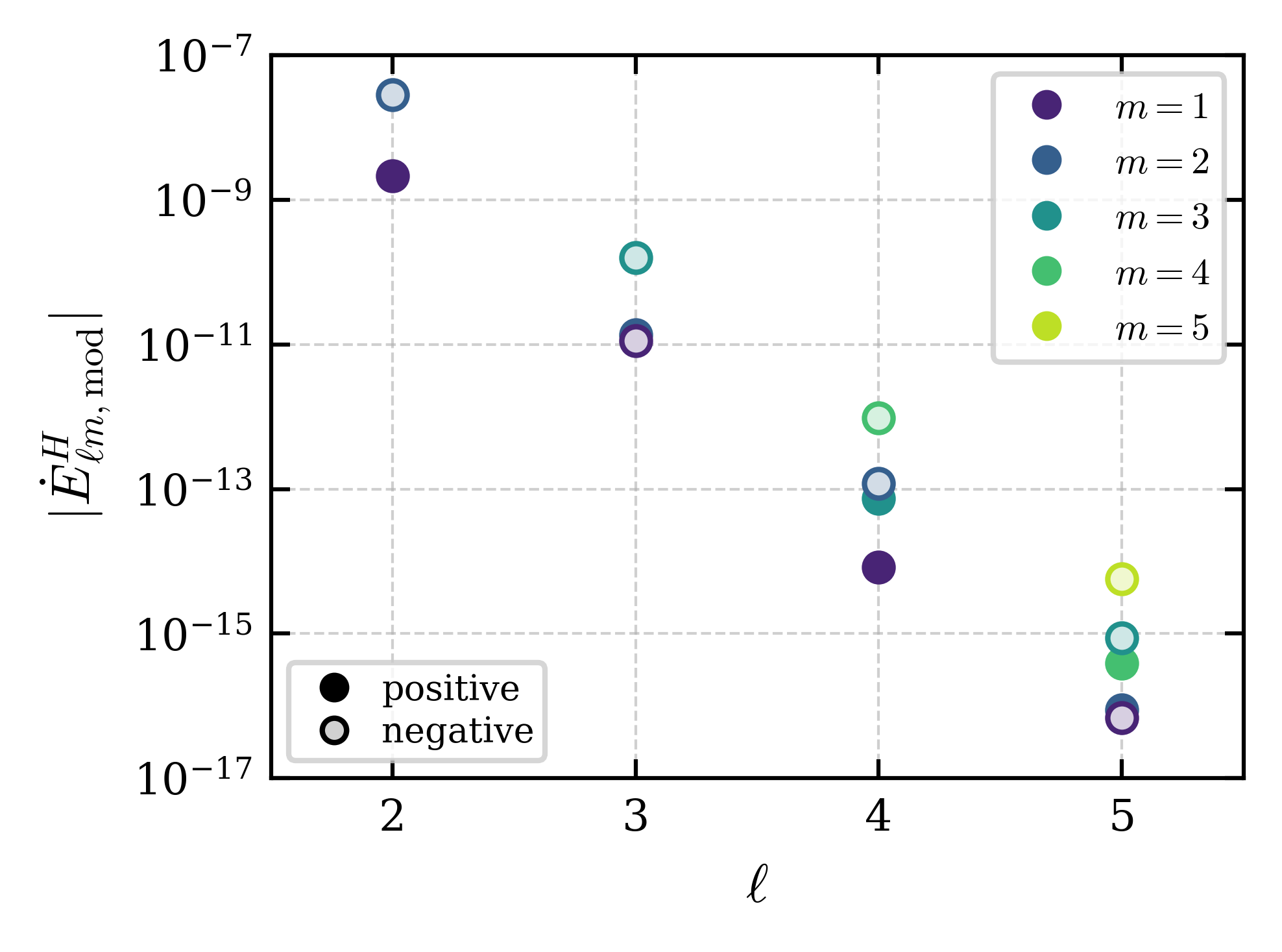}
  \caption{Absolute values of the modal contributions to the cubic-gravity correction to the horizon energy flux as a function of the multipole
number $\ell$, evaluated at $r_0=10M$, with the overall factor $\zeta\eta^2$ removed. Different colors denote the azimuthal modes $m=1,\ldots,\ell$. Filled and open markers indicate positive and negative values of $\dot E^{H}_{\ell m,\rm mod}$, respectively. Only the $m>0$ modes are shown, since the $m<0$ modes give identical signed contributions to their positive-$m$ counterparts. The stationary $m=0$ modes do not contribute at the perturbative order considered.}
\label{fig:horizon_flux_l}
\end{figure}
As shown in Fig.~\ref{fig:horizon_flux_l}, the magnitudes of the cubic-gravity corrections decrease rapidly with increasing multipole number $\ell$. The individual modes can have either sign. At $r_0=10M$, the modes shown here follow a simple parity pattern: $\ell+m$ odd modes give positive corrections, whereas $\ell+m$ even modes give negative corrections. Here and below, the sign refers to the $\mathcal{O}(\zeta\eta^2)$ correction rather than to the total physical energy flux. Despite these opposite signs, the multipolar sum converges rapidly; truncating the multipolar sum at $\ell=3$ is sufficient to achieve better than $1\%$ accuracy.

In Fig.~\ref{fig:energy_flux}, we show the mode-summed GR fluxes and the corresponding cubic-gravity corrections at the horizon and at null infinity as functions of $r_0$, including multipoles through $\ell=4$. The $m=0$ modes do not contribute at the perturbative order considered here, since both the corresponding $\mathcal{O}(\zeta^0,\eta^1)$ metric perturbations and the $\mathcal{O}(\zeta^1,\eta^0)$ background deformation are stationary and therefore cannot generate a
time-averaged energy flux. The GR contributions are shown with the overall factor $\eta^2$ removed, while the cubic-gravity corrections are shown with $\zeta\eta^2$ removed. Since $\dot E_{H,\rm mod}<0$ throughout the displayed range, we plot $-\dot E_{H,\rm mod}$ on the logarithmic scale, whereas $\dot E_{\infty,\rm mod}>0$. The magnitudes of all four contributions increase as the orbital radius decreases. Figure~\ref{fig:energy_flux} also shows that the correction is much larger relative to the GR flux at the horizon than at null infinity. On the other hand, $|\dot E_{H,\rm mod}|$ and $\dot E_{\infty,\rm mod}$ themselves are of comparable magnitude over the range considered. Because the two corrections have opposite signs, they partially cancel in the cubic-gravity correction to $\dot E_H+\dot E_\infty$.

For the horizon flux, the correction receives contributions from both $\Psi_0^{(1,1)}$ and $\Phi_{00}^{(1,2)}$. The integral of the regular part of the source from the horizon to the particle position gives the main contribution to $\Psi_0^{(1,1)}$. The contribution from $\Phi_{00}^{(1,2)}$, on the other hand, arises from the background metric correction, as discussed in Sec.~\ref{sec:horizon flux}. Its explicit form in Eq.~\eqref{eq:phi0012} involves $\Psi_4^{(0,1)}$ and $\Psi_2^{(0,1)}$. These terms help account for the relatively large correction to the horizon flux compared with its GR value.

\begin{figure}[tb]
  \centering
  \includegraphics[width=\columnwidth]{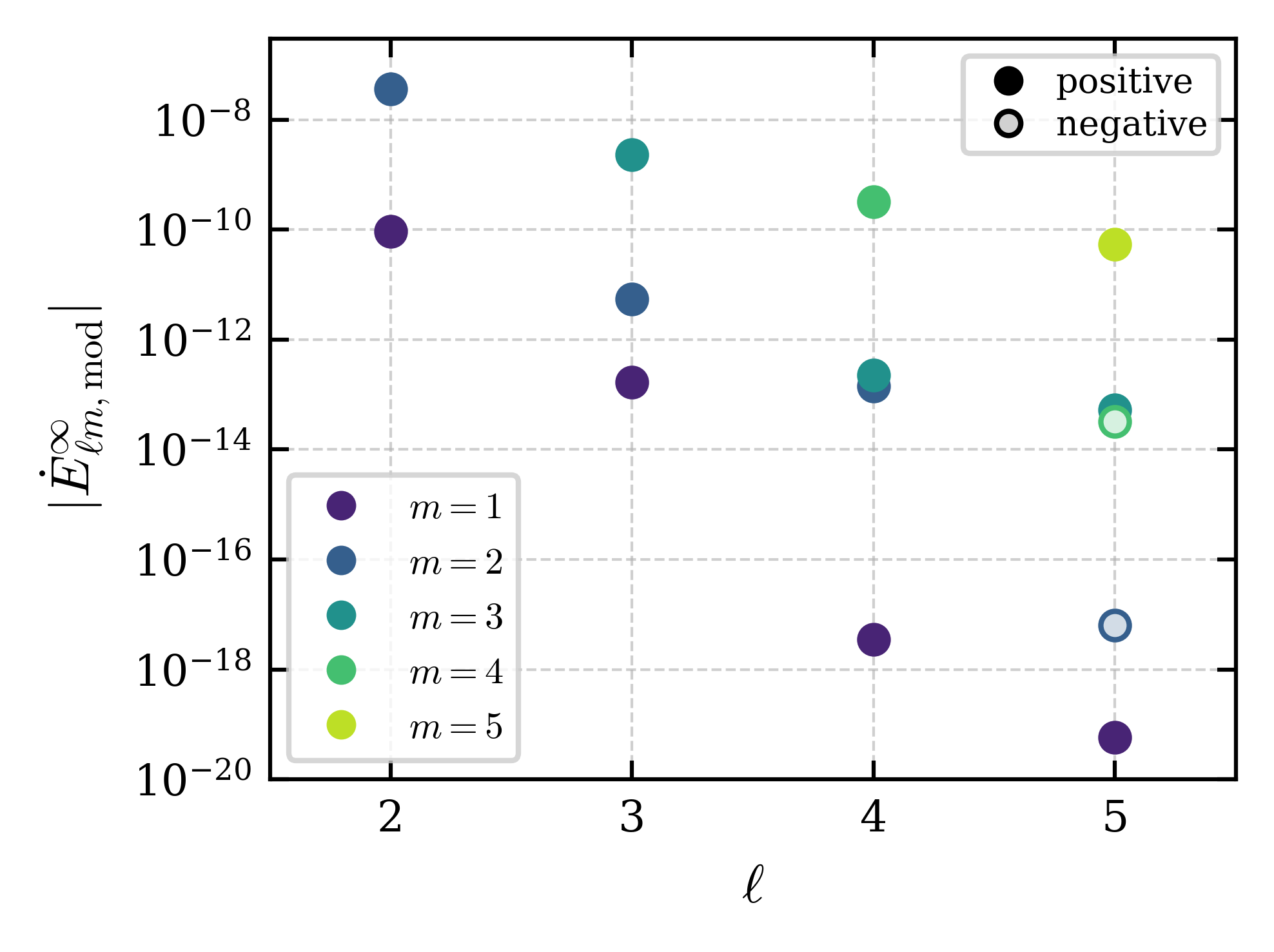}
  \caption{Absolute values of the modal contributions to the cubic-gravity correction to the energy flux at null infinity as a function of the multipole number $\ell$, evaluated at $r_0=10M$, with the overall factor $\zeta\eta^2$ removed. Different colors denote the azimuthal modes $m=1,\ldots,\ell$. Filled and open markers indicate positive and negative values of $\dot E^{\infty}_{\ell m,\rm mod}$, respectively. Only the $m>0$ modes are shown, since the $m<0$ modes give identical signed contributions to their positive-$m$ counterparts. The stationary $m=0$ modes do not contribute at the perturbative order considered.}
  \label{fig:infinity_flux_l}
\end{figure}

For the flux at null infinity, the magnitudes of the modal corrections also decrease rapidly with increasing multipole number $\ell$. Most of the modes shown in Fig.~\ref{fig:infinity_flux_l} are positive. At $r_0=10M$, only the $(5,2)$ and $(5,4)$ modes are negative, and their contributions are negligible compared with the total infinity-flux correction. The sign of each modal correction is determined by the interference between the GR and modified amplitudes; for these two modes the interference is destructive.  Including multipoles through $\ell=4$ is sufficient to achieve approximately $1\%$ accuracy. Figure~\ref{fig:energy_flux} shows that $\dot E_{\infty,\rm mod}$ is positive over the range considered, while remaining small relative to the GR infinity flux. As a further consistency check, we compare the weak-field behavior of the infinity-flux correction with the independent PN result of Ref.~\cite{Cano:2026hlv}. For the parity-preserving cubic theory, the leading relative correction enters at 5PN order with coefficient $C_5=112$. We find that our numerical result approaches this prediction as $r_0/M$ increases. A detailed comparison is presented in Appendix~\ref{ap:PN_comparison}.

Overall, the relative cubic-gravity correction is larger at the horizon than at null infinity, while the horizon and infinity corrections themselves are comparable in magnitude.

\section{Conclusion}\label{sec:conclusion}

In this work, we have developed a formalism to compute gravitational waves generated by a point particle moving in a spherically symmetric black-hole spacetime in higher-derivative gravity theories. The framework is based on the modified Teukolsky equations and is therefore naturally suited for future extension to rotating black holes. As a concrete example, we considered the parity-preserving cubic gravity and demonstrated the complete calculation of the modified Teukolsky equations together with the associated horizon and infinity fluxes.Our results show that the relative cubic-gravity correction is larger at the horizon than at null infinity, while the two corrections have comparable magnitudes and opposite signs, leading to a partial cancellation in the correction to the total  energy flux.

There are two immediate applications of this work. First, the formalism provides a key building block for constructing EMRI waveforms in higher-derivative gravity and, more generally, a broad class of EFT extensions of GR. Such waveforms could offer a powerful probe of strong-field modifications of gravity with future space-based detectors. One important ingredient that remains to be incorporated is the tidal response of the smaller compact object. In cubic gravity theories, black holes generally acquire a nonvanishing tidal Love number. Since the combination $\lambda_{\rm ev}l_c^4$ multiplying the cubic-curvature operator in Eq.~\eqref{eq:action} has dimensions of $[L]^4$, dimensional analysis implies that the leading quadrupolar tidal deformability of the smaller black hole scales as $\lambda_{\rm ev}l_c^4 m_p\sim[L]^5$. The corresponding induced quadrupole moment is therefore $\Delta Q_{ij}\propto \lambda_{\rm ev}l_c^4 m_p E_{ij}$, which generates an additional contribution to the gravitational-wave flux. A simple Newtonian estimate shows that this tidal contribution enters at the same perturbative order in the cubic coupling and the mass ratio as the modified flux computed in the present work, although with a potentially different frequency dependence. Our infinity-flux result agrees in the weak-field regime with the independent post-Newtonian calculation of Ref.~\cite{Cano:2026hlv}, which shows that, for a structureless secondary, the leading 5PN correction is proportional to the quadrupolar Love number of the primary black hole. Together with the finite-size tidal contribution of the secondary discussed above, this suggests that the leading corrections associated with the two bodies may share a common tidal origin. Therefore, both effects should be included simultaneously in a complete EMRI waveform model.

Second, as discussed in the Introduction, the EMRI-based framework developed here may also serve as the foundation for constructing waveform models for comparable-mass binary black holes through suitable mass-ratio resummation techniques. At present, however, our analysis is restricted to the adiabatic inspiral regime and therefore can only be extended reliably to the late-inspiral stage of comparable-mass systems. Constructing a complete inspiral-merger-ringdown waveform requires extending the MTF to sources moving on generic, non-adiabatic trajectories. A central challenge in this program is the reconstruction of metric perturbations generated by particles following such trajectories, although some recent work \cite{Green:2019nam, Toomani:2021jlo, Bourg:2024vre, Hollands:2024iqp, Wardell:2024yoi, Li:2026rkf} shows promise of conducting metric reconstruction for generic orbits. Addressing this problem will be an important step toward building complete binary-black-hole waveform models in higher-derivative gravity theories, and we leave this investigation to future work.

\begin{acknowledgments}
We are grateful to Pablo Cano and Luis Lehner for insightful discussions. This work makes use of the Black Hole Perturbation Toolkit. H.~Y. is supported by the Natural Science Foundation of China (Grant 12573048). N.~K. is supported by the Shuimu fellowship of Tsinghua University. D.~L. acknowledges support from the Simons Foundation (via Award No. 896696), the Simons Foundation International (via Grant No. SFI-MPS-BH-00012593-01), and the NSF (via Grants No.~PHY-2512423).
\end{acknowledgments}  

\onecolumngrid 
\appendix
\section{Explicit expressions for the source terms}\label{app:Sourceterms}

This appendix presents the explicit form of the source terms in the modified Teukolsky equations of $\Psi_{0}^{(1,1)}$ and $\Psi_{4}^{(1,1)}$, as summarized in Figs.~\ref{fig:flowchart of psi0} and \ref{fig:flowchart of psi4}, respectively. To facilitate the presentation, we first summarize the relevant NP quantities at $\mathcal{O}(\zeta^0,\eta^0)$, $\mathcal{O}(\zeta^1,\eta^0)$, and $\mathcal{O}(\zeta^0,\eta^1)$, from which the source terms are constructed.  For notational convenience, we introduce $f(r)=1-2M/r$, which will be used throughout this appendix.

At $\mathcal{O}(\zeta^0,\eta^0)$, the only nonzero Weyl scalars and spin coefficients are
\begin{align}
    \begin{split}
    \rho&=-\frac{1}{2r}f(r)\,,\quad 
    \mu=-\frac{1}{r}\,,\quad 
    \varepsilon=\frac{M}{2r^2},\quad \alpha
    =-\beta=-\frac{\cot\theta}{2\sqrt{2}r}\,,\quad
    \Psi_2=-\frac{M}{r^3}\,.
    \end{split}
\end{align}
The nonzero spin coefficients at $\mathcal{O}(\zeta^1,\eta^0)$ are
\begin{align}
    \begin{split}
    \rho^{(1,0)}&=-\frac{1}{4r}f(r)\left(-2\mathcal{H}_1(r)+r \mathcal{H}_3^\prime(r)\right)\,,\quad 
    \mu^{(1,0)}=-\frac{1}{2r}\left(-2\mathcal{H}_1(r)+r \mathcal{H}_3^\prime(r)\right)\,,\\ 
    \varepsilon^{(1,0)}
    &=-\frac{M}{2r^2}\mathcal{H}_1(r)\,,\quad
    \alpha^{(1,0)}=-\beta^{(1,0)}=\frac{\cot\theta}{4\sqrt{2}r}\mathcal{H}_3(r)\,.
    \end{split}
\end{align}
At $\mathcal{O}(\zeta^1,\eta^0)$, the only nonzero Weyl scalars are
\begin{align}
\begin{split}
    \Psi_2^{(1,0)}
    &=-\frac{1}{3r^2}\left(1-\frac{6M}{r}\right)\mathcal{H}_1(r)
    +\frac{1}{6r^2}\mathcal{H}_3(r)
    +\frac{1}{6r}\left(1-\frac{3M}{r}\right)\mathcal{H}^\prime_1(r)
    -\frac{M}{6r^2}\mathcal{H}^\prime_3(r)
    -\frac{1}{12}f(r)\mathcal{H}^{\prime\prime}_3(r)\,,
\end{split}
\end{align}
while the nonzero $\Phi_{ij}^{(1,0)}$ are given by
\begin{align}
\begin{split}
    \Phi_{00}^{(1,0)}=\frac{162M^6}{r^8}f^2(r)\,,\quad 
    \Phi_{11}^{(1,0)}=-\frac{54M^6}{r^8}f(r)\,,\quad \Phi_{22}^{(1,0)}=\frac{648 M^6}{r^8}\,,\quad \Lambda^{(1,0)}=\frac{-212 M^7+90M^6r}{r^9}\,.
\end{split} 
\end{align}
The $\Phi_{ij}$ at $\mathcal{O}(\zeta^0,\eta^1)$ are
\begin{subequations}
\begin{align}
    \Phi_{00}^{(0,1)}
    &=\frac{\pi m_p u^t(r-2 M)^2}{r^4}\delta(r-r_0)
    \delta\left(\theta-\frac{\pi }{2}\right)
    \delta\!\left[\phi-\omega_z\left(v-r_\star(r_0)\right)\right]\,,\\
    \Phi_{01}^{(0,1)}
    &=-\frac{i\sqrt{2}\pi m_p\omega_z u^t(r-2M)}{r^2}
    \delta(r-r_0) \delta\left(\theta -\frac{\pi }{2}\right)
    \delta\!\left[\phi-\omega_z\left(v-r_\star(r_0)\right)\right]\,,\\
     \Phi_{10}^{(0,1)}
     &=\frac{i\sqrt{2}\pi m_p\omega_z u^t(r-2 M)}{r^2}
     \delta(r-r_0)\delta\left(\theta-\frac{\pi}{2}\right)
     \delta\!\left[\phi-\omega_z\left(v-r_\star(r_0)\right)\right]\,,\\ 
    \Phi_{11}^{(0,1)}
    &=\frac{\pi m_p u^t\left(-2M+\omega_z^2r^3+r\right)}{r^3}
    \delta(r-r_0)\delta\left(\theta-\frac{\pi}{2}\right)
    \delta\!\left[\phi-\omega_z\left(v-r_\star(r_0)\right)\right]\,, \\
    \Phi_{02}^{(0,1)}
    &=-2\pi m_p \omega_z^2 u^t\delta(r-r_0)
    \delta\left(\theta-\frac{\pi}{2}\right)
    \delta\!\left[\phi-\omega_z\left(v-r_\star(r_0)\right)\right]\,, \\
    \Phi_{12}^{(0,1)}
    &=-\frac{2i\sqrt{2}\pi m_p \omega_z u^t}{r}\delta(r-r_0) \delta\left(\theta-\frac{\pi}{2}\right)
    \delta\!\left[\phi-\omega_z\left(v-r_\star(r_0)\right)\right]\,, \\
    \Phi_{20}^{(0,1)}
    &=-2\pi m_p \omega_z^2 u^t\delta(r-r_0)
    \delta\left(\theta-\frac{\pi }{2}\right) 
    \delta\!\left[\phi-\omega_z\left(v-r_\star(r_0)\right)\right]\,, \\
    \Phi_{21}^{(0,1)}
    &=\frac{2i\sqrt{2}\pi m_p\omega_z u^t}{r}\delta(r-r_0) \delta\left(\theta-\frac{\pi }{2}\right) 
    \delta\!\left[\phi-\omega_z\left(v-r_\star(r_0)\right)\right]\,, \\
    \Phi_{22}^{(0,1)}
    &=\frac{4\pi m_p u^t }{r^4}\delta(r-r_0)
    \delta\left(\theta-\frac{\pi}{2}\right)\delta\!\left[\phi-\omega_z\left(v-r_\star(r_0)\right)\right]\,,\\
    \Lambda^{(0,1)}
    &=\frac{\pi m_p u^t\left(-2M-\omega_z^2r^3+r\right)}{3r^3}
    \delta(r-r_0)\delta\left(\theta-\frac{\pi}{2}\right) 
    \delta\!\left[\phi-\omega_z\left(v-r_\star(r_0)\right)\right]\,. \\
\end{align}    
\end{subequations}

Given the complexity of the Teukolsky equation, we adopt the notation in Eq.~\eqref{eq:auxiliary_operators} and simplify it using properties of the spin coefficients to make the expressions more compact and transparent. One convenient notation we use is introduced below:
\begin{align}
    D_{[a,b]}=D+a\varepsilon+b\rho\,, \quad
    \Delta_{[a]}=\Delta+a\mu\,, \quad
    \delta_{[a]}=\delta+a\alpha\,, \quad
    \bar{\delta}_{[a]}=\bar{\delta}+a\alpha\,,
\end{align}
We now provide the explicit source terms appearing in the modified Teukolsky equations of $\Psi_{0}^{(1,1)}$ and $\Psi_{4}^{(1,1)}$. As illustrated in Fig.~\ref{fig:flowchart of psi0}, the modified Teukolsky equation of $\Psi_0^{(1,1)}$ is
\begin{align}
    H_0\Psi_0^{(1,1)}
    =\mathcal{S}_{\rm geo}^{(1,1)} + \mathcal{S}^{(1,1)}\,,
\end{align}
where the operator $H_0$ for a Schwarzschild black hole reduces to
\begin{align}
    H_0=D_{[-2,-5]}\Delta_{[1]}-\delta_{[2]}\bar{\delta}_{[-4]}-3\Psi_{2}\,.
\end{align}
The operator $H_0^{(1,0)}$ in the geometric source term $\mathcal{S}_{\rm geo}^{(1,1)}=-H_0^{(1,0)}\Psi_0^{(0,1)}$ takes the form:
\begin{align}
    H_0^{(1,0)}
    =&\; \left[D_{[-2,-2]}^{(1,0)}-D\left(\Psi_2^{-1}\Psi_2^{(1,0)}\right)
    -\Psi_2^{-1}D^{(1,0)}\Psi_2\right]
    \Delta_{[1]}+D_{[-2,-5]}\Delta_{[1,-4]}^{(1,0)}-\delta_{[2]}^{(1,0)}\bar{\delta}_{[-4]}-\delta_{[2]}\bar{\delta}_{[-4]}^{(1,0)}-3\Psi_{2}^{(1,0)}\,, 
\end{align}
while different parts of the source term $\mathcal{S}^{(1,1)}$ driven by the stress-energy tensor are
\begin{subequations}
\begin{align}  
    \mathcal{E}_2^{(1,0)}S_2^{(0,1)}
    -\mathcal{E}_1^{(1,0)}S_1^{(0,1)}
    =& \;\left[D_{[-2,-2]}^{(1,0)}
    -D\left(\Psi_2^{-1}\Psi_2^{(1,0)}\right)
    -\Psi_2^{-1}D^{(1,0)}\Psi_2\right]
    \left(\delta_{[2]}\Phi_{01}^{(0,1)}
    -D_{[0,-1]}\Phi_{02}^{(0,1)}\right) \nonumber\\
    & \;-\delta_{[2]}^{(1,0)}\left(\delta\Phi_{00}^{(0,1)}
    -D_{[-2,-2]}\Phi_{01}^{(0,1)}\right)\,, \\ 
    \mathcal{E}_2S_{2A}^{(1,1)}
    -\mathcal{E}_1S_{1A}^{(1,1)}
    =& \;D_{[-5]}\left(-\bar{\lambda}^{(0,1)}\Phi_{00}^{(1,0)}
    +2\sigma^{(0,1)}\Phi_{11}^{(1,0)}\right)
    -\delta_{[2]}\left(\delta_{[-2,-2,1,0]}^{(0,1)}\Phi_{00}^{(1,0)}
    -2\kappa^{(0,1)}\Phi_{11}^{(1,0)}\right)\,, \\   
    \mathcal{E}_2S_{2B}^{(1,1)}
    -\mathcal{E}_1S_{1B}^{(1,1)}
    =& \;D_{[-5]}\left(\delta_{[2]}^{(1,0)}\Phi_{01}^{(0,1)}
    -D_{[-1]}^{(1,0)}\Phi_{02}^{(0,1)}\right)
    -\delta_{[2]}\left(\delta^{(1,0)}\Phi_{00}^{(0,1)}
    -D_{[-2]}^{(1,0)}\Phi_{01}^{(0,1)}\right)\,, \\
    \mathcal{E}_2S_{2C}^{(1,1)}
    -\mathcal{E}_1S_{1C}^{(1,1)}
    =& \;D_{[-5]}\left(\delta_{[2]}\Phi_{01}^{(1,1)}
    -D_{[-1]}\Phi_{02}^{(1,1)}\right)
    -\delta_{[2]}\left(\delta\Phi_{00}^{(1,1)}
    -D_{[-2]}\Phi_{01}^{(1,1)}\right)\,.  
\end{align}    
\end{subequations}

Similarly, as illustrated in Fig.~\ref{fig:flowchart of psi4}, the modified Teukolsky equation of $\Psi_4^{(1,1)}$ is
\begin{align}
    H_4\Psi_4^{(1,1)}
    = \mathcal{T}_{\rm geo}^{(1,1)}+\mathcal{T}^{(1,1)}\,,
\end{align}
where the operator $H_4$ is 
\begin{align}
    H_4
    =\Delta_{[5]}D_{[4,-1]}-\bar{\delta}_{[-4]}\delta_{[-4]}-3\Psi_{2}\,,
\end{align}
and $H_4$ in the geometric source term $\mathcal{T}_{\rm geo}^{(1,1)}=-H_4^{(1,0)}\Psi_4^{(0,1)}$ is 
\begin{align}
    H_4^{(1,0)}
    =&\left[\Delta_{[2]}^{(1,0)}-\Delta\left(\Psi_2^{-1}\Psi_2^{(1,0)}\right) 
    -\Psi_2^{-1}\Delta^{(1,0)}\Psi_2\right]D_{[4,-1]}
    +\Delta_{[5]}D_{[4,-1]}^{(1,0)}-\bar{\delta}_{[2]}^{(1,0)}\delta_{[-4]}-\bar{\delta}_{[2]}\delta_{[-4]}^{(1,0)}-3\Psi_{2}^{(1,0)}\,. 
\end{align}
The different parts of the source term $\mathcal{T}^{(1,1)}$ driven by the stress-energy tensor are
\begin{subequations}
\begin{align}
    \mathcal{E}_4^{(1,0)}S_4^{(0,1)}
    -\mathcal{E}_3^{(1,0)}S_3^{(0,1)}
    =& \;\left[\Delta_{[2]}^{(1,0)}
    -\Delta\left(\Psi_2^{-1}\Psi_2^{(1,0)}\right)
    -\Psi_2^{-1}\Delta^{(1,0)}\Psi_2\right]
    \left(\bar{\delta}_{[2]}\Phi_{21}^{(0,1)}
    -\Delta_{[1]}\Phi_{20}^{(0,1)}\right) \nonumber\\
    & \;-\bar{\delta}_{[2]}^{(1,0)}
    \left(\bar{\delta}\Phi_{22}^{(0,1)}
    -\Delta_{[2]}\Phi_{21}^{(0,1)}\right)\,, \\
    \mathcal{E}_4S_{4A}^{(1,1)}
    -\mathcal{E}_3S_{3A}^{(1,1)}
    =& \;\Delta_{[5]}\left(-2\lambda^{(0,1)}\Phi_{11}^{(1,0)}
    +\bar{\sigma}^{(0,1)}\Phi_{22}^{(1,0)}\right)
    -\bar{\delta}_{[-4]}
    \left(\bar{\delta}_{[2,2,0,-1]}^{(0,1)}\Phi_{22}^{(1,0)}
    +2\nu^{(0,1)}\Phi_{11}^{(1,0)}\right)\,, \\   
    \mathcal{E}_4S_{4B}^{(1,1)}
    -\mathcal{E}_3S_{3B}^{(1,1)}
    =& \;\Delta_{[5]}\left(\bar{\delta}_{[2]}^{(1,0)}\Phi_{21}^{(0,1)}
    -\Delta_{[1]}^{(1,0)}\Phi_{20}^{(0,1)}\right)
    -\bar{\delta}_{[-4]}\left(\bar{\delta}^{(1,0)}\Phi_{22}^{(0,1)}
    -\Delta_{[2]}^{(1,0)}\Phi_{21}^{(0,1)}\right)\,, \\
    \mathcal{E}_4S_{4C}^{(1,1)}
    -\mathcal{E}_3S_{3C}^{(1,1)}
    =& \;\Delta_{[5]}\left(\bar{\delta}_{[-4]}\Phi_{21}^{(1,1)}
    -\Delta_{[1]}\Phi_{20}^{(1,1)}\right)
    -\bar{\delta}_{[-4]}\left(\bar{\delta}\Phi_{22}^{(1,1)}
    -\Delta_{[2]}\Phi_{21}^{(1,1)}\right)\,. 
\end{align}   
\end{subequations}

\section{Logarithmic gauge artifacts in the asymptotic expansion}
\label{app:Logarithmic }

In this appendix, we provide an intuitive explanation for the divergence of the source terms by examining the asymptotic structure of the modified Teukolsky equations. As we shall show, the divergent source terms generate logarithmic contributions in the asymptotic expansion of the solutions.

We begin with the behavior of $\Psi_0^{(1,1)}$ near the horizon in Schwarzschild coordinates. The modified Teukolsky equation takes the form
\begin{align}
   H_0^{(0,0)}\Psi_0^{(1,1)}
   = -H_0^{(1,0)}\Psi_0^{(0,1)}+\mathcal{S}_{A}^{(1,1)}
   +\mathcal{S}_{B}^{(1,1)}\,,
\end{align}
where we restore the superscript $(0,0)$ of $\mathcal{O}(\zeta^0,\eta^0)$ quantities for clarity. Since the operator structure of
$\mathcal{S}_{\rm geo}=-H_0^{(1,0)}\Psi_0^{(0,1)}$
is representative of the divergent source terms, it is sufficient to consider this contribution as an illustrative example. The same analysis applies directly to $\mathcal{S}_A^{(1,1)}$.

Expanding the operators near the horizon yields
\begin{align}
    H_0^{(0,0)}
    =& \;\left[-\frac{r-2 M}{4 M}+\frac{(r-2 M)^2}{8 M^2}
    -\frac{(r-2 M)^3}{16 M^3}+\mathcal{O}\left((r-2 M)^4\right)\right]\partial_r^2 \notag\\
    & \;+\left[-\frac{3}{4 M}+\frac{3 (r-2 M)^2}{16 M^3}
    -\frac{3 (r-2 M)^3}{16 M^4}+\mathcal{O}\left((r-2 M)^4\right)\right]\partial_r \notag\\
    & \;+\left[\frac{-M \omega ^2+i \omega}{r-2 M}
    +\frac{\ell^2+\ell-4 M^2 \omega ^2-12 i M \omega -6}{8 M^2}
    +\mathcal{O}\left((r-2 M)^1\right)\right]\,,\\[1ex]
    H_0^{(1,0)}
    =& \;\left[(-\frac{5 (r-2 M)}{32 M^5}
    +\frac{29 (r-2 M)^2}{64 M^6}
    -\frac{59 (r-2 M)^3}{64 M^7}
    +\mathcal{O}\left((r-2 M)^4\right)\right]\partial_r^2 \notag\\
    & \;+\left[-\frac{15}{32 M^5}
    -\frac{53 (r-2 M)}{64 M^6}
    +\frac{227 (r-2 M)^2}{32 M^7}
    +\mathcal{O}\left((r-2 M)^3\right)\right]\partial_r \notag\\
    & \;+\left[\frac{-3 M \omega ^2+4 i \omega}{8 M^4 (r-2 M)}
    +\frac{5 \ell^2+5 \ell+20 M^2 \omega ^2-266 i M \omega -240}{64 M^6}
    +\mathcal{O}\left((r-2 M)^2\right)\right]\,.
\end{align}

Correspondingly, we define the series expansion of $\Psi_0$:
\begin{align}
    & \Psi_0^{(0,1)}
    =B_{\rm trans}(r-2M)^{-2-2 i M \omega }
    \left[1+a_1(r-2M)+a_2(r-2M)^2
    +\mathcal{O}\left((r-2M)^3\right)\right]\,, \\ 
    & \Psi_0^{(1,1)}
    =(r-2M)^{-\gamma}\left[c_0+c_1(r-2M)+c_2(r-2M)^2
    +\mathcal{O}\left((r-2M)^3\right)
    +L_1\log(r-2M)+L_2 (r-2M) \log(r-2M) \right]\,.
\end{align}
Substituting these expansions into the modified Teukolsky equation and matching coefficients order by order, one finds that a consistent Frobenius expansion requires the inclusion of logarithmic contributions. The Frobenius indicial equation yields $\gamma = 2 + 2iM\omega$. The asymptotic analysis determines only the recursive structure of the series and leaves the overall amplitude $c_0$ undetermined.

The appearance of logarithmic terms signals that the Schwarzschild coordinate formulation is not adapted to the physical boundary conditions. Rather than attempting to remove these contributions by imposing additional constraints, it is more convenient to work in a coordinate system and tetrad that are regular on the horizon. After transforming to ingoing Eddington--Finkelstein coordinates and adopting the Hawking--Hartle tetrad, the logarithmic contributions disappear from the asymptotic expansion.

An analogous phenomenon occurs in the asymptotic analysis of $\Psi_4^{(1,1)}$ at future null infinity before performing the $\mathcal{O}(\zeta^1,\eta^0)$ coordinate transformation. The corresponding modified Teukolsky equation is
\begin{align}
    H_4^{(0,0)}\Psi_4^{(1,1)}
    =-H_4^{(1,0)}\Psi_4^{(0,1)}+\mathcal{T}_{A}^{(1,1)}
    +\mathcal{T}_{B}^{(1,1)}\,.
\end{align}
As with $\Psi_0^{(1,1)}$, we consider only $\mathcal{T}_{\rm geo}=-H_4^{(1,0)} \Psi_4^{(0,1)}$.
\begin{align}
H_4^{(0,0)}
={}&\left(-\frac{1}{2}+\frac{M}{r}\right)\partial_r^2
+\left(\frac{7 M}{r^2}-\frac{3}{r}\right)\partial_r\notag\\
&+\left(
-\frac{\omega ^2}{2}
+\frac{-M \omega ^2+2 i \omega}{r}
+\frac{\ell^2+\ell-4 M^2 \omega ^2-4 i M \omega -6}{2 r^2}
+\mathcal{O}\!\left(\frac{1}{r^3}\right)
\right),
\\[1ex]
H_4^{(1,0)}={}&
\left(
-\frac{32}{231 M^4}
+\frac{32}{77 M^3 r}
\right)\partial_r^2
+\left(
\frac{256}{77 M^3 r^2}
-\frac{64}{77 M^4 r}
\right)\partial_r
\notag\\
&+\left(-\frac{32\left(M\omega^2-2 i\omega\right)}
       {231 M^4 r}-\frac{32\left(-\ell^2-\ell+3 M^2\omega^2+6 i M\omega+6\right)}
       {231 M^4 r^2}
+\mathcal{O}\!\left(\frac{1}{r^3}\right)
\right)\,.
\end{align}
Correspondingly, we choose the following series expansion for $\Psi_4$:
\begin{align}
    &\Psi_4^{(0,1)}=C_{trans}e^{i\omega r}r^{-1+2 i M \omega }\left(1+a_1\frac{1}{r}+a_2\frac{1}{r^2}+\mathcal{O}\left(\frac{1}{r^3}\right)\right)\,, \\ 
    &\Psi_4^{(1,1)}=e^{a r} r^\gamma\left(c_0+c_1 \frac{1}{r}+L_1\frac{\log(r)}{r}+c_2 \frac{1}{r^2}+L_2\frac{\log(r)}{r^2}+\mathcal{O}\left(\frac{1}{r^3}\right)\right)\,.
\end{align}
Substituting the asymptotic ansatz into the equation and matching powers of $1/r$, one again finds that logarithmic contributions are required for consistency. The leading asymptotic behavior is characterized by $a=i\omega, \gamma=2iM\omega$, while the coefficient $c_1$ remains undetermined by the asymptotic expansion alone. These logarithmic terms are removed after performing the $\mathcal{O}(\zeta^1,\eta^0)$ coordinate transformation. Their origin can be traced to the slow asymptotic falloff of the $\mathcal{O}(\zeta^1,\eta^0)$ metric perturbation and therefore represents a coordinate- and tetrad-dependent artifact rather than a physical feature of the spacetime.

This asymptotic analysis provides a simple explanation for the coordinate and tetrad choices adopted in the main text: in a regular gauge, the modified Teukolsky variables admit the standard power-series expansions both at the horizon and at null infinity, without logarithmic contamination.

\section{The Taylor expansion of the source term}\label{ap:series_expansion}

In this appendix, we describe how the Taylor expansion coefficients of the source terms are obtained. As discussed in Sec.~\ref{sec:solve equation} , the evaluation of the lower incomplete Gamma function requires a near-horizon expansion of the source terms. Since these source terms are constructed from the $\mathcal{O}(\zeta^0,\eta^1)$ metric perturbations, we first derive the corresponding near-horizon expansions of the metric amplitudes.

Following Refs.~\cite{Barack:2005nr, Barack:2007tm}, the perturbation system consists of ten field equations supplemented by four Lorenz gauge conditions. These fourteen equations naturally decompose into even- and odd-parity sectors. The even-parity sector contains seven field equations and three Lorenz gauge conditions for the perturbation amplitudes $h^{(1-7)}$, while the odd-parity sector contains three field equations and one Lorenz gauge condition for $h^{(8-10)}$. The two sectors can therefore be treated independently. Within each sector, the equations are arranged hierarchically in~\cite{Barack:2005nr}: at each step, the equation to be solved contains only one unknown amplitude, with coupling terms involving other amplitudes determined at earlier steps. This structure yields a straightforward sequential procedure for finding the metric perturbation components directly.

Because our calculation is performed in ingoing Eddington-Finkelstein coordinates, an additional factor $e^{-i\omega r_\star}$ must be extracted from the frequency-domain solutions. Accordingly, the perturbation amplitudes of Refs.~\cite{Barack:2005nr, Barack:2007tm} are rewritten in the form $h^{(i)}(r)e^{-i\omega v}$, and the corresponding field equations and Lorenz gauge conditions are modified consistently.

Let us Taylor expand the perturbation amplitudes about the horizon as
\begin{align}
    h^{(i)}(r)= \sum_{n} b_n^{(i)} (r-2M)^n\,, 
\end{align}
Substituting these expansions into the field equations and matching powers of $(r-2M)$ order by order yields a hierarchy of algebraic equations for the coefficients $b_n^{(i)}$. These equations can be solved recursively, expressing higher-order coefficients in terms of lower-order ones. At first sight, the resulting expansion is characterized by the ten leading coefficients $b_0^{(i)}$. Furthermore, we note that the coefficients $b_2^{(2)}$ and $b_4^{(1)}$ cannot be determined from the field equations alone. The four Lorenz gauge conditions remove this apparent redundancy. They impose the relations:
\begin{align}
    b_0^{(1)}=b_0^{(2)}\,,\qquad
    b_0^{(4)}=b_0^{(5)}\,,\qquad
    b_0^{(8)}=b_0^{(9)}\,,
\end{align}
and also determine the coefficients $b_2^{(2)}$ and $b_4^{(1)}$. Together with the field equations, they further imply
\begin{align}
b_1^{(1)}=b_1^{(2)}\,.
\end{align}
This provides an additional constraint on the leading coefficients $b_0^{(i)}$. Consequently, all Taylor coefficients can be expressed in terms of the horizon values $b_0^{(i)}$. These coefficients are not all independent because of the gauge constraints above, but they provide a convenient parametrization of the near-horizon solution. In practice, the $b_0^{(i)}$ are obtained by fitting the numerical solutions near the horizon.

Since the lower incomplete Gamma function decays rapidly for a suitably chosen matching radius $r_1$, only the first few terms in the near-horizon expansion are required in practice. Therefore, it is unnecessary to derive closed-form recursion relations. Instead, the coefficients are computed iteratively order by order using the procedure described above.

The above procedure can also be applied to compute the near-horizon series expansions of the homogeneous solutions ${}_{\pm 2}R^{\rm in}_{\ell m}$ and ${}_{\pm 2}R^{\rm up}_{\ell m}$. In this case, the calculation is considerably simpler because the homogeneous Teukolsky equation involves only a single radial function rather than a coupled system of equations. For example, the ingoing solution can be expanded as
\begin{align}
    {}_{\pm 2}R^{\rm in}_{\ell m}
    =(r-2M)^{-\gamma_2}
    \left[c_0+c_1(r-2M)+c_2(r-2M)^2
    +O\!\left((r-2M)^3\right)\right]\,.
\end{align}
Substituting this ansatz into the homogeneous Teukolsky equation and matching powers of $(r-2M)$ determines the coefficients $c_n$ recursively in terms of the leading coefficient $c_0$. The value of $c_0$ is obtained from the BHPToolkit~\cite{BHPToolkit}.


\section{Consistency checks of the modified Teukolsky calculation }\label{ap:Check_calculation}

In this appendix, we present two complementary consistency checks of the modified Teukolsky calculation developed in Secs.~\ref{sec:MTF} and~\ref{sec:solve equation}. We first compare the weak-field behavior of the energy flux at null infinity with independent post-Newtonian (PN) results. We then use the $RK$ theory as a benchmark for the source
construction and solution procedure. In this theory, a field redefinition provides two independent descriptions of the same dynamics, allowing the corresponding Weyl-scalar solutions to be compared directly. These two checks probe complementary aspects of the calculation: the PN comparison tests the weak-field scaling and relative normalization of the modified contribution to the infinity flux, while the $RK$ comparison tests the underlying source construction and Weyl-scalar solutions.

\subsection{Post-Newtonian comparison}\label{ap:PN_comparison}

In the weak-field regime, the energy flux at null infinity obtained from the modified Teukolsky calculation can be compared with independent PN results~\cite{Cano:2026hlv,Wang:2026qst}. With the overall expansion parameters factored out as in the main text, a leading relative correction at $n$ PN order takes the form
\begin{align}
    \frac{\dot E^\infty_{\rm mod}}{\dot E^\infty_{\rm GR}}= C_n\left(\frac{M}{r_0}\right)^n +\mathcal{O}\!\left[\left(\frac{M}{r_0}\right)^{n+1}\right],
    \label{eq:PN_flux_general}
\end{align}
where $C_n$ is the corresponding leading PN coefficient. For circular orbits, $M/r_0\sim v^2$, so a factor $(M/r_0)^n$ corresponds to a relative $n$PN correction.

For the parity-preserving cubic theory considered in the main text, Ref.~\cite{Cano:2026hlv} gives $n=5$ and $C_5=112$. For the $RK$ theory considered below, the leading correction instead appears at 6PN order, with $n=6$ and $C_6=288$~\cite{Wang:2026qst}. To compare these coefficients directly with our numerical results, we define $\mathcal{C}(r_0)\equiv(r_0/M)^n (\dot E^\infty_{\rm mod}/\dot E^\infty_{\rm GR})$, with $n=5$ for cubic gravity and $n=6$ for the $RK$ theory. Accordingly, in the weak-field limit $r_0/M\rightarrow\infty$, $\mathcal{C}(r_0)$ approaches $112$ for cubic gravity and $288$ for the $RK$ theory.

\begin{figure}[t]
    \centering
    \begin{minipage}[c]{0.48\columnwidth}
        \centering
        \includegraphics[width=\linewidth]
        {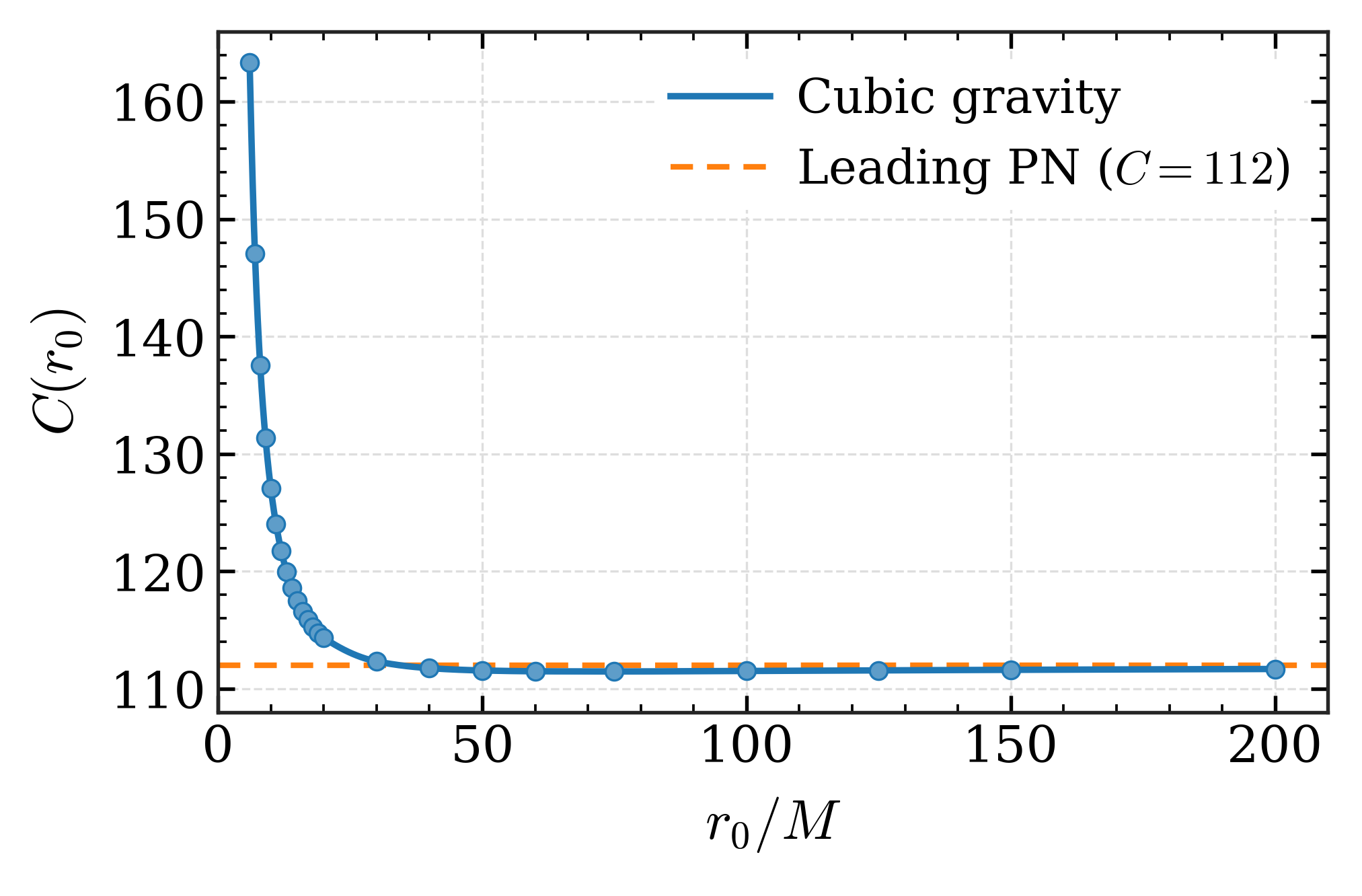}
    \end{minipage}
    \hfill
    \begin{minipage}[c]{0.48\columnwidth}
        \centering
        \includegraphics[width=\linewidth]
        {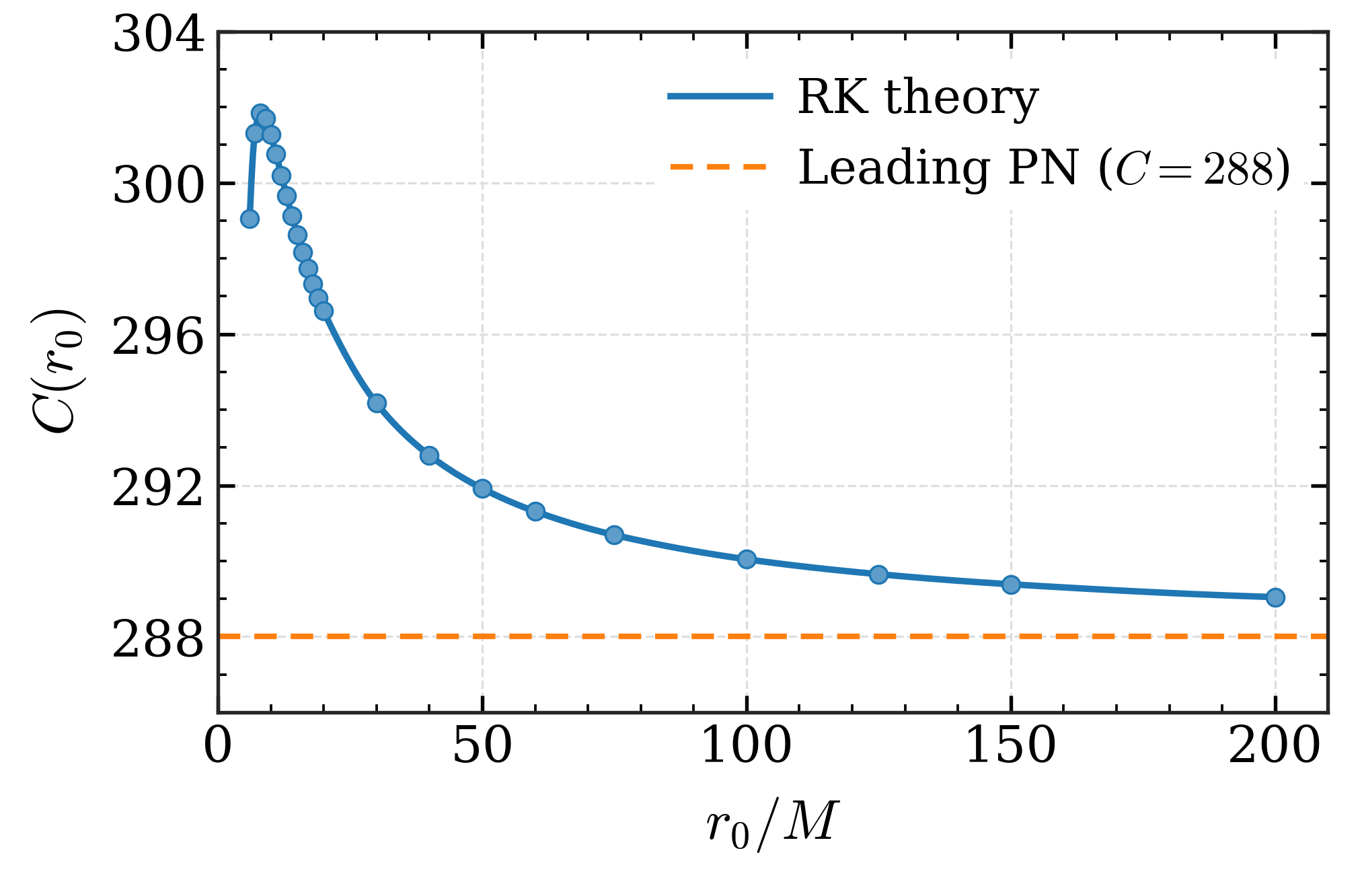}
    \end{minipage}
    \caption{
    Comparison of the modified energy flux at null infinity with the leading PN predictions. The left and right panels show the rescaled quantity $\mathcal{C}(r_0)$ for the parity-preserving cubic and $RK$ theories, respectively. The blue curves and markers denote the results obtained from the modified Teukolsky calculation, while the orange dashed lines denote the corresponding leading PN coefficients, $C_5=112$ and $C_6=288$.}
    \label{fig:PN_flux_comparison}
\end{figure}

As shown in Fig.~\ref{fig:PN_flux_comparison}, in both theories the numerical results approach the corresponding leading PN values as $r_0/M$ increases. This agreement provides an independent check of the weak-field scaling and relative normalization of the modified contribution to the infinity flux.

\subsection{Field-redefinition check in the $RK$ theory} \label{ap:RK_FR}

We next turn to a complementary check of the source construction and the solution procedure using the $RK$ theory. In this theory, the higher-curvature interaction can be removed by a field redefinition. Together with consistent tetrad choices in the two descriptions, this field redefinition establishes a direct relation between the corresponding Weyl-scalar perturbations. We can therefore construct and solve the modified Teukolsky equations independently in the two descriptions and directly compare the resulting solutions.

The action of the $RK$ theory is
\begin{align}
  S_{\rm RK}={}&\frac{1}{16\pi}\int  d^4x\sqrt{|g|}(R+\lambda_{\rm ev}l_c^4 R R_{\mu\nu\rho\sigma}R^{\mu\nu\rho\sigma}) -m_p \int d\tau .\label{eq:RK_action}
\end{align}
Throughout this appendix, as in the main text, we use the dimensionless coupling $\zeta=\lambda_{\rm ev}l_c^4/M^4$ as the perturbative expansion parameter. At linear order in $\zeta$, the higher-curvature interaction in Eq.~\eqref{eq:RK_action} can be removed by the field redefinition $g_{\mu\nu}\rightarrow (1-\lambda_{\rm ev}l_c^4 R_{\mu\nu\rho\sigma}R^{\mu\nu\rho\sigma})g_{\mu\nu}$. In the following, we refer to the descriptions before and after the field redefinition as the $RK$ and $FR$ descriptions, respectively. When necessary, the superscripts ${\rm RK}$ and ${\rm FR}$ are used to distinguish quantities in the two descriptions; otherwise, these labels are left implicit.

Under the field redefinition, the higher-curvature interaction is transferred from the bulk action to the point-particle action. Keeping terms to linear order in $\zeta$, the resulting action is
\begin{align}
    S_{\rm FR}={}& \frac{1}{16\pi}\int d^4x\,\sqrt{|g|}\,R-m_p\int\left(1-\frac{1}{2}\lambda_{\rm ev}l_c^4 R_{\mu\nu\rho\sigma}R^{\mu\nu\rho\sigma}\right)d\tau ,\\
    ={}&\frac{1}{16\pi}\int d^4x\,\sqrt{|g|}\,R-m_p\int d\tau+4m_p\lambda_{\rm ev}l_c^4\int d\tau\,E_{\mu\nu}E^{\mu\nu}-4 m_p\lambda_{\rm ev}l_c^4\int d\tau\,B_{\mu\nu}B^{\mu\nu},
\label{eq:RK_FR_action}
\end{align}
where we have used $R_{\mu\nu\rho\sigma}R^{\mu\nu\rho\sigma}=C_{\mu\nu\rho\sigma}C^{\mu\nu\rho\sigma}=8\left(E_{\mu\nu}E^{\mu\nu}-B_{\mu\nu}B^{\mu\nu}\right)$ on the Ricci-flat GR background.

We first construct the modified Teukolsky equations directly in the $RK$ description, following the procedure adopted in Sec.~\ref{sec:MTF}. Varying the higher-curvature interaction in Eq.~\eqref{eq:RK_action} with respect to the metric gives the effective stress-energy tensor
\begin{align}
    T_{\mu\nu}^{\rm RK}={}\lambda_{\rm ev}l_c^4\Big[-2R\,R_{\mu\alpha\beta\gamma}R_{\nu}{}^{\alpha\beta\gamma}-\left(R_{\mu\nu}-\frac{1}{2}g_{\mu\nu}R\right)R_{\alpha\beta\gamma\delta}R^{\alpha\beta\gamma\delta}+\nabla_{\mu}\nabla_{\nu}\left(R_{\alpha\beta\gamma\delta}R^{\alpha\beta\gamma\delta}\right)-g_{\mu\nu}\nabla_{\alpha}\nabla^{\alpha}\left(R_{\beta\gamma\delta\epsilon}R^{\beta\gamma\delta\epsilon}\right)-4\nabla^{\alpha}\nabla^{\beta}\left(R\,R_{\mu\alpha\nu\beta}\right)\Big].
\label{eq:RK_effective_stress_energy}
\end{align}

The corresponding static and spherically symmetric background metric at $\mathcal{O}(\zeta)$ is
\begin{align}
    ds^2=\left[1+\zeta  \mathcal{H}(r)\right]\left[-f(r)dt^2+\frac{dr^2}{f(r)}+r^2d\Omega^2\right],\label{eq:RK_metric}
\end{align}
where $\mathcal{H}(r)=-\frac{48M^6}{r^6}$. In these coordinates, a convenient choice for the $\mathcal{O}(\zeta^1,\eta^0)$ tetrad perturbations is
\begin{align}
\begin{split}
    & l^{\mu(1,0)}
   =-\mathcal{H}(r)\left(1,-\frac{1}{2}f(r),0,0\right)\,, \label{eq:RK_10l} \\
   & n^{\mu(1,0)}
   =\left(0,0,0,0\right)\,, \\
   & m^{\mu(1,0)}
   =-\frac{1}{2\sqrt{2}r}\mathcal{H}(r)\left(0,0,1,i \csc\theta\right )\,.
\end{split}
\end{align}

The nonvanishing spin coefficients at $\mathcal{O}(\zeta^1,\eta^0)$ are
\begin{align}
    \begin{split}
    \rho^{(1,0)}&=-\frac{1}{4r}f(r)\left(-2\mathcal{H}(r)+r \mathcal{H}^\prime(r)\right)\,,\quad 
    \mu^{(1,0)}=-\gamma^{(1,0)}=-\frac{1}{2}\mathcal{H}^\prime(r)\,,\\ 
    \varepsilon^{(1,0)}
    &=-\frac{M}{2r^2}\mathcal{H}(r)\,,\quad
    \alpha^{(1,0)}=-\beta^{(1,0)}=\frac{\cot\theta}{4\sqrt{2}r}\mathcal{H}(r)\,.
    \end{split}
\end{align}
At $\mathcal{O}(\zeta^0,\eta^1)$, we adopt the same tetrad choice and NP quantities as in the cubic-gravity calculation in Sec.~\ref{sec:MTF}. To obtain the contribution from $T_{\mu\nu}^{\rm RK}$ at $\mathcal{O}(\zeta^1,\eta^1)$, we follow the procedure described in the main text: we first project $T_{\mu\nu}^{\rm RK}$ onto the NP basis, express the resulting components entirely in terms of NP quantities, and then expand them to $\mathcal{O}(\zeta^1,\eta^1)$. For the point-particle contribution, the form of the four-velocity remains the same as at $\mathcal{O}(\zeta^0,\eta^1)$, while its components, together with the orbital radius, receive corrections. For the $RK$ background, we find
\begin{align}
    \tilde{r}_0 &= r_0 +\zeta \frac{r_0^2(r_0-3M)}{6M}\partial_r\mathcal{H}(r_0) ,\\
    \tilde{u}^t &= u^t-\zeta\frac{1}{2}\mathcal{H}(r_0)u^t .
\end{align}
Combining the geometric, higher-curvature, and point-particle contributions, the modified Teukolsky equations in the $RK$ description take the form
\begin{align}
    H_0\Psi_0^{(1,1),{\rm RK}}&=\mathcal{S}_{\rm geo}^{(1,1),\rm RK}+\mathcal{S}_{A}^{(1,1),\rm RK}+\mathcal{S}_{B}^{(1,1),\rm RK},\\
    H_4\Psi_4^{(1,1),{\rm RK}}&=\mathcal{T}_{\rm geo}^{(1,1),\rm RK}+\mathcal{T}_{A}^{(1,1),\rm RK}+\mathcal{T}_{B}^{(1,1),\rm RK}.
\end{align}
This completes the direct construction in the $RK$ description. We next repeat the calculation in the $FR$ description. Since the field redefinition also induces a transformation of the tetrad, the consistency check is most naturally performed at the level of the Weyl-scalar solutions rather than by directly comparing the two equations. We therefore postpone this comparison until the $FR$ construction is complete.

We now turn to the field-redefined description. Since the bulk action in Eq.~\eqref{eq:RK_FR_action} reduces to the Einstein-Hilbert action, there is no higher-curvature effective stress-energy tensor in the bulk, and hence the background metric receives no $\mathcal{O}(\zeta^1,\eta^0)$ correction. We accordingly choose the $\mathcal{O}(\zeta^1,\eta^0)$ tetrad perturbation to vanish, so that all NP quantities at this order vanish as well. At $\mathcal{O}(\zeta^0,\eta^1)$, we again adopt the same tetrad choice and NP quantities as in Sec.~\ref{sec:MTF}.

The particle sector, however, becomes more involved after the field redefinition, since the higher-curvature interaction is now encoded in the worldline action. Varying the point-particle terms in Eq.~\eqref{eq:RK_FR_action} with respect to the metric gives the corresponding stress-energy tensor,
\begin{align}
    T_{\mu\nu}^{\rm FR}={}&-\frac{1}{2}m_p\lambda_{\rm ev}l_c^4\int\Big[\big(-u_\mu u_\nu C_{\alpha\beta\gamma\delta}C^{\alpha\beta\gamma\delta}+4C_\mu{}^{\alpha\beta\gamma}C_{\nu\alpha\beta\gamma}\big)\delta^{(4)}\big(x-z(\tau)\big)+8C_{\mu\alpha\nu\beta}\nabla^\alpha\nabla^\beta\delta^{(4)}\!\big(x-z(\tau)\big)\Big]\frac{d\tau}{\sqrt{-g}} .
\end{align}
To extract its $\mathcal{O}(\zeta^1,\eta^1)$ contribution, one must account for both the explicit curvature-dependent worldline interaction and the $\mathcal{O}(\zeta)$ corrections to the particle motion, including those to the orbital radius and four-velocity. Unlike the calculation in the main text, these corrections are not determined by geodesic motion on a deformed background. Instead, they follow from the modified equation of motion implied by the field-redefined worldline action.

To determine these corrections, we derive the particle equation of motion directly from the worldline action in Eq.~\eqref{eq:RK_FR_action}. For this purpose, it is convenient to parametrize the worldline by an arbitrary parameter $\lambda$, such that $d\tau=\sqrt{-g_{\mu\nu}\dot{x}^{\mu}\dot{x}^{\nu}}\,d\lambda$, where $\dot{x}^{\mu}\equiv dx^{\mu}/d\lambda$. The point-particle action can then be written as $S_{p}^{\rm FR}=\int d\lambda\,L_{\rm FR},$ with
\begin{align}
    L_{\rm FR}=-m_p\left(1-\frac{1}{2}\lambda_{\rm ev}l_c^4 R_{\mu\nu\rho\sigma}R^{\mu\nu\rho\sigma}\right)\sqrt{-g_{\mu\nu}\dot{x}^{\mu}\dot{x}^{\nu}} .
\end{align}
For compactness, we define $K\equiv R_{\mu\nu\rho\sigma}R^{\mu\nu\rho\sigma}$. The derivatives of the Lagrangian with respect to $x^\mu$ and $\dot{x}^\mu$ are then
\begin{align}
    \begin{split}
    \frac{\partial L_{\rm FR}}{\partial \dot{x}^{\mu}}={}& m_p\left(1-\frac{1}{2}\lambda_{\rm ev}l_c^4 K\right)\frac{g_{\mu\nu}\dot{x}^{\nu}}{\sqrt{-g_{\alpha\beta}\dot{x}^{\alpha}\dot{x}^{\beta}}},\\
    \frac{\partial L_{\rm FR}}{\partial x^{\mu}}={}&\frac{m_p}{2}\left(1-\frac{1}{2}\lambda_{\rm ev}l_c^4 K\right)\frac{\partial_{\mu}g_{\alpha\beta}\dot{x}^{\alpha}\dot{x}^{\beta}}{\sqrt{-g_{\rho\sigma}\dot{x}^{\rho}\dot{x}^{\sigma}}}+\frac{m_p}{2}\lambda_{\rm ev}l_c^4 \sqrt{-g_{\alpha\beta}\dot{x}^{\alpha}\dot{x}^{\beta}}\,\partial_{\mu}K .
    \end{split}
\end{align}
Substituting these expressions into the Euler--Lagrange equation and subsequently choosing the proper-time parametrization $\lambda=\tau$, such that $u^\mu=dx^\mu/d\tau$, we obtain, to linear order in $\zeta$,
\begin{align}
    u^\alpha\nabla_\alpha u^\mu =\frac{1}{2}\lambda_{\rm ev}l_c^4\left(g^{\mu\nu}+u^\mu u^\nu\right)\nabla_\nu K .
\label{eq:FR_EOM}
\end{align}
Since the right-hand side of Eq.~\eqref{eq:FR_EOM} is already of $\mathcal{O}(\zeta)$, the curvature quantities appearing there can be evaluated on the zeroth-order GR background.

Solving Eq.~\eqref{eq:FR_EOM} together with the normalization condition $u^\mu u_\mu=-1$ for a circular equatorial orbit, while keeping the orbital frequency $\omega_z$ fixed, we find
\begin{align}
    \tilde{r}_0 &= r_0 +\zeta\frac{48M^5}{r_0^4 (u^{t})^2} ,\\
    \tilde{u}^t &= u^t .
\end{align}
Thus, in the $FR$ description, the orbital radius receives an $\mathcal{O}(\zeta)$ correction, whereas the time component of the four-velocity does not.

Taking these corrections into account, we can evaluate the point-particle stress-energy tensor at $\mathcal{O}(\zeta^1,\eta^1)$. As an additional check, we have verified that the resulting stress-energy tensor satisfies the covariant conservation equation at the same perturbative order, $\nabla^\mu T_{\mu\nu}^{\rm FR}=0$. Projecting this stress-energy tensor onto the NP basis then determines the corresponding $\Phi_{ij}^{(1,1),{\rm FR}}$, from which the source terms of the modified Teukolsky equations are constructed following the same procedure as in Sec.~\ref{sec:MTF}.

Since the $\mathcal{O}(\zeta^1,\eta^0)$ background-metric correction and NP quantities vanish in the $FR$ description, the resulting modified Teukolsky equations are considerably simpler than their $RK$ counterparts. In particular, there is no regular source generated by a higher-curvature bulk effective stress-energy tensor. The source is instead entirely supported on the particle worldline and consists of Dirac delta functions and their derivatives. The resulting equations for $\Psi_0^{(1,1),{\rm FR}}$ and $\Psi_4^{(1,1),{\rm FR}}$ are
\begin{align}
    \mathcal{H}_0\Psi_0^{(1,1),{\rm FR}} &= \mathcal{S}_{B}^{(1,1),\rm FR},\\
    \mathcal{H}_4\Psi_4^{(1,1),{\rm FR}} &= \mathcal{T}_{B}^{(1,1),\rm FR} .
\end{align}

Having constructed the modified Teukolsky equations independently in the $RK$ and $FR$ descriptions, we now compare the corresponding solutions. Using the field redefinition together with the $\mathcal{O}(\zeta^1,\eta^0)$ tetrad choices introduced above, the Weyl scalars in the two descriptions are related at $\mathcal{O}(\zeta^1,\eta^1)$ by
\begin{align}
    \Psi_0^{(1,1),{\rm RK}}&=\Psi_0^{(1,1),{\rm FR}}-2\mathcal{H}(r) \Psi_0^{(0,1)},
    \label{eq:RK_FR_Psi0_relation}\\
    \Psi_4^{(1,1),{\rm RK}} &=\Psi_4^{(1,1),{\rm FR}}.
    \label{eq:RK_FR_Psi4_relation}
\end{align}
If the source construction and the solution of the modified Teukolsky equations are correct, the solutions obtained independently in the two descriptions must satisfy these relations. We note that Eqs.~\eqref{eq:RK_FR_Psi0_relation} and~\eqref{eq:RK_FR_Psi4_relation} depend on the $\mathcal{O}(\zeta^1,\eta^0)$ tetrad choice, as expected since the Weyl scalars are tetrad dependent. For example, with the alternative tetrad choice $l^{\mu(1,0)}=0,n^{\mu(1,0)}=-\mathcal{H}(r)n^{\mu(0,0)},m^{\mu(1,0)}=-\frac{1}{2}\mathcal{H}(r)m^{\mu(0,0)}$, the roles of $\Psi_0$ and $\Psi_4$ in the above relations are interchanged: the field-redefinition correction appears in $\Psi_4$ rather than in $\Psi_0$. In the following, we use the tetrad choice introduced above and verify Eqs.~\eqref{eq:RK_FR_Psi0_relation} and~\eqref{eq:RK_FR_Psi4_relation} separately.

\begin{figure}[t]
    \centering
    \begin{minipage}[c]{0.48\columnwidth}
        \centering
        \includegraphics[width=\linewidth]{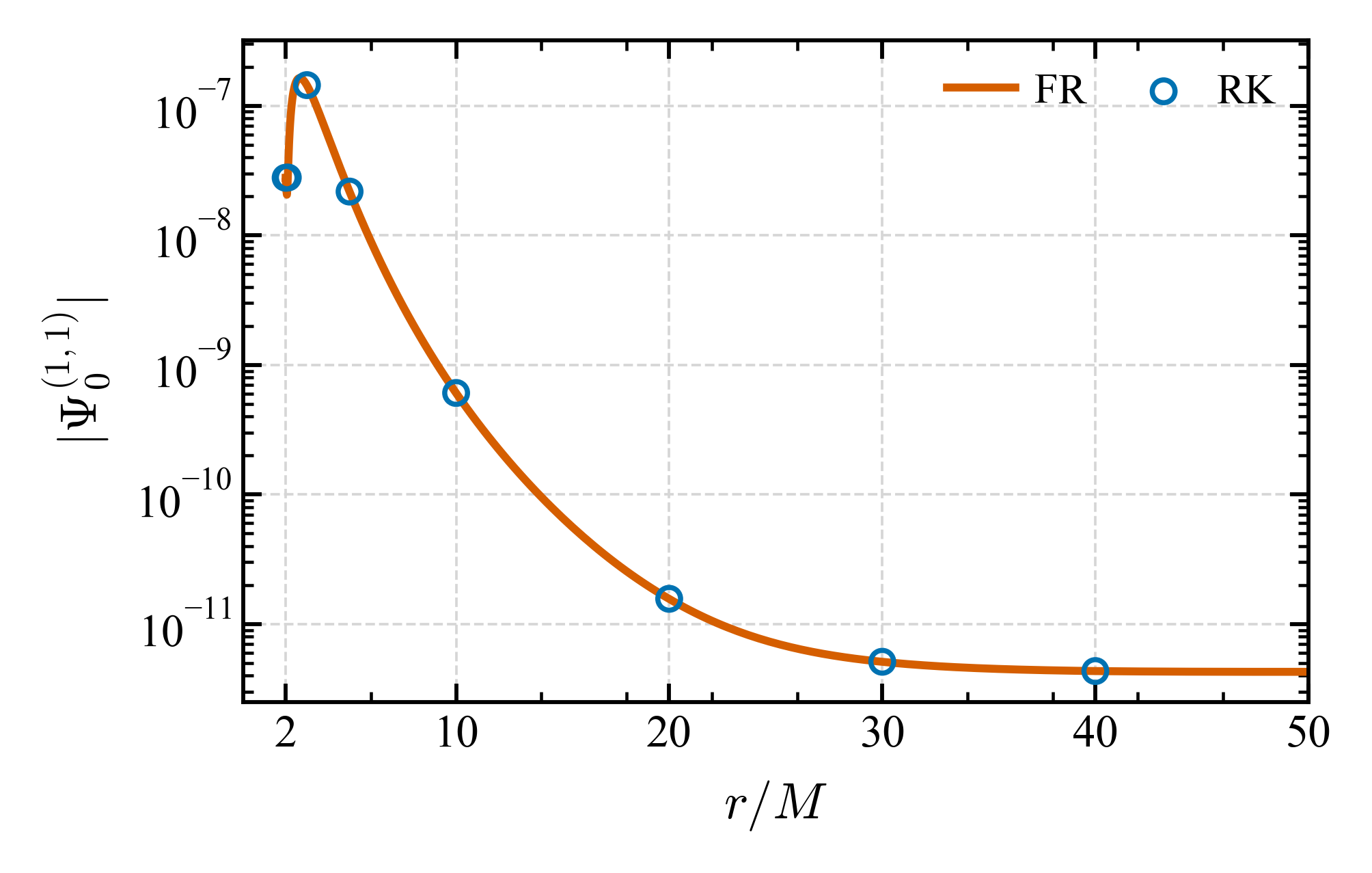}
    \end{minipage}
    \hfill
    \begin{minipage}[c]{0.48\columnwidth}
        \centering
        \includegraphics[width=\linewidth]{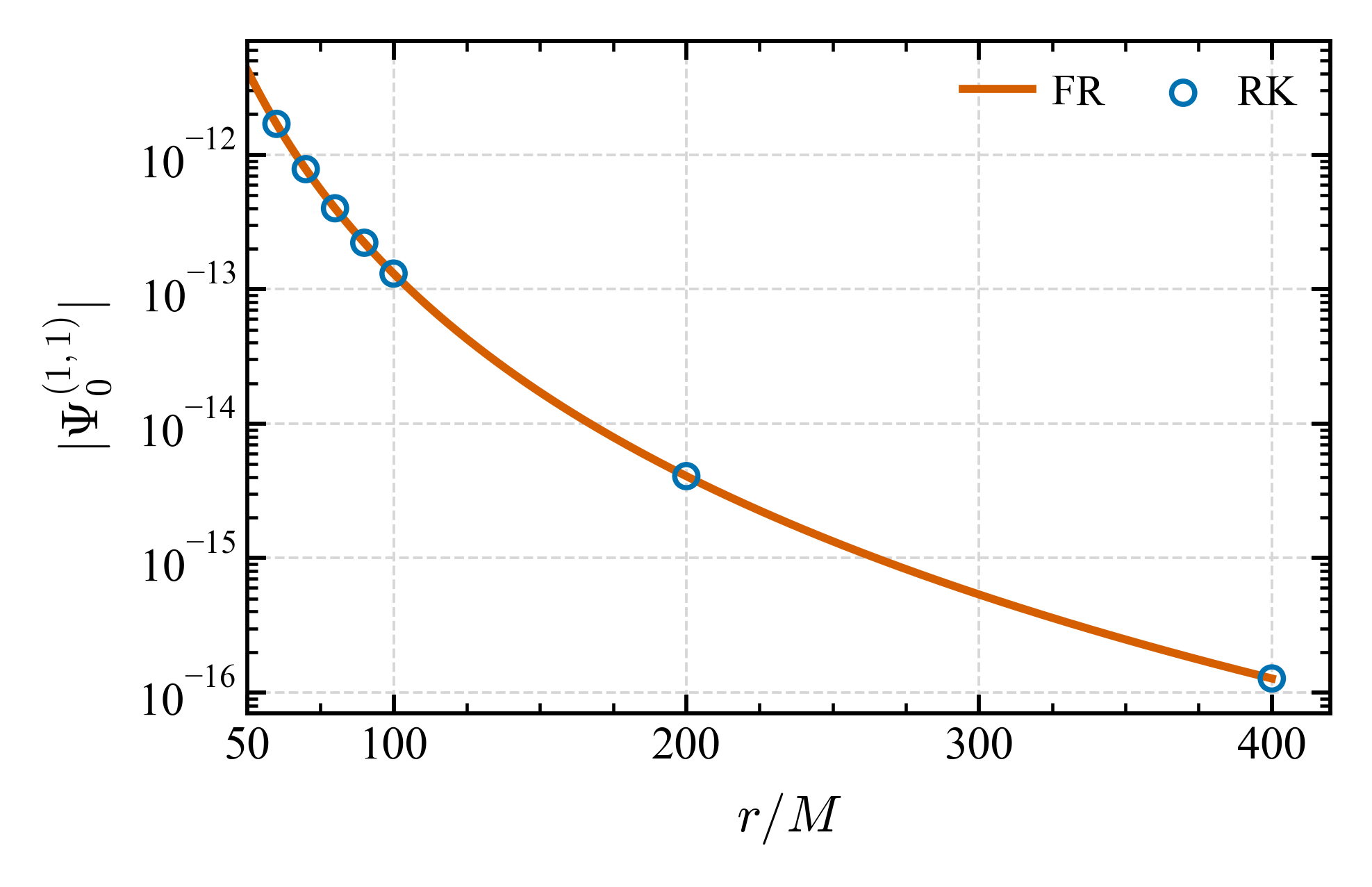}
    \end{minipage}
    \caption{
    Verification of the field-redefinition relation for $\Psi_0^{(1,1)}$. The figure compares the two sides of Eq.~\eqref{eq:RK_FR_Psi0_relation}: the blue circles denote the direct $RK$ calculation of $\Psi_0^{(1,1),{\rm RK}}$, while the orange curve denotes the transformed $FR$ result, $\Psi_0^{(1,1),{\rm FR}}-2\mathcal{H}(r)\Psi_0^{(0,1)}$. The left panel shows the radial region $2M\lesssim r\lesssim r_0$, while the right panel shows the large-radius region $r_0\lesssim r\lesssim400M$.
    }
    \label{fig:RK_FR_Psi0}
\end{figure}

We first consider $\Psi_0$. Using the independently constructed solutions in the $RK$ and $FR$ descriptions, we evaluate the two sides of Eq.~\eqref{eq:RK_FR_Psi0_relation}. The agreement between
$\Psi_0^{(1,1),{\rm RK}}$ and $\Psi_0^{(1,1),{\rm FR}}-2\mathcal{H}(r)\Psi_0^{(0,1)}$ provides a direct verification of the field-redefinition relation. The comparison is shown in Fig.~\ref{fig:RK_FR_Psi0}.

We next turn to $\Psi_4$. A useful simplification occurs in the source of the $RK$ equation. For the $\mathcal{O}(\zeta^1,\eta^0)$ tetrad adopted above, we find a cancellation between $\mathcal{T}_{\rm geo}^{(1,1),{\rm RK}}$ and $\mathcal{T}_{A}^{(1,1),{\rm RK}}$. Although $\mathcal{T}_{\rm geo}^{(1,1),{\rm RK}}$ and $\mathcal{T}_{A}^{(1,1),{\rm RK}}$ individually belong to the regular sector in the decomposition used in the main text, after applying the Ricci and Bianchi identities their sum can be rewritten entirely in terms of Dirac delta functions and their derivatives.

This simplification depends on the $\mathcal{O}(\zeta^1,\eta^0)$ tetrad choice. If the alternative tetrad introduced above is adopted instead, the analogous cancellation occurs in the source of the $\Psi_0$ equation rather than that of $\Psi_4$. This behavior is likely related to the particularly simple structure of the $RK$ background, which is conformally related to the Schwarzschild metric, together with the corresponding $\mathcal{O}(\zeta^1,\eta^0)$ tetrad choice. Correspondingly, the relations between the $\mathcal{O}(\zeta^1,\eta^0)$ and $\mathcal{O}(\zeta^0,\eta^0)$ NP quantities take a particularly simple form.

For the tetrad adopted in our calculation, the complete sources of the $\Psi_4$ equations in both the $RK$ and $FR$ descriptions are therefore supported entirely on the particle worldline. Since the two equations involve the same Teukolsky operator and Eq.~\eqref{eq:RK_FR_Psi4_relation} requires $\Psi_4^{(1,1),{\rm RK}}=\Psi_4^{(1,1),{\rm FR}}$, consistency requires the corresponding source terms to agree. We find explicitly that
\begin{align}
    \mathcal{T}_{geo}^{(1,1),\rm RK}+\mathcal{T}_{A}^{(1,1),\rm RK}+\mathcal{T}_{B}^{(1,1),\rm RK}= \mathcal{T}_B^{(1,1),{\rm FR}},
\end{align}
providing a direct check of the $\Psi_4$ source construction.

The agreement between the independently constructed $RK$ and $FR$ descriptions for both $\Psi_0$ and $\Psi_4$ therefore provides a nontrivial consistency check of the source construction and the solution procedure used in the modified Teukolsky calculation. Together with the PN comparison above, these results provide complementary checks of the modified Teukolsky construction and of the resulting energy flux at null infinity.

\section{Expression for $\Phi_{00}$}
\label{app:phi00}

In this appendix, we present the explicit expression for $\Phi_{00}$ relevant to the horizon-flux calculation and demonstrate how the Ricci and Bianchi identities imply the vanishing of the $\mathcal{O}(\zeta^1,\eta^1)$ contribution. Although the detailed form of $\Phi_{00}$ depends on the underlying modified gravity theory, the argument presented below illustrates how the potentially nonvanishing $\mathcal{O}(\zeta^1,\eta^1)$ terms cancel on the horizon. In more general theories, both the $\mathcal{O}(\zeta^1,\eta^1)$ and $\mathcal{O}(\zeta^1,\eta^2)$ contributions may be nonzero, but only the latter contributes to the averaged horizon flux after oscillatory terms are discarded.

Using the notation introduced in Appendix~\ref{app:Sourceterms}, the $\mathcal{O}(\zeta^1,\eta^1)$ contribution to $\Phi_{00}$ is
\begin{align}
\begin{split}
    \Phi_{00}^{(1,1)}
    =&-12\Psi_2\left[
    \delta_{[-2]}\delta_{[-4]}\bar{\Psi}_0^{(0,1)}
    +\bar{\delta}_{[-2]}\bar{\delta}_{[-4]}\Psi_0^{(0,1)}
    -2D_{[-2,0]}D\left(\Psi_2^{(0,1)}+\bar{\Psi}_2^{(0,1)}\right)
    +\delta_{[-2]}D_{[-2,0]}\bar{\Psi}_1^{(0,1)}
    +\bar{\delta}_{[-2]}D_{[-2,0]}\Psi_1^{(0,1)}\right]\,.
\end{split}
\end{align}
Since the horizon flux is evaluated on the event horizon, we may use the horizon condition $\kappa=0$. Furthermore, the quantities $\Phi_{ij}^{(0,1)}$ do not contribute on the horizon. The Ricci and Bianchi identities, therefore, reduce to
\begin{align}
    & D\tau^{(0,1)}=\Psi_1^{(0,1)}\,, \\
    & \left(\bar{\delta}-2\alpha\right) \tau^{(0,1)}=\Psi_2^{(0,1)}\,, \\
    &\left( D-2\varepsilon \right)\Psi_1^{(0,1)}-\left(\delta-4\alpha\right)\Psi_0^{(0,1)}=0\,, \\
    &  \left(\bar{\delta}-2\alpha\right)\Psi_1^{(0,1)}-D\Psi_2^{(0,1)}=0\,.
\end{align}
Substituting these relations into the expression for $\Phi_{00}^{(1,1)}$, we find that all terms cancel identically, implying $\Phi_{00}^{(1,1)}=0$ and consequently $\rho^{(1,1)}=0$.

At $\mathcal{O}(\zeta^1,\eta^2)$, the corresponding expression is more complicated. After repeatedly applying the Ricci and Bianchi identities, $\Phi_{00}^{(1,2)}$ can be reduced to
\begin{align}
    \begin{split} 
    \Phi_{00}^{(1,2)}=& 12\left(D_{[-4]}\Psi_0^{(0,1)}D_{[4]}\Psi_4^{(0,1)}+D_{[-4]}\bar{\Psi}_0^{(0,1)}D_{[4]}\bar{\Psi}_4^{(0,1)}\right)\\& +12\left[3\left(D\Psi_2^{(0,1)}\right)^2+3\left(D\bar{\Psi}_2^{(0,1)}\right)^2-4\left(\bar{\delta}\Psi_2^{(0,1)}+\bar{\delta}^{(0,1)}\Psi_2\right)\bar{\delta}_{[-4]}\Psi_0^{(0,1)}-4\left(\delta\Psi_2^{(0,1)}+\delta^{(0,1)}\Psi_2\right)\delta_{[-4]}\bar{\Psi}_0^{(0,1)}\right]\,.  \label{eq:phi0012}
    \end{split}
\end{align}
Unlike the $\mathcal{O}(\zeta^1,\eta^1)$ contribution, the above expression does not vanish identically and therefore contributes to the horizon-flux correction discussed in the main text.

\section{Transformation to the horizon adapted tetrad}
\label{app:gauge transformation}

As discussed in Sec.~\ref{sec:horizon flux}, the tetrad employed in the computation of the modified Teukolsky equation does not satisfy the requirement that $l^\mu$ coincide with a generator of the event horizon. Consequently, the horizon-flux calculation must be performed in a horizon-adapted tetrad. In this appendix, we construct the required transformation explicitly.

To this end, we first determine the generator of the perturbed event horizon. Since our calculation is performed in ingoing Eddington-Finkelstein coordinates using the Hawking-Hartle tetrad, the background vector $l^{\mu}$ already generates the event horizon. Furthermore, Eq.~\eqref{eq:10l} implies that $l^{\mu(1,0)}$ vanishes on the horizon. Because the $\mathcal{O}(\zeta^1,\eta^0)$ correction does not shift the location of the event horizon, which remains at $r=2M$, only the $\mathcal{O}(\zeta^0,\eta^1)$ perturbation contributes to the correction of the horizon generator.

To determine this correction, we first identify the location of the perturbed event horizon. Since the $\mathcal{O}(\zeta^0,\eta^1)$ perturbation does not change either the mass or the spin of the black hole, the event horizon must reduce to $r=2M$ when the perturbation is removed. We therefore parameterize the perturbed horizon as
\begin{align}
 f(v,r,\theta,\phi)=r-2M-F(v,\theta,\phi)^{(0,1)}=0\,,
\end{align}
where $F^{(0,1)}$ is linear in the metric perturbation $h_{ab}^{(0,1)}$. The normal covector to this hypersurface is then 
\begin{align}
    N_\mu=(0,1,0,0)+\left(-\frac{\partial F(v,\theta,\phi)^{(0,1)}}{\partial v},0,-\frac{\partial F(v,\theta,\phi)^{(0,1)}}{\partial \theta},-\frac{\partial F(v,\theta,\phi)^{(0,1)}}{\partial \phi}\right)\,.
\end{align}
Imposing the null condition on $N^\mu$ yields
\begin{align}
    \partial_v F(v,\theta,\phi)^{(0,1)}-\frac{F(v,\theta,\phi)^{(0,1)}}{4M}=-\frac{1}{2}h^{rr(0,1)}|_{r=2M}=-\frac{m_p}{8M}\sum_{\ell=0}^{\infty}\sum_{m=-\ell}^{\ell}h^{(1)}_{\ell m}|_{r=2M}Y^{\ell m}(\theta,\phi)e^{-i \omega v}\,.
\end{align}
Integrating this equation yields
\begin{align}
    F(v,\theta,\phi)^{(0,1)}=\frac{m_p }{2(4i\omega M+1)}\sum_{\ell=0}^{\infty}\sum_{m=-\ell}^{\ell}h^{(1)}_{\ell m}|_{r=2M}Y^{\ell m}(\theta,\phi)e^{-i \omega v}\,.
\end{align}
Here, the integration constant is fixed by requiring that the perturbed horizon reduce to $r=2M$ in the limit $h_{ab}^{(0,1)}\to0$.

The corresponding generator of the perturbed event horizon is therefore
\begin{align}
\begin{split}
    l^\mu=&(1,0,0,0)\\+&\left(-h^{vr(0,1)}|_{r=2M},-\frac{1}{2}h^{rr(0,1)}|_{r=2M}+\frac{ F(v,\theta,\phi)^{(0,1)}}{4M},-h^{\theta r(0,1)}|_{r=2M}-\frac{1}{4M^2} \frac{\partial F(v,\theta,\phi)^{(0,1)}}{\partial \theta},-h^{\phi r(0,1)}|_{r=2M}-\frac{1}{4M^2 \sin^2\theta}\frac{\partial F(v,\theta,\phi)^{(0,1)}}{\partial \phi}\right)\,.\label{eq:generator}
\end{split}
\end{align}
Comparing Eq.~\eqref{eq:01l} with the above expression, using the on-horizon identity $h^{rr(0,1)}|_{r=2M}=h^{(0,1)}_{ll}|_{r=2M}$, together with the form of $n^{\mu}$ given in Eq.~\eqref{eq:n00}, we find that the $l^{r(0,1)}$ component in Eq.~\eqref{eq:generator} coincides with that in Eq.~\eqref{eq:01l}. Consequently, no additional coordinate transformation is required, and the horizon-adapted tetrad may be obtained solely through a tetrad rotation.

We therefore seek a tetrad rotation that maps the tetrad used in the modified Teukolsky calculation to one whose null vector $l^\mu$ coincides with the horizon generator given in Eq.~\eqref{eq:generator}. Recall that, in deriving the modified Teukolsky equation, the $\mathcal{O}(\zeta^0,\eta^1)$ tetrad was chosen so that $\Psi_{1,3}^{(0,1)} = 0$ [see Eq.~\eqref{eq:psi13rotation}]. Although there are several equivalent ways to obtain a horizon-adapted tetrad, a particularly convenient approach is to first undo this transformation at the level of the Weyl scalars and then apply the horizon-adapting rotation given below:
\begin{align}
\begin{split}
    l^{\mu(0,1)}
    \rightarrow l^{\mu(0,1)}+\bar b^{(0,1)}m^{\mu}+b^{(0,1)}\bar{m}^{\mu}
    -\delta A^{(0,1)}l^{\mu}\,,\quad 
    m^{\mu(0,1)}\rightarrow m^{\mu(0,1)}+b^{(0,1)}n^{\mu}\,.
\end{split}
\end{align}
For the present problem, it is sufficient to apply the above null rotation, chosen so that the transformed $l^\mu$ coincides with the perturbed horizon generator. The remaining classes of tetrad rotations either leave the horizon generator unchanged or do not contribute to the final expression for the horizon flux. The corresponding rotation parameters are determined by requiring that the transformed tetrad reproduce the horizon generator in Eq.~\eqref{eq:generator}, which yields

\begin{align}
    &\delta A^{(0,1)}=h^{vr(0,1)}|_{r=2M}\,, \\
    &\frac{1}{2\sqrt{2}M}\left(\overline{b}^{(0,1)}+b^{(0,1)}\right)=-h^{\theta r(0,1)}|_{r=2M}-\frac{1}{4M^2} \frac{\partial F(v,\theta,\phi)^{(0,1)}}{\partial \theta}\,, \\
    &\frac{i}{2\sqrt{2}M \sin\theta}\left(\overline{b}^{(0,1)}-b^{(0,1)}\right)=-h^{\phi r(0,1)}|_{r=2M}-\frac{1}{4M^2 \sin^2\theta} \frac{\partial F(v,\theta,\phi)^{(0,1)}}{\partial \phi}\,.
\end{align}

As a consistency check, we evaluate $\kappa^{(0,1)}$ in the transformed tetrad. After accounting for the perturbative displacement of the event horizon, we find that $\kappa^{(0,1)}$ vanishes on the horizon, confirming that the transformed tetrad is indeed adapted to the horizon generators. This calculation also fixes the required $m^{r(0,1)}$ component entering the horizon-flux computation.

Having determined the complete transformation, we now examine its effect on the horizon-flux formula in Eq.~\eqref{eq:horizonflux}. After expressing $\rho$, $\sigma$, and $\Phi_{00}$ in terms of Weyl scalars and $\varepsilon$, and recalling that both the $\mathcal{O}(\zeta^0,\eta^0)$ and $\mathcal{O}(\zeta^1,\eta^0)$ backgrounds are of Petrov type D, the only quantities affected by the transformation are $\delta^{(0,1)}\Psi_2$ and $\bar{\delta}^{(0,1)}\Psi_2$.
Under the above tetrad transformation, they transform as
\begin{align}
   & \delta^{(0,1)}\Psi_2\rightarrow \delta^{(0,1)}\Psi_2+b^{(0,1)}\Delta\Psi_2\,,\\
   & \bar{\delta}^{(0,1)}\Psi_2\rightarrow \bar{\delta}^{(0,1)}\Psi_2+\bar{b}^{(0,1)}\Delta\Psi_2\,.
\end{align}
The above construction completely determines the horizon-adapted tetrad used in the flux calculation.

\twocolumngrid 
 
\bibliography{reference_new}

\end{document}